\documentclass[12pt]{article}

\usepackage{a4}
\usepackage{wrapfig}
\usepackage{boxedminipage}
\usepackage{boites} 
\usepackage{pifont}
\usepackage{euscript}
\usepackage{boxedminipage}
\usepackage{dcolumn}
\usepackage{float}
\usepackage{fancybox}
\usepackage{rotating}
\usepackage{multirow}
\usepackage{latexsym}
\usepackage{amssymb}
\usepackage{amsmath}
\usepackage{epsfig} 
\usepackage{verbatim}
\usepackage{subfigure}
\usepackage{mathrsfs}
\usepackage{url}
\usepackage{lscape}
\usepackage{psfrag}
\usepackage[all]{xy}
\usepackage{enumitem}

\usepackage{graphicx}
\usepackage[english]{babel}

\newcommand{\ix}{\mathrm{i}}
\newcommand{\matlab}{\mathrm{M}\mathrm{{\scriptstyle ATLAB}}}

\newtheorem{example}{Example}

\newtheorem{alg}{Algorithm}[section]
\newtheorem{theorem}{Theorem}[section]

\newcommand{\ealg}{\end{alg}}
\newcommand{\balg}{\begin{alg}}

\newcommand{\bigO}{\EuScript{O}}

\newcommand{\bmu}{\vect{\mu}}

\newcommand{\bs}{\mathbf{s}}

\newcommand{\T}{{\top}}

\newcommand{\bX}{\mathbf{X}}
\newcommand{\bx}{\mathbf{x}}

\newcommand{\br}{\mathbf{r}}
\newcommand{\by}{\mathbf{y}}
\newcommand{\bZ}{\mathbf{Z}}
\newcommand{\bu}{\mathbf{u}}

\newcommand{\bY}{\mathbf{Y}}

\newcommand{\bW}{\mathbf{W}}

\newcommand{\bt}{\mathbf{t}}

\renewcommand{\tilde}{\widetilde}

\newcommand{\ben}{\begin{enumerate}}
\newcommand{\een}{\end{enumerate}}
\newcommand{\beq}{\begin{equation}}
\newcommand{\eeq}{\end{equation}}
\newcommand{\ei}{\end{itemize}}
\newcommand{\bi}{\begin{itemize}}
\newcommand{\bex}{\begin{example}}
\newcommand{\eex}{\end{example}}
\newcommand{\berem}{\begin{remark}}
\newcommand{\erem}{\end{remark}}
\newcommand{\beprop}{\begin{proposition}}
\newcommand{\eprop}{\end{proposition}}

\newcommand{\bh}{\mathbf{h}}

\newcommand{\Var}{\text{Var}}

\newcommand{\Cov}{\text{Cov}}
\newcommand{\cov}{\Cov}

\newcommand{\bC}{\mathbf{C}}

\renewcommand{\epsilon}{\varepsilon}
\renewcommand{\rho}{\varrho}
\renewcommand{\log}{\ln}

\renewcommand{\leq}{\leqslant}
\renewcommand{\geq}{\geqslant}

\newcommand{\Poi}{{\sf Poi}}
\newcommand{\poi}{\Poi}
\newcommand{\Po}{\Poi}

\newcommand{\U}{{\sf U}}

\newcommand{\Nor}{{\sf N}}
\newcommand{\nor}{\Nor}

\newcommand{\Em}{\mathbb E}
\newcommand{\Pm}{\mathbb P}
\newcommand{\R}{\mathbb R}

\newcommand{\cE}{{\cal E}}

\newcommand{\scN}{\mathscr{N}}

\newcommand{\scT}{\mathscr{T}}

\newcommand{\scX}{\mathscr{X}}

\newcommand{\gvn}{\,|\,}

\newcommand{\e}{\text{e}}

\newcommand{\vect}[1]{\boldsymbol #1}

\newcommand{\di}{\text{d}}

\newcommand{\cG}{{{\cal G}}}

\def\acro#1#2{\vskip4pt\hbox to\textwidth{\normalsize
\hbox to5pc{#1\hfill}\vtop{\advance\hsize by
-5pc\raggedright\noindent#2}}}

\def\symbol#1#2{\vskip4pt\hbox to\textwidth{\normalsize
\hbox to5pc{#1\hfill}\vtop{\advance\hsize by
-5pc\raggedright\noindent#2}}}

\newcommand{\I}{\text{I}}

\newcommand{\iidsim}{\stackrel{\text{iid}}{\sim}}
\newcommand{\simiid}{\iidsim}

\newcommand{\diag}{\text{diag}}

\newcommand{\chk}[1]{}

\newcommand{\idef}{\stackrel{\text{def}}{=}}

\begin{document}
\title{\vspace*{-1cm}Spatial Process Generation}
\author{Dirk P. Kroese\footnote{School of Mathematics and Physics, The
University of Queensland, Brisbane 4072, Australia} \and Zdravko
I. Botev\footnote{School of Mathematics and Statistics, The University
of New South Wales, Sydney 2052, Australia}}
\date{}
\maketitle

\section{Introduction}
Spatial processes are mathematical models for spatial data; that is,
spatially arranged measurements and patterns. The collection and
analysis of such data is of interest to many scientific and
engineering disciplines, including the earth sciences, materials
design, urban planning, and astronomy. Examples of spatial data are
geo-statistical measurements, such as groundwater contaminant
concentrations, temperature reports from different cities, maps of the
locations of meteorite impacts or geological faults, and satellite
images or demographic maps. The availability of fast computers and
advances in Monte Carlo simulation methods have greatly enhanced the
understanding of spatial processes. The aim of this chapter is to give
an overview of the main types of spatial processes, and show how to
generate them using a computer.

 From a mathematical point of view, a spatial process is a collection
of random variables $\{X_{\bt}, \bt \in \scT\}$ where the {\bf index set}
$\scT$ is some subset of the $d$-dimensional Euclidean space
$\R^d$. Thus, $X_\bt$ is a random quantity associated with a spacial
position $\bt$ rather than time. The set of possible values of $X_\bt$
is called the {\bf state space} of the spatial process. Thus, spatial
processes can be classified into four types, based on whether the
index set and state space are continuous or discrete. An example of a
spatial process with a discrete index set and a discrete state space
is the Ising model in statistical mechanics, where sites arranged on a
grid are assigned either a positive or negative ``spin''; see, for
example, \cite{mcmc:swendsen}.  Image
analysis, where a discrete number of pixels is assigned a continuous
gray scale, provides an example of a process with a discrete index set
and a continuous state space. A random configuration of points in
$\R^d$ can be viewed as an example of a spatial process with a
continuous index sets and discrete state space $\{0,1\}$. If, in
addition, continuous measurements are recorded at these points (e.g.,
rainfall data), one obtains a process in which both the index set and
the state space are continuous.

Spatial processes can also be classified according to their
distributional properties. For example, if the random variables of a
spatial process jointly have a multivariate normal distribution, the
process is said to be {\bf Gaussian}. Another important property is
the {\bf Markov} property, which deals with the local conditional
independence of the random variables in the spatial process. A
prominent class of spatial processes is that of the {\bf point
processes}, which are characterized by the random positions of points
in space. The most important example is the {\bf Poisson process},
whose foremost property is that the (random) numbers of points in
nonoverlapping sets are independent of each other. {\bf L\'evy
fields} are spatial processes
that generalize this independence property.



The rest of this chapter is organized as follows. We discuss in
Section \ref{sec:GMRF} the generation of spatial processes that are
both Gaussian and Markov.  In Section~\ref{sec:SPP} we introduce various
spatial point processes, including the Poisson, compound Poisson,
cluster, and Cox processes, and explain how to simulate these. Section~\ref{sec:GPBOTWP} looks at ways of generating spatial processes based
on the Wiener process. Finally, Section \ref{sec:LP} deals with the
generation of L\'evy processes and fields.
  
\section{Gaussian Markov Random Fields}
\label{sec:GMRF}
\begin{figure}[H]
\includegraphics[width=0.31\linewidth]{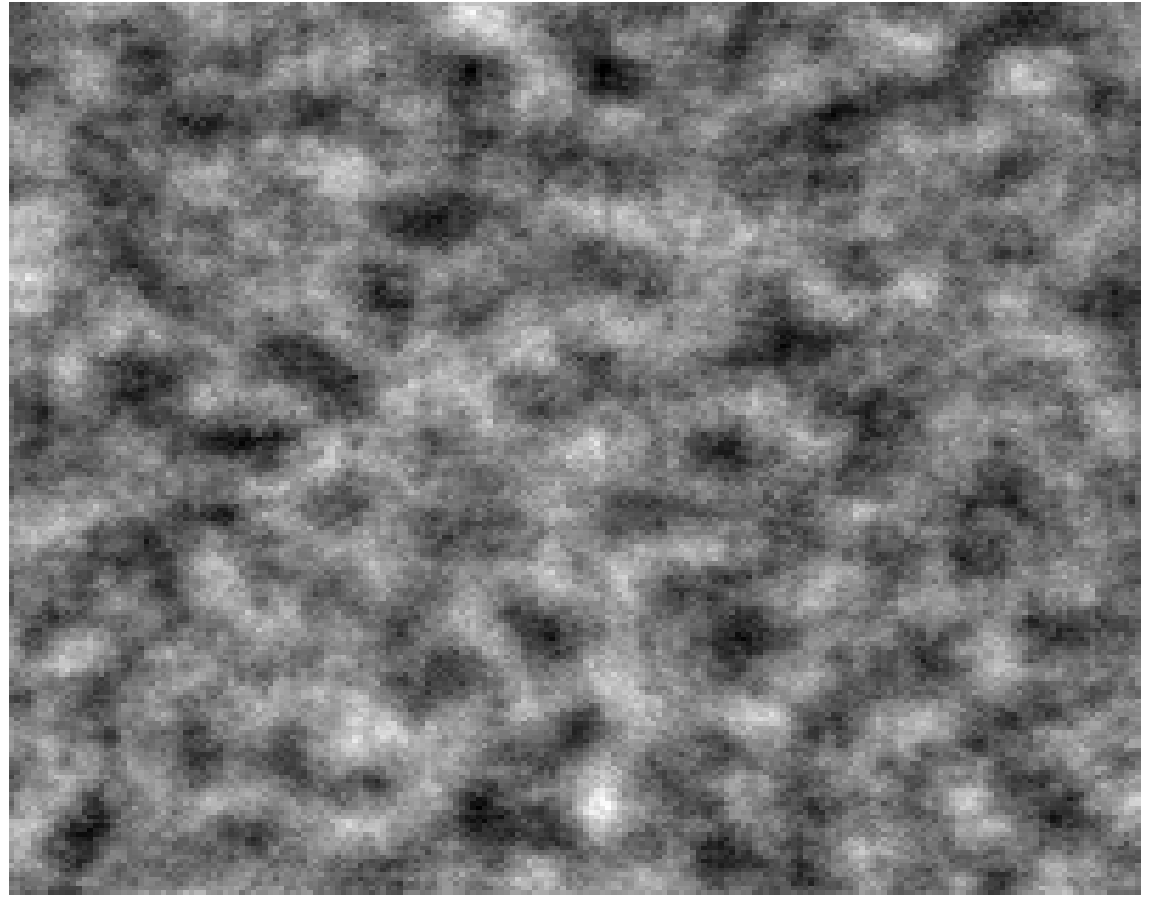}~
\includegraphics[width=0.31\linewidth]{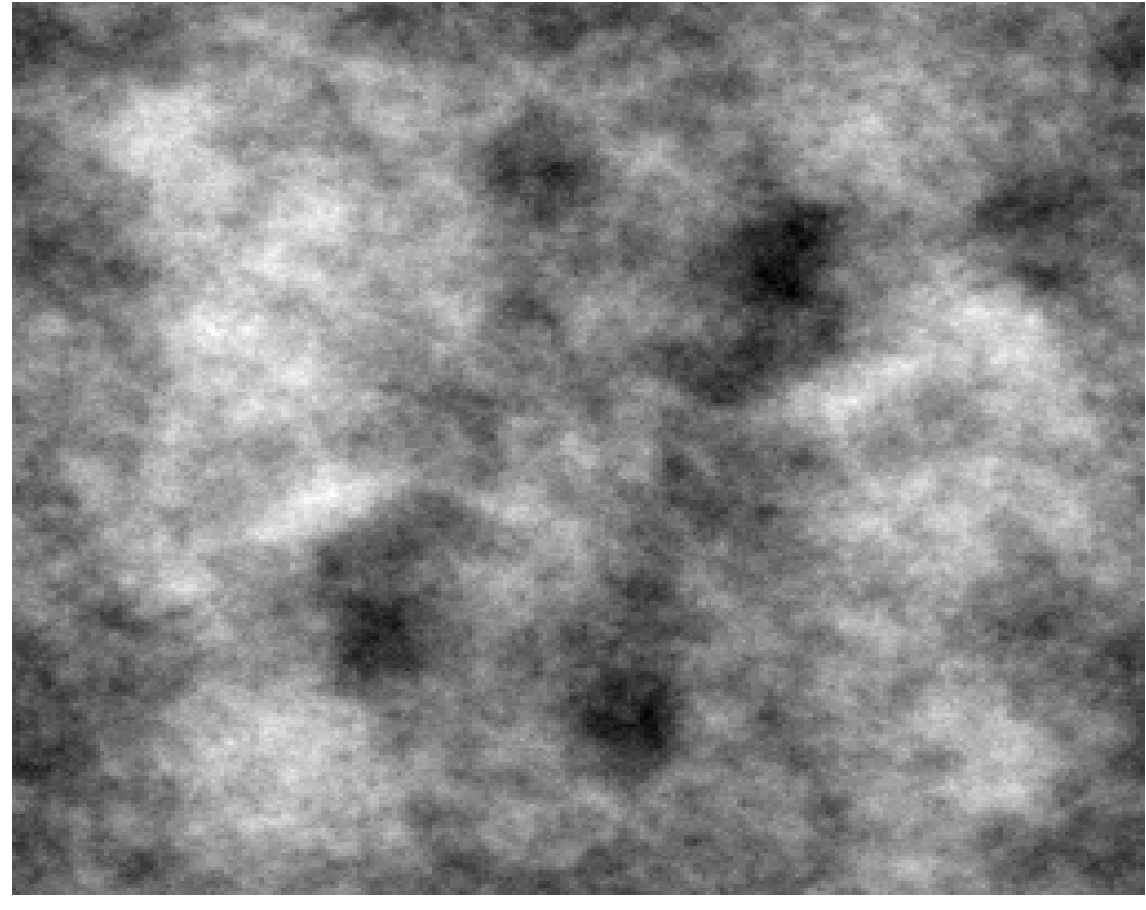}~
\includegraphics[width=0.32\linewidth]{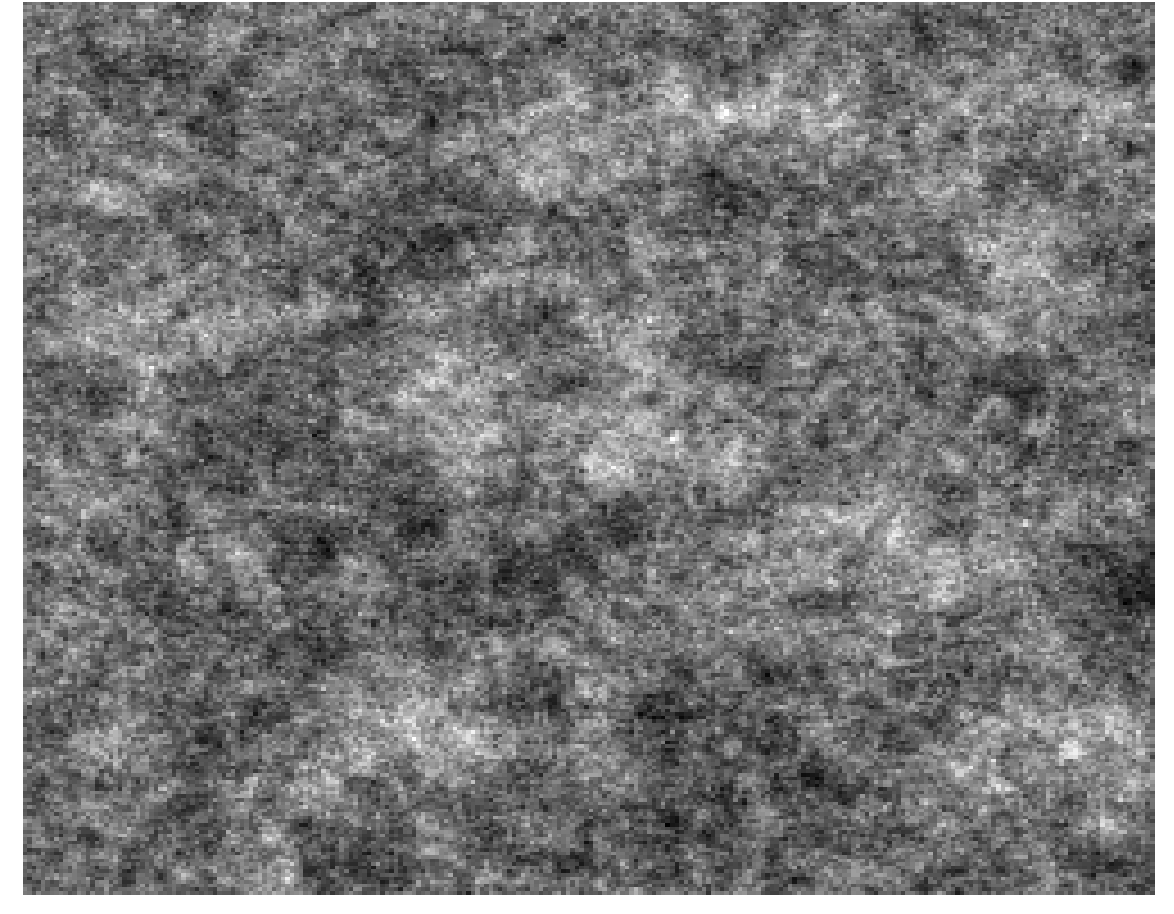}

\caption{Illustrations  of zero-mean Gaussian random fields. 
Left: Moving average spatial process. 
Middle: Stationary Gaussian random field on the torus. 
Right:  Gaussian Markov random field. 
}
\label{fig:stationary_gauss}
\end{figure}

A spatial stochastic process on $\R^2$ or $\R^3$ is often called a {\bf random
field}. Figure~\ref{fig:stationary_gauss} depicts realizations of three
different types of random fields that are characterized by  {\em Gaussian}
and {\em Markovian} properties, which are discussed below.

\subsection{Gaussian Property}
\label{sec:Gaussian property}
A stochastic process $\{\tilde X_{\bt},\bt \in \scT\}$ is said to be
{\bf Gaussian} if all its finite-dimensional distributions are
Gaussian (normal). That is, if for any choice of $n$ and $\bt_1,
\ldots,\bt_n\in\scT$, we have 
\begin{equation}\label{mvnormvect}
\bX \idef
(X_1,\ldots,X_n)^\T \idef (\tilde X_{\bt_1}, \ldots,\tilde X_{\bt_n})^\T \sim
\nor(\bmu, \Sigma)
\end{equation}
for some {\bf expectation vector} $\bmu$ and {\bf covariance matrix}
$\Sigma$. Hence, any linear combination $\sum_{i=1}^n b_i \tilde
X_{\bt_i}$ has a normal distribution. A Gaussian process is determined
completely by its {\bf expectation function} $\tilde \mu_{\bt} = \Em
\tilde X_{\bt}$ and {\bf covariance function} $\tilde \Sigma_{\bs,\bt}
= \cov(\tilde X_\bs,\tilde X_\bt)$.  To generate a Gaussian process
with expectation function $\tilde \mu_\bt$ and covariance function
$\tilde \Sigma_{\bs,\bt}$ at positions $\bt_1,\ldots,\bt_n$,  one can sample the
multivariate normal random vector in \eqref{mvnormvect} via the following algorithm.
\begin{alg}[Gaussian Process Generator] \label{alg.gausproc}  ~
\ben
\item  Construct the mean vector ${\bmu} = ({\mu}_1,\ldots,{\mu}_n)^\T$ and covariance matrix ${\Sigma} = ({\Sigma}_{ij})$ by setting ${\mu}_i = \tilde \mu_{\bt_i},
i=1,\ldots,n$ and ${\Sigma}_{ij} = \tilde\Sigma_{\bt_i, \bt_j}$, $i,j =
1,\ldots,n$.
\item Find a square root $A$ of $\Sigma$, so that $\Sigma = A A^\T$.
\item Generate $Z_1, \ldots, Z_n \simiid {\Nor}(0,1).$ Let $\bZ =
(Z_1,\ldots,Z_n)^\T$.
\item Output ${\bX} = \bmu + A \bZ$.
\een
\end{alg}
Using Cholesky's square-root method, it is always possible to find a
real-valued lower triangular matrix $A$ such that $\Sigma = A
A^\T$. Sometimes it is easier to work with a decomposition of the form
$\Sigma = B B^*$, where $B = B_1 + \ix B_2$ is a complex matrix with 
conjugate transpose $B^* = B_1^\T - \ix B_2^\T$. Let $\bZ = \bZ_1 + \ix \bZ_2$, where
$\bZ_1$ and $\bZ_2$ are independent standard normal random vectors, as in Step~3
above. Then, the random vector $\bX = \Re(B \bZ) = B_1 \bZ_1 - B_2
\bZ_2$ has covariance matrix $\Sigma$.  

A Gaussian vector $\bX \sim \Nor(\bmu, \Sigma)$ can also be simulated
 using its {\bf precision matrix} $\Lambda = \Sigma^{-1}$. Let
 $\bZ=D^\T\bY$, where $\bZ \simiid {\Nor}(0,1)$. If $DD^\T$ is the
 (lower) Cholesky factorization of $\Lambda$, then $\bY$ is a zero-mean
 multivariate normal vector with covariance matrix 
\[
\Em \bY \bY^\T =  (D^{-1})^\T \Em \bZ \bZ^\T D^{-1}
= (D D^\T)^{-1} = \Lambda^{-1} = \Sigma\;. 
\]
The following algorithm describes how a Gaussian process can be generated using  a precision matrix. 

\begin{alg}[Gaussian Process Generator Using a Precision Matrix]\label{alg.gauss.precision}~
\ben
\item Derive the Cholesky decomposition $ \Lambda = D D^\T$ of the precision matrix.
\item Generate $Z_1,\ldots,Z_n \simiid\Nor(0,1)$. Let $\bZ =
(Z_1,\ldots,Z_n)^\T$. 
\item Solve $\bY$ from $\bZ = D^\T \bY$, using forward substitution. 
\item Output $ \bX = \bmu + \bY$.
\een
\end{alg}

The Cholesky decomposition of a general $n\times n$
covariance or precision matrix takes $\bigO(n^3)$ floating point operations. The
generation of large-dimensional Gaussian vectors becomes very time-consuming for large $n$, unless some extra structure is
introduced. 
In some cases the Cholesky decomposition can be altogether
avoided by utilizing the fact that any Gaussian vector can be written 
as an {\em affine 
transformation} $\bX = \bmu + A \bZ$ of  a  ``white
noise''  vector $\bZ \simiid \Nor(0,1)$; as in Step~4 of Algorithm~\ref{alg.gausproc}.
An example where such a transformation can be carried out directly is the
following {\bf moving average} Gaussian process $ \bX = \{X_{\bt}, \bt \in
\scT\}$, 
where $\scT$ is a two-dimensional grid of equally-spaced points. Here
each $X_{\bt}$ is equal to the average of all white noise terms
$Z_{\bs}$ with $\bs$ lying in a disc of radius $r$ around $\bt$. That is, 
\[
X_{\bt} = \frac{1}{N_r} \sum_{\bs \: : \: \| \bt - \bs\| \leq r} Z_{\bs}\;, 
\]
where $N_r$ is the number of grid points in the disc.
Such spatial processes have been used to describe rough energy
surfaces for charge transport \cite{brereton_etal,Kendall_Ord_1990}.
 The following $\matlab$ program
produces a realization of this process on a $200\times 200$ grid, using
a radius $r = 6$. A typical outcome is depicted in the left pane of
Figure~\ref{fig:stationary_gauss}. 
\medskip
\begin{breakbox}
\begin{verbatim}
n = 300;
r = 6; % radius (maximal 49)
noise =  randn(n);
[x,y]=meshgrid(-r:r,-r:r);
mask=((x.^2+y.^2)<=r^2);  %(2*r+1)x(2*r+1) bit mask
x = zeros(n,n);
nmin = 50; nmax = 250;
for i=nmin:nmax
    for j=nmin:nmax
        A = noise((i-r):(i+r), (j-r):(j+r));
        x(i,j) = sum(sum(A.*mask));
    end
end
Nr = sum(sum(mask)); x = x(nmin:nmax, nmin:nmax)/Nr;
imagesc(x); colormap(gray)
\end{verbatim}
\end{breakbox}
\bigskip

\subsection{Generating Stationary Processes via Circulant embedding}
\label{sec:stationary gaussian}
Another approach to efficiently generate Gaussian spatial
processes is to exploit the structural properties of  {\em stationary}
Gaussian processes.
 A
Gaussian process $\{\tilde{X}_\bt, \bt \in \R^d\}$ is said to be {\bf stationary}
if the expectation function, $\Em\tilde{X}_\bt$, is constant and the
covariance function, Cov$(\tilde{X}_\bs,\tilde{X}_\bt)$, is invariant
under translations; that is 
$
\Cov(\tilde{X}_{\bs + \bu},\tilde{X}_{\bt
+ \bu}) = \text{Cov}(\tilde{X}_\bs,\tilde{X}_\bt)\,.$

An illustrative example is the generation of a stationary Gaussian
on the unit {\em torus}; that is, the unit square $[0,1]\times[0,1]$
in which points on opposite sides are identified with each other.
In particular, we wish to generate a zero-mean Gaussian random process
$\{\tilde{X}_\bt\}$ on each
of the grid points $\{(i,j)/n, i=0,\ldots,n-1, j = 0, \ldots,n-1\}$
corresponding to a covariance function of the form
\begin{equation}\label{cov1}
\cov(\tilde{X}_\bs,\tilde{X}_\bt) = \exp\{-c\, \|\bs - \bt\|_T^\alpha\}
, 
\end{equation}
where $\| \bs - \bt \|_T = \|(s_1 - t_1, s_2 - t_2) \|_T \idef
\sqrt{\sum_{k=1}^2 (\min\{|s_k - t_k|, 1 -
 |s_k - t_k|\})^2}$ is the Euclidean distance on the torus.  
Notice that this renders the process not only stationary but also 
{\bf isotropic} (that is, the distribution remains the same under rotations).
We can arrange the grid points in the order
$(0,0),(0,1/n),\ldots,(0, 1 -1/n),(1/n,0), \ldots, (1-1/n, 1 -
1/n)$. The values of the Gaussian process can be accordingly gathered
in an $n^2 \times 1$ vector $\bX$ or, alternatively an $n\times n$
matrix $\underline{\bX}$.  
Let $\Sigma$ be the $n^2 \times n^2$ covariance matrix of $\bX$.
The key to efficient generation of $\bX$ is that $\Sigma$ is a
symmetric {\em
block-circulant matrix with circulant blocks}. That is, $\Sigma$ has
the form 
\[
\Sigma = \begin{pmatrix}
C_{1} & C_{2} & C_{3} & \ldots&  C_{n} \\
C_{n} & C_{1} & C_{2} & \ldots & C_{n-1} \\
  & & \ldots &  &  \\
C_{2} & C_{3} & \ldots & C_{n}& C_{1}
\end{pmatrix}, 
\]
where each $C_i$ is a circulant matrix
$\text{circ}(c_{i1},\ldots,c_{in})$. The matrix $\Sigma$ is thus completely
specified by its first row, which we gather in an $n\times n$ matrix
$G$. The eigenstructure of block-circulant matrices with circulant
blocks is well known; see, for example, \cite{Barnett90}. Specifically, 
$\Sigma$ is of the form 
\begin{equation}
\label{diagonalization}
\Sigma = P^* \diag(\vect{\gamma})P,
\end{equation} 
where  $P^*$ denotes the complex conjugate transpose of 
 $P$, and $P$ is  the Kronecker product of two {\em discrete Fourier transform} matrices; that is,
$P = F \otimes F$, where 
$
F_{jk} = \exp(-2\pi \ix j k/n)/\sqrt{n}, j,k
= 0,1,\ldots,n-1.
$
 The vector of eigenvalues $\vect{\gamma} = (\gamma_{1},
\ldots,\gamma_{n^2})^\T$ ordered as an $n \times n$ matrix $\Gamma$ satisfies
$\Gamma = n \, F^* G F$. Since $\Sigma$ is a covariance
matrix, the component-wise square
root
 $\sqrt{\Gamma}$ is well-defined and real-valued. The matrix 
$B = P^*
\diag(\sqrt{\vect{\gamma}})$ is a complex square root of
$\Sigma$, so that $\bX$ can be generated by  drawing  $\bZ = \bZ_1 + \ix \bZ_2$, where
$\bZ_1$ and $\bZ_2$ are independent standard normal random vectors,
and returning the real part of $B \bZ$. 
It will be convenient to gather $\bZ$ into an $n\times n$ matrix 
$\underline{\bZ}$.

The evaluation of both $\Gamma$ and $\underline{\bX}$ can be done
efficiently by using the (appropriately scaled) 
{\em two-dimensional Fast Fourier Transform}
(FFT2).  In particular,
$\Gamma$ is the FFT2 of the matrix $G$ and $\underline{\bX}$ is the real part of the FFT2 of the
matrix $\sqrt{\Gamma}\odot\underline{\bZ}$,  where 
$\odot$ denotes component-wise multiplication. The following $\matlab$ program generates
the outcome of a stationary Gaussian random field on a 
$256\times 256$ grid, for a covariance function of the form \eqref{cov1},
with $c = 8$ and $\alpha = 1$. The realization is shown in the middle 
pane of Figure~\ref{fig:stationary_gauss}.
\medskip

\begin{breakbox}
\begin{verbatim}
n = 2^8; 
t1 = [0:1/n:1-1/n]; t2 = t1;
for i=1:n   % first row of cov. matrix, arranged in a matrix
  for j=1:n
       G(i,j)=exp(-8*sqrt(min(abs(t1(1)-t1(i)), ...
       1-abs(t1(1)-t1(i)))^2 + min(abs(t2(1)-t2(j)), ...
       1-abs(t2(1)-t2(j)))^2));
  end;    
end;
Gamma = fft2(G);   % the eigenvalue matrix n*fft2(G/n)
Z = randn(n,n) + sqrt(-1)*randn(n,n);
X = real(fft2(sqrt(Gamma).*Z/n));
imagesc(X); colormap(gray)
\end{verbatim}
\end{breakbox}
\medskip
Dietrich and Newsam \cite{Dietrich_Newsam_1997} and Wood and Chan
 \cite{chan_wood,wood_chan} discuss the generation of general stationary
Gaussian processes in $\R^d$ with  covariance  function
\[
 \text{Cov}(\tilde{X}_\bs,\tilde{X}_\bt)=\rho(\bs-\bt)\;.
\]
Recall that if $\rho(\bs-\bt)=\rho(\|\bs-\bt\|)$, then the random field is not only
stationary, but isotropic. 

Dietrich and Newsam propose the method of \textbf{circulant embedding}, which allows the efficient generation of a stationary Gaussian field via the FFT. The idea is to  {\em
 embed} the covariance matrix into a block circulant matrix with each block being circulant itself (as in the last example), and then construct the matrix square root of the block circulant matrix using
 FFT techniques. The FFT permits the   fast  generation of the Gaussian field with this block circulant 
covariance matrix. Finally, the marginal distribution of appropriate sub-blocks 
of this Gaussian field have the desired covariance structure.

Here we consider the two-dimensional case. The aim is to generate a zero-mean stationary Gaussian field 
over the $n\times m$ rectangular grid 
\[
\cG=\{(i\Delta_x  ,j\Delta_y),\;i=0,\ldots,n-1, \;j=0,\ldots,m-1\},
\]
 where $\Delta_x$ and $\Delta_y$ denote the corresponding
 horizontal and vertical spacing along the grid. The algorithm can broadly be described as follows.
 
\begin{figure}[htb]
\hspace{-0.5cm}\includegraphics[width=.55\linewidth]{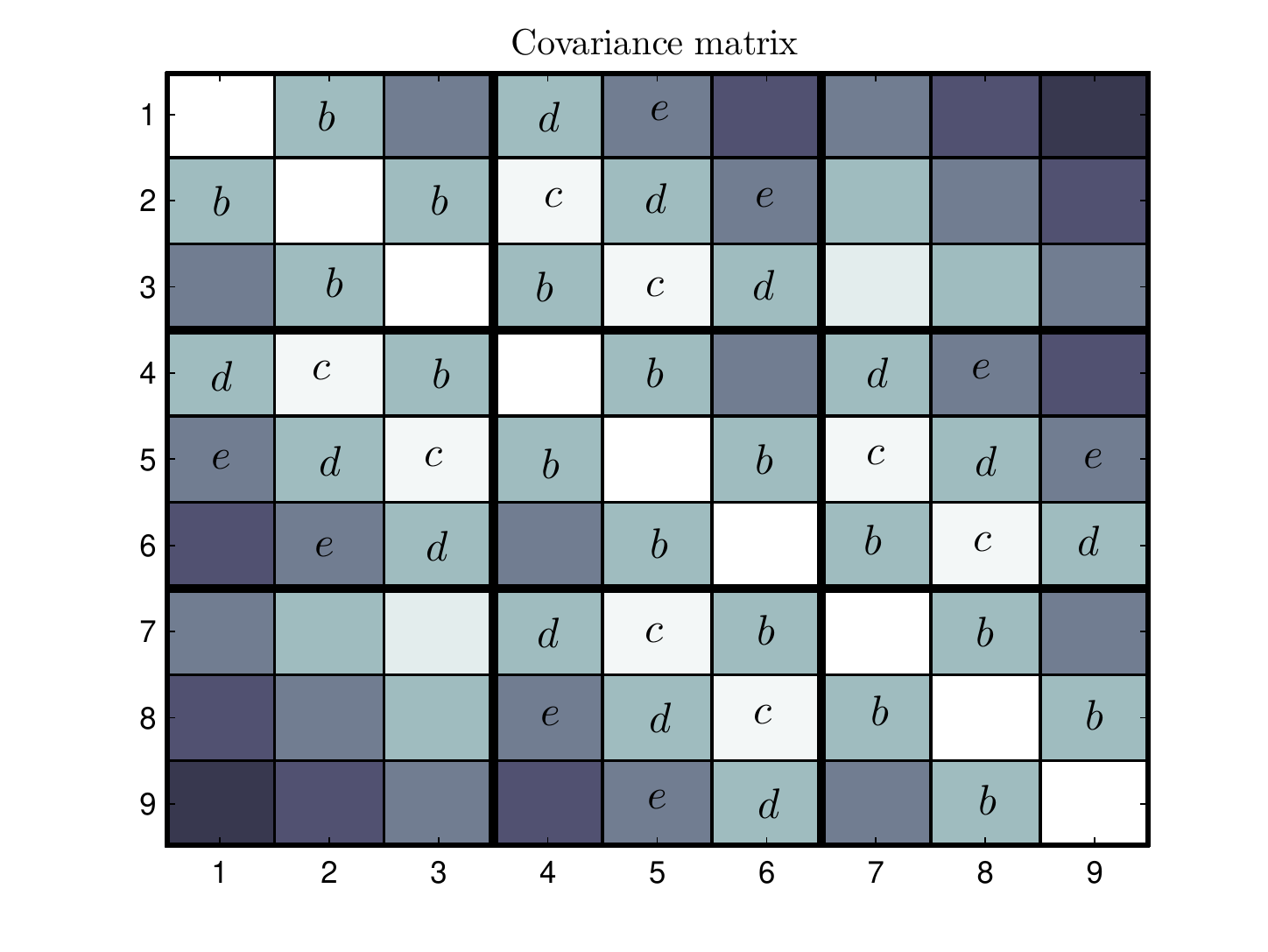}
\hspace{-0.6cm}~\includegraphics[width=.55\linewidth]{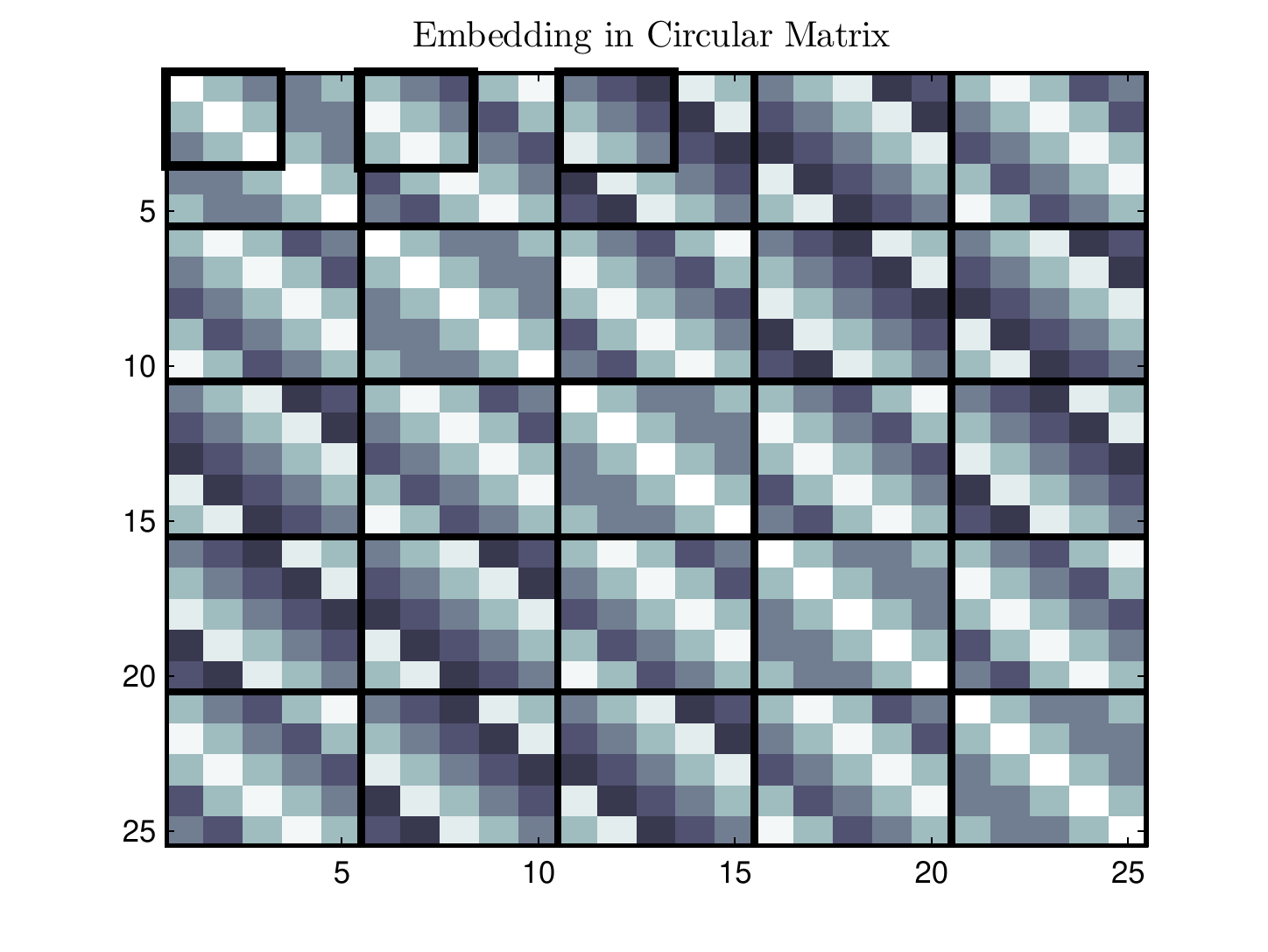}
\caption{Left panel: the symmetric block-Toeplitz covariance matrix $\Omega$ with $n=m=3$.  Right panel: block-circulant matrix $\Sigma$ of size $((2n-1)\times (2m-1))^2 =25\times 25$.
The first three block rows  of  $\Omega$ are embedded in the upper left corners of the first three block columns of the circulant matrix $\Sigma$.}
\label{fig:cov}
\end{figure}

\textbf{Step 1. Building and storing the covariance matrix.}  The grid points can be arranged into 
a column vector of size $n\,m$ to yield the
 $m^2 n^2$ covariance matrix $\Omega_{i,j}=\rho(\bs_i-\bs_j),\;i,j=1,\ldots,m\,n$, where
\[
 \bs_k\idef\left((k-1)\mod m,\quad \left\lfloor \frac{k}{m} \right\rfloor \right), \quad k=1,\ldots,m\,n\;.
\] 
The matrix $\Omega$ has
symmetric block-Toeplitz structure, where each block is a Toeplitz (not necessarily symmetric) matrix.  For example, the left panel of Figure~\ref{fig:cov} shows
the block-Toeplitz covariance matrix with $m=n=3$. Each $3\times 3$ block is itself a Toeplitz matrix with entry values coded in color. For instance, we have $\Omega_{2,4}=\Omega_{3,5}=c$.

Matrix $\Omega$ is thus uniquely characterized
by its first block row
$
(R_1,\ldots,R_{n})
$
where each of the $n$ blocks is an $m\times m$ Toeplitz matrix.
The $k$-th $m\times m$ Toeplitz matrix  consists of the sub-block of $\Omega$ with entries  
\[
 \Omega_{i,j},\quad   i=1,\ldots,m,\;j=(km+1),\ldots,(k+1)m\;.
\]
 Notice that each block $R_k$ is itself characterized by its first row and column. 
(Each Toeplitz block $R_k$ will be characterized by the first row only provided the covariance function has the form
$\rho(\bs-\bt)=\rho(\|\bs\|,\|\bt\|)$, in which case each $R_k$ is symmetric.) Thus, in general the covariance matrix can be completely characterized by 
the entries of a pair of $m\times n$ and $n\times m$ matrices storing the first columns and rows of all $n$ Toeplitz blocks $
(R_1,\ldots,R_{n})
$. In typical applications, the computation
of these two matrices is the most time-consuming step. 

\textbf{Step 2. Embedding in block circulant matrix.}
Each Toeplitz matrix $R_k$ is embedded in
the upper left corner of an $2m+1$ circulant matrix $C_k$. For example, in the right panel of Figure~\ref{fig:cov} the embedded blocks $R_1,R_2,R_3$ are shown within bold rectangles.
Finally, let $\Sigma$ be the $(2n-1)(2m-1)\times (2n-1)(2m-1)$ block circulant matrix with first block row given 
by
$
(C_1,\ldots,C_n,C_n^\T,\ldots,C_2^\T).
$ 
This gives the \emph{minimal embedding} in the sense that there is no  block circulant matrix of smaller size that  embeds $\Omega$.

\textbf{Step 3. Computing the square root of the block circulant matrix.}
After the embedding we are essentially generating a Gaussian process on a torus with covariance matrix $\Sigma$ as in the last  example. 
The block circulant matrix $\Sigma$ can  be diagonalized as in
\eqref{diagonalization} to yield $\Sigma=P^*\diag(\vect{\gamma})P$, where
 $P$ is  the $(2n-1)(2m-1)\times (2n-1)(2m-1)$ two-dimensional discrete Fourier transform matrix. The vector of eigenvalues $\vect{\gamma}$ is of length $(2n-1)(2m-1)$ and is   arranged in a $(2m-1)\times (2n-1)$ matrix 
 $\Gamma$ so that the first column of $\Gamma$ consists of the
 first $2m-1$ entries of $\vect{\gamma}$, the second column of $\Gamma$
 consists of the next $2m-1$ entries and so on. It follows that if $G$ is 
 an $(2m-1)\times (2n-1)$ matrix storing the entries of the first block row of $\Sigma$, then 
 $\Gamma$ is the FFT2 of $G$. Assuming that $\vect{\gamma}>0$, we obtain the square root factor $B=P^*\diag(\sqrt{\vect{\gamma}})$, so that $\Sigma=B^*B$.


\textbf{Step 4. Extracting the appropriate sub-block.}
Next, we compute the FFT2 of the
array $\sqrt{\Gamma}\odot \underline{\bZ}$, where the square root is applied 
component-wise to $\Gamma$ and $\underline{\bZ}$ is an $(2m-1)\times(2n-1)$ complex Gaussian matrix with entries $\underline{\bZ}_{j,k}=U_{j,k}+\ix V_{j,k}$, $U_{j,k}, V_{j,k}\simiid \Nor(0,1)$ for all $j$ and $k$. Finally,  the first
$m\times n$ sub-blocks of the real and imaginary parts of $\mathrm{FFT2}(\sqrt{\Gamma}\odot \underline{\bZ})$
represent two independent realization of a stationary Gaussian field   with covariance $\Sigma$ on the grid $\cG$.  If more realizations are required, we store the values $\sqrt{\Gamma}$ and 
we repeat Step 4 only.  The complexity of the circulant embedding inherits the complexity of the
 FFT approach, which is of order $\bigO(mn\log(m+n))$, and compares very favorably with the standard Cholesky decomposition method of order $\bigO(m^3n^3)$.

\begin{figure}[H]
\centerline{\includegraphics[width=.5\linewidth]{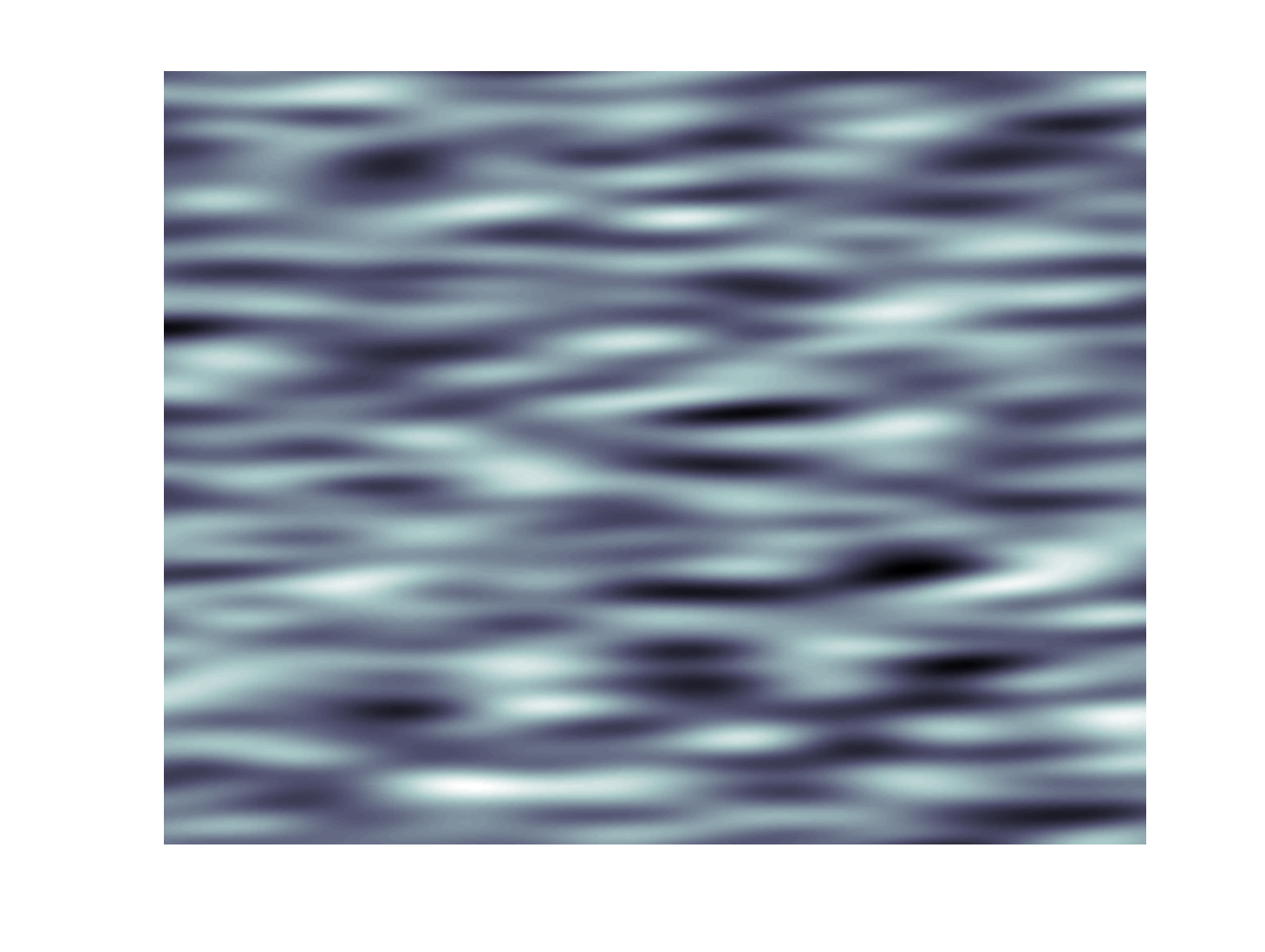}}
\caption{Realization of a stationary nonisotropic Gaussian field with covariance function \eqref{wavy_field}.}
\label{fig:gauss_waves}
\end{figure}

 As a numerical example (see \cite{Dietrich_Newsam_1997}), Figure~\ref{fig:gauss_waves} shows a realization of a stationary nonisotropic Gaussian field  with  $m=512,n=384,\; \Delta_x=\Delta_y=1$, and covariance function 
 \begin{equation}
 \label{wavy_field}
\rho(\bs-\bt)= \rho(\bh)=\left(1-\frac{h_1^2}{50^2}-\frac{h_1h_2}{50\times 15}-\frac{h_2^2}{15^2}\right)\exp\left(-\frac{h_1^2}{50^2}-\frac{h_2^2}{15^2}\right)\;.
 \end{equation}

The following $\matlab$ code implements the procedure described above with covariance function \eqref{wavy_field}. 

\medskip

\begin{breakbox}
\begin{verbatim}
n=384; m=512; % size of grid is m*n
% size of covariance matrix is m^2*n^2
tx=[0:n-1]; ty=[0:m-1]; % create grid for field
rho=@(x,y)((1-x^2/50^2-x*y/(15*50)-y^2/15^2)...
      *exp(-(x^2/50^2+y^2/15^2)));
Rows=zeros(m,n); Cols=Rows;
for i=1:n
    for j=1:m
        Rows(j,i)=rho(tx(i)-tx(1),ty(j)-ty(1)); % rows of blocks 
        Cols(j,i)=rho(tx(1)-tx(i),ty(j)-ty(1)); % columns 
    end
end
% create the first row of the block circulant matrix 
% with circulant blocks and store it as a matrix suitable for fft2
BlkCirc_row=[Rows, Cols(:,end:-1:2);
    Cols(end:-1:2,:), Rows(end:-1:2,end:-1:2)];
% compute eigen-values
lam=real(fft2(BlkCirc_row))/(2*m-1)/(2*n-1);

if abs(min(lam(lam(:)<0)))>10^-15
    error('Could not find positive definite embedding!')
else
    lam(lam(:)<0)=0; lam=sqrt(lam);
end
% generate field with covariance given by block circulant matrix
F=fft2(lam.*complex(randn(2*m-1,2*n-1),randn(2*m-1,2*n-1)));
F=F(1:m,1:n); % extract sub-block with desired covariance
field1=real(F); field2=imag(F); % two independent fields 
imagesc(tx,ty,field1), colormap bone
\end{verbatim}
\end{breakbox}
\medskip

Extensions to three and more dimensions  are possible, see \cite{Dietrich_Newsam_1997}. For example, in the three dimensional case the correlation matrix $\Omega$ will be symmetric block Toeplitz matrix with each block
 satisfying the properties of a two dimensional covariance matrix.

Throughout the discussion so far we have always assumed that the block circulant matrix $\Sigma$ is a covariance matrix itself. If this is  the case, then we say that we have a \emph{nonnegative definite embedding} of $\Omega$. A nonnegative definite embedding ensures that the square root of $\vect{\gamma}$ is real.
Generally, if the correlation between points on the grid that are sufficiently far apart
is zero, then a non-negative embedding will exist, see \cite{Dietrich_Newsam_1997}.  
A method that exploits this observation is the  \emph{intrinsic embedding}  method
proposed by Stein \cite{stein02} (see also \cite{Gneiting2012}).  
 Stein's method depends on the
construction of a compactly supported covariance  function that yields to a nonnegative circulant embedding. The idea is to  modify the original covariance  function so that  it decays smoothly to zero. 
In more detail, suppose we wish to  simulate a process with covariance  $\rho$ over the set $\{\bh: \|\bh\|\leq 1, \bh>\mathbf{0}\}$. To achieve this, we simulate a process with covariance function
  \begin{equation}
  \label{stein}
  \psi(\bh)=\begin{cases} c_0+c_2\|\bh\|^2+\rho(\bh), & \|\bh\|\leq 1\\
   \phi(\bh), & 1\leq\|\bh\|\leq R\\
   0,& \|\bh\|\geq R
  \end{cases},
  \end{equation} 
 where the constants $c_0,c_2, \;R\geq 1$ and function $\phi$ are selected so that  $\psi$ is a continuous (and as many times differentiable as possible), stationary and isotropic covariance function on $\mathbb{R}^2$. The process will have
  covariance structure of $c_0+c_2\|\bh\|^2+\rho(\bh)$ in the disk  $\{\bh: \|\bh\|\leq 1, \bh>\mathbf{0}\}$, which can then be easily
 transformed into a process with covariance  $\rho(\bh)$. We give an example of this in Section~\ref{sec:FBS}, where we generate fractional Brownian surfaces via the intrinsic embedding technique. Generally, the smoother the original covariance function, the harder it is to embed via Stein's method, because
a covariance function that is smoother close to the origin
has to be even smoother elsewhere. 



 
\subsection{Markov Property}\label{ssec:Markovprop}
A {\bf Markov random field} is a spatial stochastic process $\bX
=\{X_\bt, \bt \in \scT\}$ that possesses a {\em  Markov property}, in the sense
that
\[
(X_\bt\gvn X_\bs, \bs \in \scT \setminus\{\bt\}) \sim (X_\bt \gvn
X_\bs, \bs \in \scN_\bt), 
\]
where $\scN_\bt$ is the set of ``neighbors''
of $\bt$. Thus, for each $\bt \in \scT$ the conditional distribution
of  $X_{\bt}$ given all other values $X_{\bs}$ is equal to the  conditional distribution
of  $X_{\bt}$ given only the neighboring values.  

\begin{figure}[H]
\centerline{\includegraphics[width=0.3\linewidth]{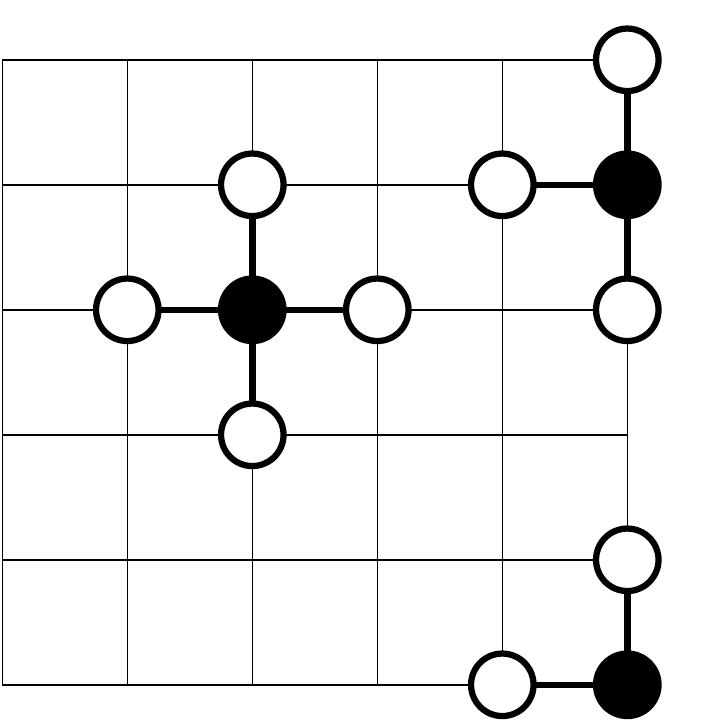}}
\caption{A graphical model for 2D spatial process. Each vertex has at
most four neighbors.}
\label{fig:gmrf}
\end{figure}

Markov random fields are often defined via an undirected graph  ${\cal G} = ({V},{E})$. 
In such a {\em graphical model} the vertices of the graph correspond
to the indices $\bt \in \scT$ of the random field, and the edges 
describe the dependencies between the random variables. In particular,
there is {\em no} edge between nodes $\bs$ and $\bt$ in the graph if and
only if $X_{\bs}$ and $X_{\bt}$ are conditionally independent, given
all other values $\{X_{\bu}, \bu \neq i,j\}$.  In a Markov random
field described by a graph $\cal G$, the set of neighbors $\scN_\bt$
of $\bt$ corresponds to the set of vertices that share an edge with
$\bt$; that is, those vertices that are {\em adjacent} to $\bt$.
An example of a graphical model for a 2-dimensional Markov random
field is shown in Figure \ref{fig:gmrf}. In this case vertex
corner nodes have two neighbors and interior nodes have four neighbors.
 
Of particular importance are Markov random fields that are also
Gaussian. Suppose $\bX$ is such a {\bf Gaussian Markov random field}
(GMRF), with corresponding graphical model ${\cal G} = (V,E)$. 
 We may
think of $\bX$ as a column vector of $n=|V|$ elements, or as a spatial
arrangement of random variables (pixel values), as in
Figure~\ref{fig:gmrf}. Without loss of generality we assume that $\Em
\bX  = \vect{0}$, and that the index set $\scT$ is identified with
the vertex set $V$, whose elements are labeled as $1,2,\ldots,n$. 
Because $\bX$ is Gaussian, its pdf  is 
given by 
\[
f(\bx) = (2 \pi)^{-n/2} \sqrt{\det(\Lambda)}\: \e^{-\frac{1}{2} \bx^\T \Lambda \bx }\;, 
\]
where $\Lambda = (\lambda_{ij})$ is the precision
matrix. 
In particular, the conditional joint pdf of   $X_{i}$ and
$X_{j}$ is 
\[
f(x_i,x_j \gvn x_k, k \neq i,j) \propto \exp\left(-\frac{1}{2}
(\lambda_{ii} x_i^2 + 2x_i  \,a +
 \lambda_{ij} x_i x_j + \lambda_{jj} x_j^2 + 2 x_j \, b)\right), 
\]
where $a$ and $b$ may depend on $x_k, k \neq i,j$. 
This shows that $X_i$ and $X_j$ are conditionally independent given
$\{X_k, k \neq i,j\}$, if and only if $\lambda_{ij} = 0$. Consequently,
$(i,j)$ is an edge in the graphical model if and only if $\lambda_{ij}
\neq 0$. In typical applications (for example  in image analysis)  each
vertex in the graphical model only has a small number of adjacent
vertices, as in Figure~\ref{fig:gmrf}. In such cases the precision matrix
is thus {\em sparse}, and the Gaussian vector can be generated
efficiently using, for example,  sparse Cholesky factorization \cite{golub87}.

As an illustrative example, consider the graphical model of 
Figure~\ref{fig:gmrf} on a  grid of $n = m^2$ points: $\{1,\ldots,m\} \times
\{1, \ldots,m\}$. We can efficiently generate a GMRF on this grid via 
Algorithm~\ref{alg.gauss.precision}, provided the Cholesky
decomposition can be carried out efficiently. For this we can use for example the {\em
band Cholesky} method \cite{golub87}, which takes $n(p^2 + 3 p)$ floating
point operations, where
$p$ is the bandwidth of the matrix; that is, $p = \max_{i,j}\{|i -j|:
\lambda_{ij} = 0\}$. 
The right pane of
Figure~\ref{fig:stationary_gauss} shows a realization of the GMRF on a
$250\times 250$ grid, using parameters $\lambda_{ii} = 2$ for all
$i=1,\ldots,n$ and $\lambda_{ij} = -0.5$ for all neighboring elements
$j \in {\cal N}_j$ of $i = 1,\ldots,n$. The $\matlab$ code may be
found in Section~5.1 of \cite{handbook_dirk}.  
Further details on such construction methods for GMRFs may be found, for example, in the monograph by Rue and Held \cite{rueheld}.


\section{Point Processes}
Point processes on $\R^d$ are spatial processes describing random
configurations of $d$-dimensional points.
 Spatially distributed point patterns
occur frequently in nature and in a wide variety of scientific
disciplines such as spatial epidemiology, material science, forestry,
geography. The positions of accidents on a highway during a fixed time
period, and the times of earthquakes in Japan are examples of
one-dimensional spatial point processes. Two-dimensional examples
include the positions of cities on a map, the positions of farms with
Mad Cow Disease in the UK, and the positions of broken connections in
a communications or energy network. In three dimensions, we observe
the positions of stars in the universe, the positions of mineral
deposits underground, or the times and positions of earthquakes in
Japan. Spatial processes provide excellent
models for many of these point patterns \cite{Baddeley07,CaseStudies,Mathematicalpointprocesses,Digglebook}. Spatial processes also have an
important role in stochastic modeling of complex microstructures, for example,
graphite electrodes used in Lithium-ion batteries \cite{stenzeletal}. 

 Mathematically, point processes 
can be described in three ways: (1) as random
sets of points, (2) as random-sized vectors of random positions, and
(3) as random counting measures.
In this section we discuss some of the important point processes and
their generalizations, including Poisson processes, marked point
processes, and cluster processes. 

\label{sec:SPP}
\subsection{Poisson Process}
\label{PoissonProcesses}
Poisson processes are used to model random configurations of points in
space and time. Let $E$ be some subset of $\R^d$ and let $\cE$ be the
collection of (Borel) sets on $E$. To any collection of random  points
$ \{X_1,\ldots,X_N\}$ in $E$ corresponds a {\bf random counting
measure} $\bX(A), A \in {\cal E}$ defined by 
\begin{equation}
 \bX(A) = \sum_{i=1}^N \I_{\{ X_i \in A\}}, \quad A \in \cE\;, 
\label{prm}
\end{equation}
which  counts the random number of points in $A$. 
We may {\em identify} the random measure $\bX$ defined in \eqref{prm} 
with the random set $\{X_i, i = 1,\ldots,N\}$, or with the random vector
$(X_1,\ldots,X_N)$. Note that the $d$-dimensional points $x_i, X_i$ etc.\ are {\em not} written in
boldface font here, in contrast to Section~\ref{sec:GMRF}.
The measure $\mu(A) = \Em \bX(A), A \in {\cal E}$ is called the {\bf
mean measure} of $\bX$.  
In most practical cases the mean measure $\mu$ has a density $\lambda$, called the {\bf intensity};
so that \[ \mu(A) = \Em\bX(A)= \int_A \lambda(x) \, \di x\;.\]
We will assume from now on that such an intensity function exists. 

The most important point process which holds the key to the analysis
of point pattern data is the Poisson process. A random counting
measure $\bX$ is said to be a {\bf Poisson random measure} with  mean
measure $\mu$ if the following properties hold:
\begin{enumerate}
\item  For any set $A \in {\cal E}$ the random variable 
$\bX(A)$ has a Poisson distribution with mean $\mu(A)$. We write  
$\bX(A) \sim \Poi(\mu(A))$.
\item For any disjoint sets $A_1,\dots,A_N \in \cE$,  the random
variables  
$\bX(A_1),\dots, \bX(A_N)$ are {independent}. 
\end{enumerate}
The
Poisson process is said to be {\bf homogeneous} if the intensity
function is constant. 
 An important corollary of Properties 1 and 2 is:
\begin{enumerate}
\item[3.] Conditional upon $\bX(E)=N$, the points $X_1,\ldots,X_N$ are independent of each other and have pdf $g(x) = \lambda(x)/\mu(E)$.  
\end{enumerate}
This result is the key to generating a Poisson random measure on $\R^d$.

\begin{alg}[Generating a Poisson Random Measure]\label{alg:poi_rand_measure}~
\bi
\item[1.] Generate a Poisson random variable $N \sim {\Po}(\mu(E))$.
\item[2.] Draw $X_1,\ldots,X_N \simiid g$, where $g(x) =\lambda(x)/\mu(E)$, and  return 
 these as the points of the Poisson random measure.
\ei
\end{alg}

\begin{figure}[H]
\centerline{\includegraphics[width=0.5\linewidth]{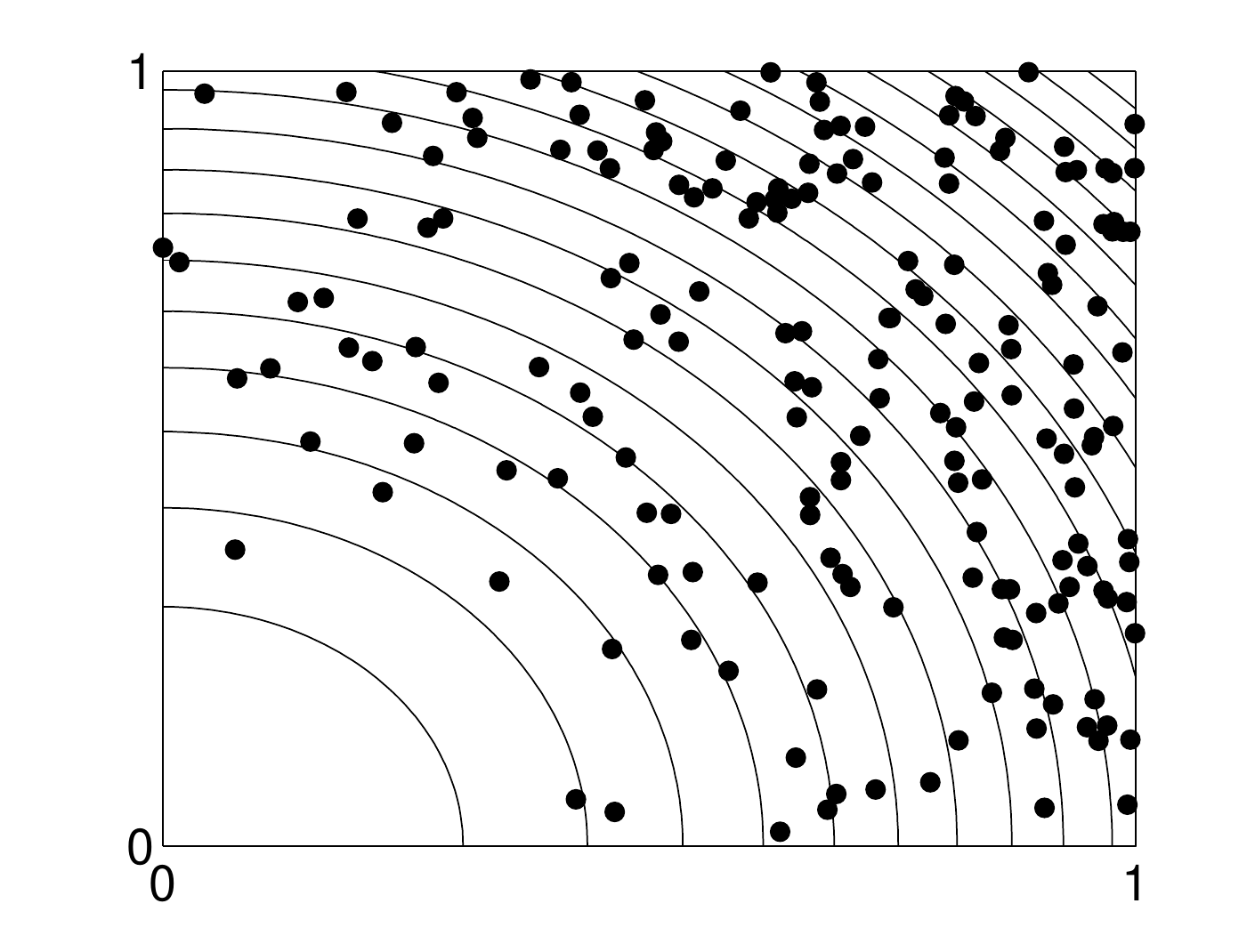}}
\caption{A realization of a nonhomogeneous Poisson process with
intensity $\lambda(x_1,x_2) = 300 (x_1^2 + x_2^2)$ (contour lines shown) on the unit
square.}
\label{fig:nonhompoi}
 \end{figure}

As a specific example, consider the generation of a 2-dimensional
Poisson process with intensity $\lambda(x_1,x_2) = 300(x_1^2 + x_2^2)$
on the unit square $E = [0,1]^2$. Since the pdf
$g(x_1,x_2) = \lambda(\bx)/\mu(E) = 3(x_1^2 + x_2^2)/2$ is bounded by 3,
drawing from $g$ can be done
simply via the acceptance--rejection method \cite{mcmc:robcas04}. That is, draw 
$(X_1,X_2)$ uniformly on $E$ and $Z$ uniformly on $[0,3]$, and  accept
$(X_1,X_2)$ 
if $g(X_1,X_2) \leq Z$; otherwise repeat.
 This acceptance--rejection step is then repeated $N \sim
\Poi(200)$ times. An alternative, but equivalent, method is to
generate a {\em homogeneous} Poisson process on $E$, with intensity
$\lambda^* = 600$, and to {\em thin out} the points by accepting each
point $\bx$ with probability $\lambda(x_1,x_2)/\lambda^*$.   
The following $\matlab$ implements this thinning procedure. A typical
realization is given in Figure~\ref{fig:nonhompoi}. 
\medskip

\begin{breakbox}
\begin{verbatim}
lambda = @(x) 300*(x(:,1).^2 + x(:,2).^2);
lamstar = 600;
N=poissrnd(lamstar); x = rand(N,2); % homogeneous PP
ind = find(rand(N,1) < lambda(x)/lamstar);
xa = x(ind,:); % thinned PP
plot(xa(:,1),xa(:,2))
\end{verbatim}
\end{breakbox}

\subsection{Marked Point Processes}
\label{prop:markedPP}
A natural way to extend the notion of a point process is to associate
with each point $X_i \in \R^d$ a \emph{mark} $Y_i \in \R^m$,
representing an attribute such as  width, velocity, weight
etc.  The
collection  $\{(X_i, Y_i)\}$ is 
called a {\bf marked point process}. 
In a marked point process with {\bf independent marking} the marks are
assumed to be  independent of each other and of the points, and
distributed according to a fixed mark distribution. 
The following gives a useful connection between marked point processes
and Poisson processes; see, for example, \cite{Mathematicalpointprocesses}.

\begin{theorem}\rm \label{the:pois} If $\{(X_i,Y_i)\}$ is a Poisson
process on $\R^d \times \R^m$ with intensity $\zeta$, and
$K(x) = \int \zeta(x,y) \, \di y < \infty$ for all $x$, then $\{X_i\}$ is
 a Poisson process on $\R^d$ with intensity $K$, and $\{(X_i,
Y_i)\}$ is a marked Poisson process with mark density
$\zeta(x,\cdot)/K(x)$ on $\R^m$.
\end{theorem}

A (spatial) marked Poisson process with independent marking is an important example
of a (spatial) {\bf L\'evy process}: a stochastic process with independent and
stationary increments (discussed in more detail in Section \ref{sec:LP}).

\begin{figure}[H]
\centerline{\includegraphics[width=0.6\linewidth]{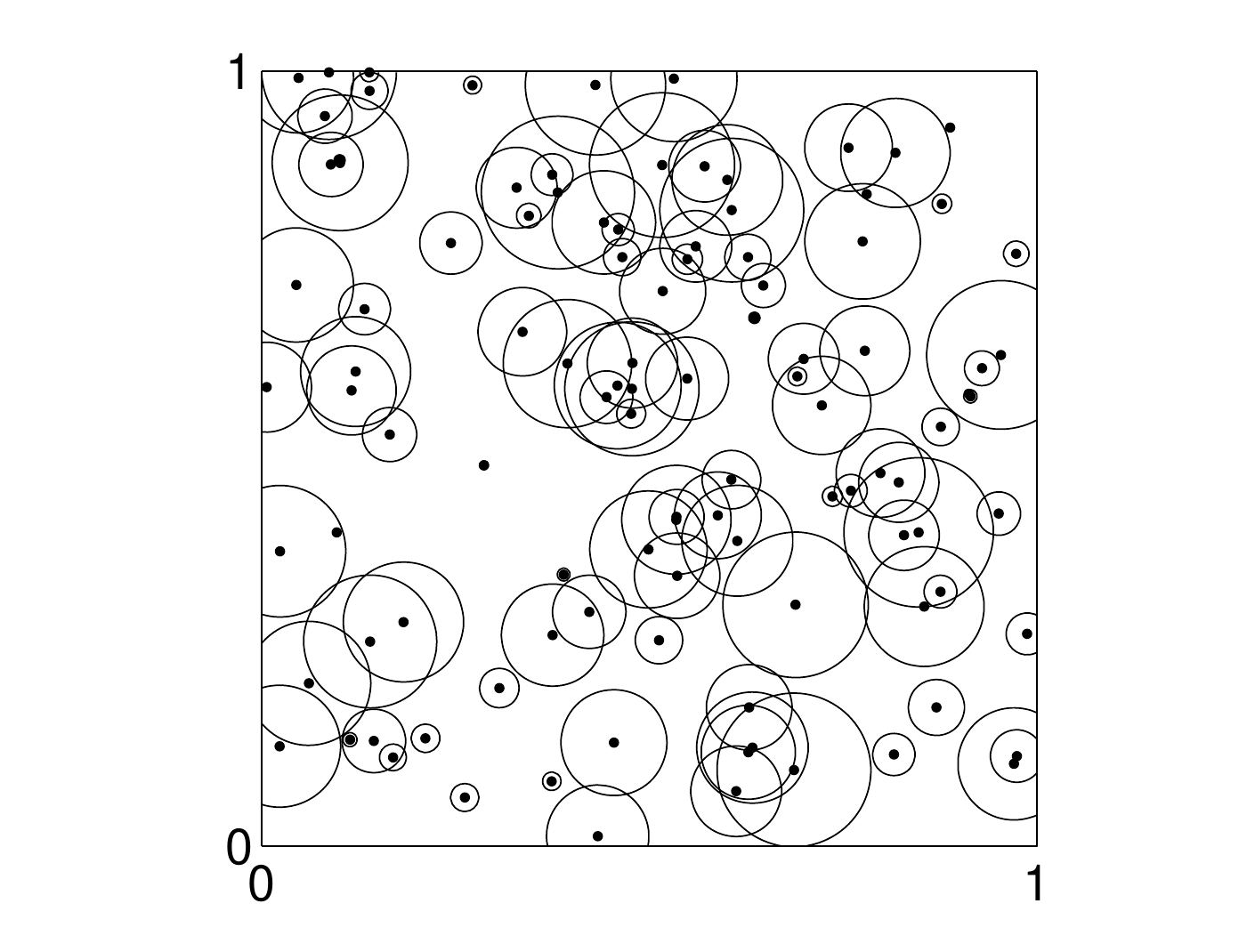}}
\caption{A realization of a marked Poisson process with independent
marking on the unit square.
 The Poisson process has intensity 100. The marks are uniformly distributed on [0,0.1]
and correspond to the radii of the circles.} 
\label{fig:markedPoisson}
\end{figure}

The generation of a marked Poisson process with independent marks is
virtually identical to that of an ordinary (that is, non-marked)
Poisson process. The only difference is that for each point the mark
has to be drawn from the mark distribution. 
An example of a realization of a marked Poisson process is given in
Figure~\ref{fig:markedPoisson}. Here the marks are uniformly
distributed on [0,0.1], and the underlying Poisson process is
homogeneous with intensity 100.

\subsection{Cluster Process}
In nature one often observes point patterns that display clustering.
An example is the spread of plants from a weed
species where the plants are initially introduced by birds at a
number of geographically dispersed locations and the plants then
spread themselves locally. 

Let  ${\bf
C}$ be a point process of ``centers'' and associate with each $c\in {\bf C}$ a
 point process $\bX^{c}$, which may include $c$. The combined set of points $\bX=\cup_{c\in {\bf
C}}\bX^{c}$ constitutes a {\bf cluster process}. If ${\bf C}$ is a Poisson process, then $\bX$ is
called a {\bf Poisson cluster process}. 

As a specific example,
consider the following {\bf Hawkes process}. Here, the center process
${\bf C}$ is a Poisson process on  $\R^d$ (or a subset thereof) with some intensity function
$\lambda(\cdot)$. The clusters are generated as follows.
 For each $c \in {\bf C}$, independently generate ``first-generation offspring'' according to a  Poisson process with intensity 
$\rho(x-c)$, where $\rho(\cdot)$ is a positive function on $\R^d$ with
integral less than unity. Then, for each first-generation offspring $c_1$ 
generate a Poisson process with intensity $\rho(x - c_1)$, and so on.
The collection of all generated points forms the Hawkes
process. The requirement that $\int \rho(y)\, \di y < 1$ simply means that 
 the expected number of offspring of each point is less than one. 

\begin{figure}[H]
\centering
\includegraphics[width=0.6\linewidth]{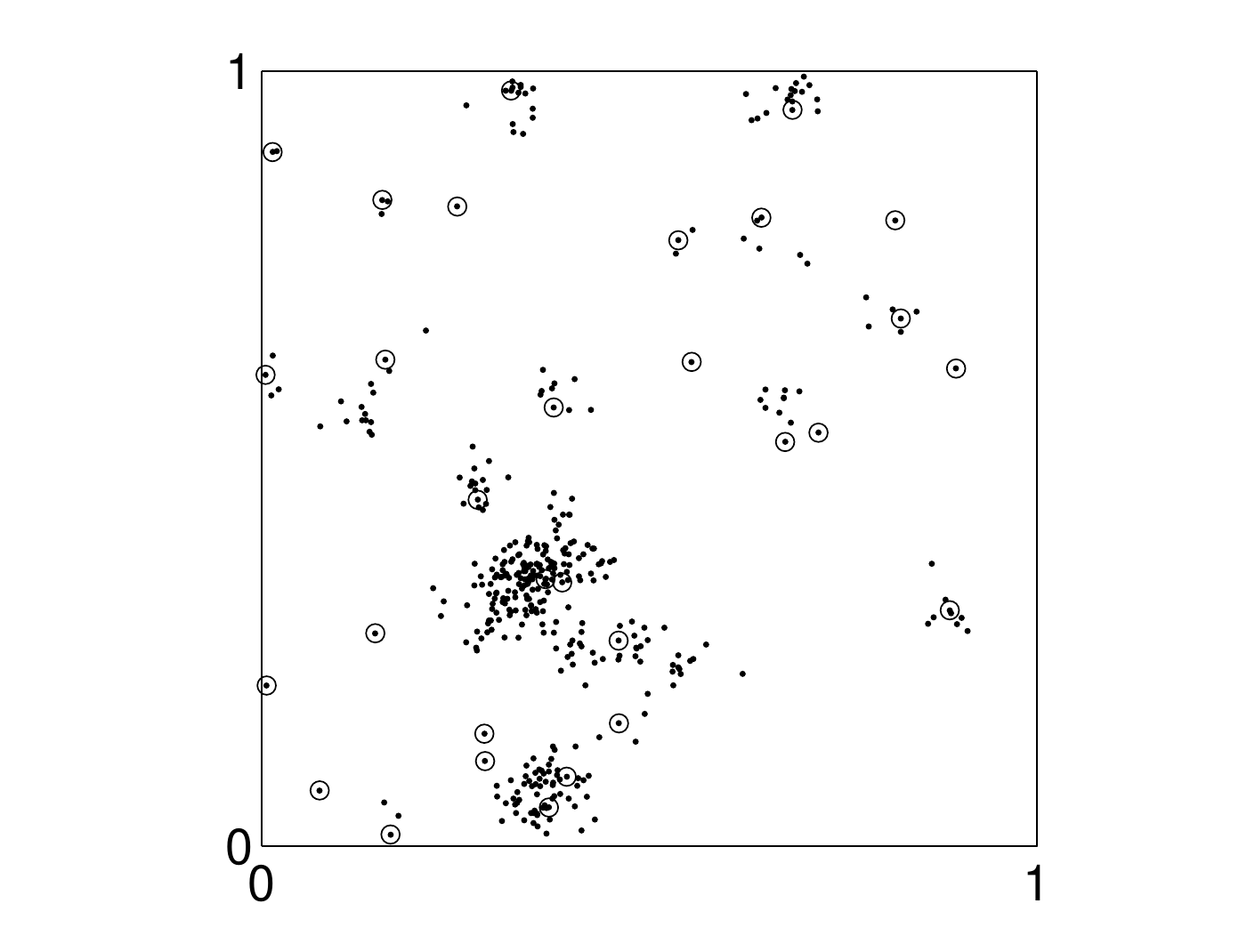}
\caption{Realization of a two-dimensional Hawkes process with
centers (encircled) forming a Poisson process on $[0,1]\times[0,1]$
with intensity $\lambda = 30$. The offspring intensity is given in \eqref{offs}.}
\label{fig:Hawkes}
\end{figure}

Figure~\ref{fig:Hawkes} displays a realization for $\R^2$ with the
cluster centers forming a Poisson process on the unit square with
intensity $\lambda = 30$.  The offspring intensity is here
 \begin{equation}\label{offs}
\rho(x_1,x_2) = \frac{\alpha}{2 \pi \sigma^2} \e^{-\frac{1}{2 \sigma^2}(x_1^2
+ x_2^2)}, \quad (x_1,x_2) \in \R^2, 
\end{equation}
with 
$\alpha = 0.9$ and 
with $\sigma = 0.02$. This means that the number $N$ of offspring for a
single generation from a parent at position $(x_1,x_2)$ 
has a Poisson distribution with parameter
$\alpha$. And given $N=n$, the offspring are iid distributed
according to a bivariate normal random vector with independent
components with means $x_1$ and $x_2$ and both variances
$\sigma^2$. In Figure~\ref{fig:Hawkes} the cluster centers are
indicated by circles. Note that the 
process possibly has points outside the displayed box. The $\matlab$ code is
given below. 


\newpage

\begin{breakbox}
\begin{verbatim}
lambda = 30;           %intensity of initial points (centers)
mean_children = 0.9;   %mean number of children of each point
X = zeros(10^5,2);     %initialise the points
N = poissrnd(lambda);  %number of centers
X(1:N,:) = rand(N,2);  %generate the centers 
total_so_far = N;      %total number of points generated
next = 1;
while next < total_so_far 
    nextX = X(next,:);   %select next point
    N_children = poissrnd(mean_children);    %number of children 
    NewX = repmat(nextX,N_children,1) + 0.02*randn(N_children,2); 
    X(total_so_far+(1:N_children),:) = NewX;  %update point list 
    total_so_far = total_so_far+N_children;
    next = next+1;
end
X=X(1:total_so_far,:); %cut off unused rows
plot(X(:,1),X(:,2),'.')
\end{verbatim}
\end{breakbox}

\subsection{Cox Process}
Suppose we wish to model a point pattern
of trees given that we know the soil quality $h(x_1,x_2)$ at each
point $(x_1,x_2)$. We
could use a non-homogeneous Poisson process with an intensity
$\lambda$ that is an increasing function of $h$; for
example, $\lambda(x_1,x_2)=\e^{\alpha + \beta h(x_1,x_2)}$ for
some known $\alpha>0$ and $\beta>0$. In practice, however, the soil quality itself could be
random, and be described via a random field. Consequently,
one could 
could try instead to model the point pattern as a Poisson process with
a {\em random} intensity function  $\Lambda$.
 Such processes were introduced by Cox as
doubly stochastic Poisson processes and are now called  Cox
processes \cite{cox_1955}. 

More precisely, we say that $\bX$ is a {\bf Cox process} driven by the random
intensity function $\Lambda$ if conditional on  $\Lambda$
the point process $\bX$ is Poisson with
intensity function $\lambda$. Simulation of a Cox process on a set
$\scT \subset \R^d$ is thus a two-step procedure.


\begin{alg}[Simulation of a Cox Process]\label{alg:coxprocess}~
\begin{enumerate}
\item Simulate a realization $\lambda = \{\lambda(x), x \in \scT\}$ of the
random intensity $\Lambda$.
\item Given $\Lambda = \lambda$, simulate $\bX$ as an inhomogeneous Poisson process with intensity $\lambda$. 
\end{enumerate}
\end{alg}

\begin{figure}[H]
\centering
\subfigure[]{
\includegraphics[width=0.45\linewidth]{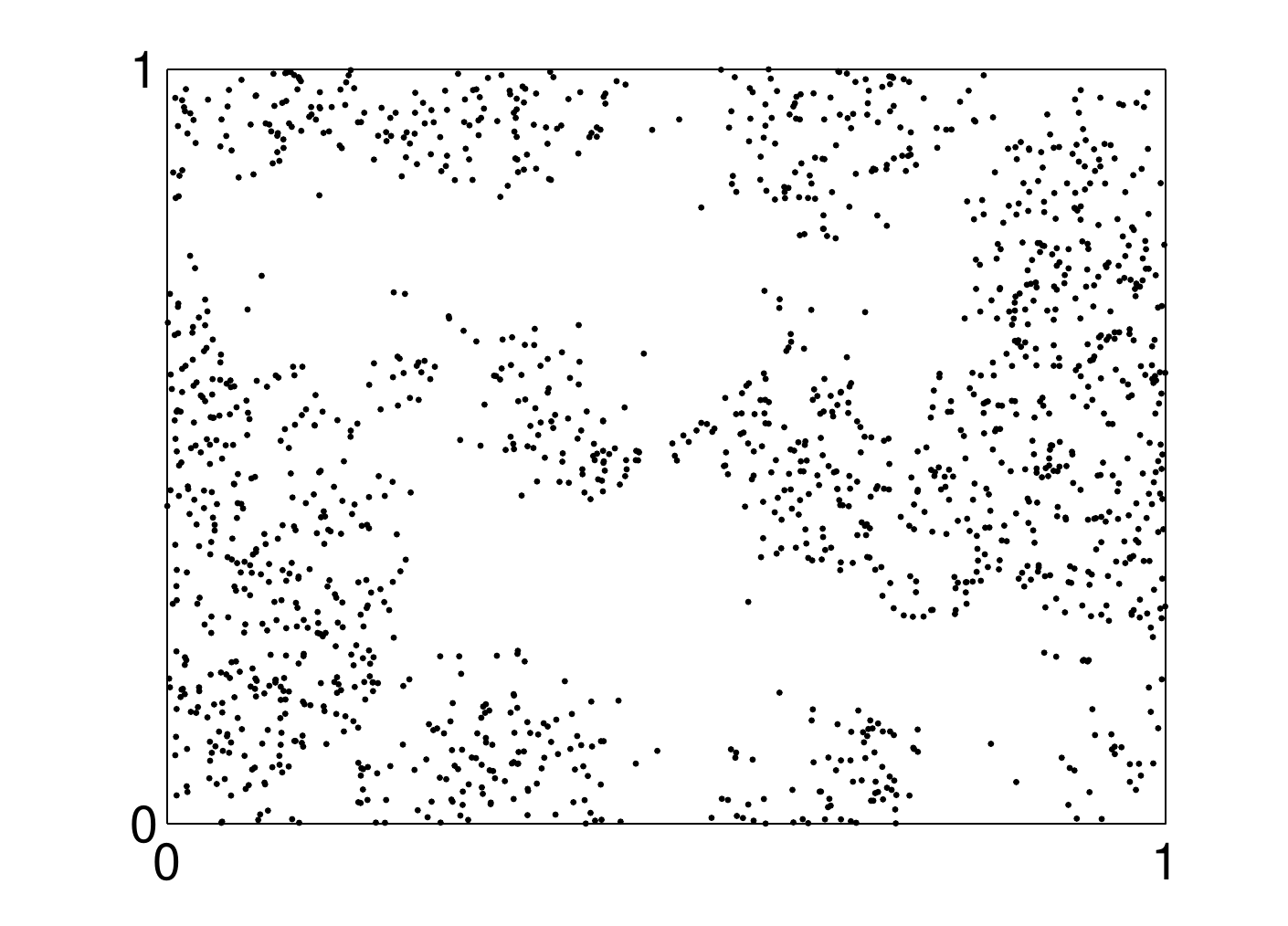}
\label{subfig:CoxProcess}
}
\subfigure[]{
\includegraphics[width=0.45\linewidth]{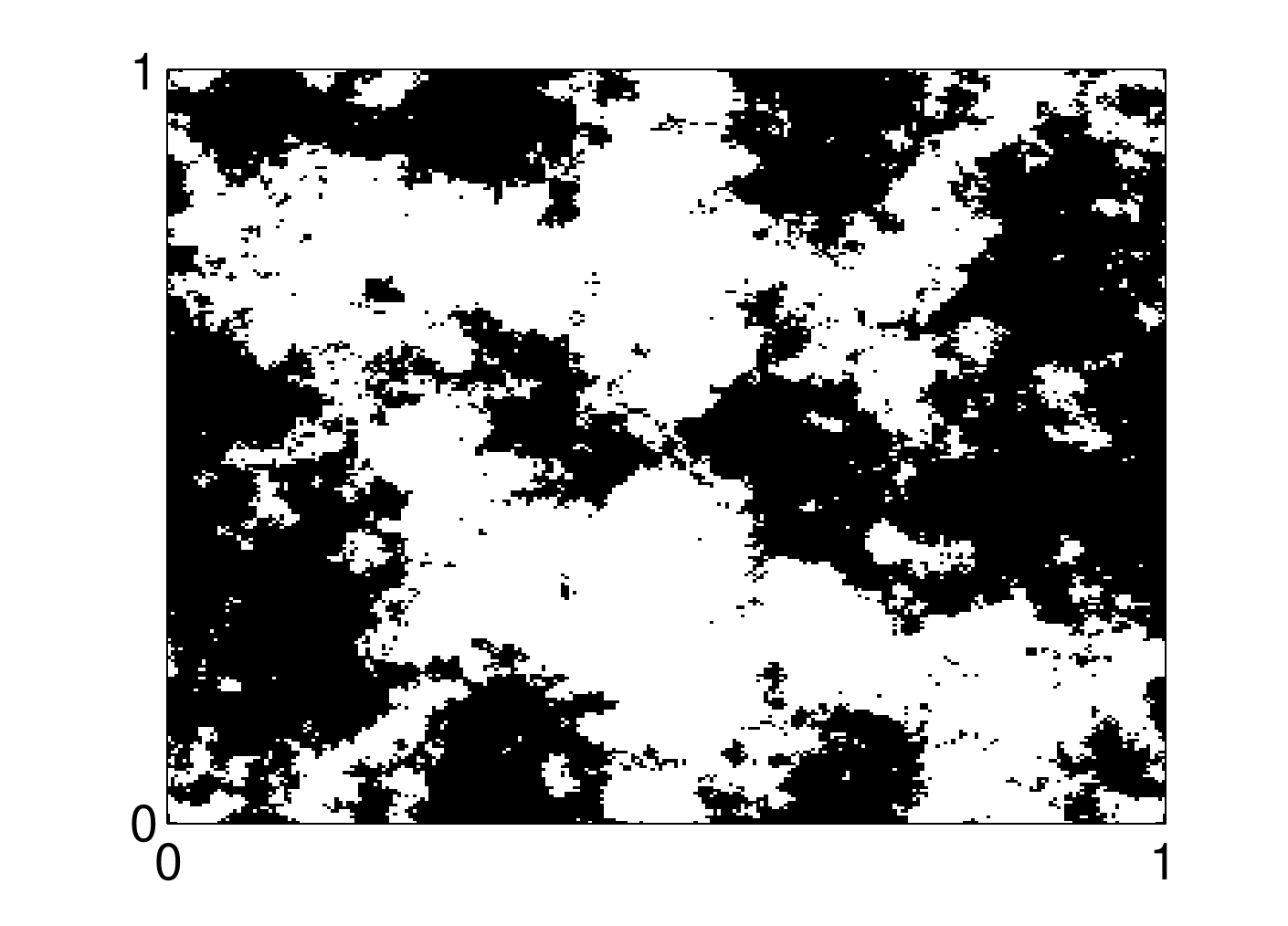}
\label{subfig:randomintensity}
}
\caption{Realization of a Cox
process: (a) on a square generated by a the random intensity  function
given in (b).  The black points have intensity 3000 and the white points have intensity
0.} 
 \label{fig:CoxBrowniansheet}
\end{figure}

Figure~\ref{subfig:CoxProcess} shows a realization of a Cox process on
the unit square $\scT = [0,1]\times [0,1]$, with a  random intensity whose realization is given in 
Figure~\ref{subfig:randomintensity}. The random intensity at position 
$(x_1,x_2)$ is either 3000 or 0, depending on whether the value of a
random field on $\scT$ is negative or not. The random field 
that we used is the stationary Gaussian process on the torus
described in Section~\ref{sec:GMRF}.

 Given an outcome of the random
intensity, the Cox process is constructed by generating a
homogeneous Poisson process with rate $3000$ on $\scT$ and accepting only
those points for which the random intensity is non-zero. The value of the
intensity is taken to be constant within each square $\{(i + u)/n, (j
+ v)/n), 0 \leq u, v \leq 1\}$, $i,j = 0, \ldots,n-1$.
The following $\matlab$ code is to be appended to the code used for
the stationary Gaussian process generation  on
the torus in Section~\ref{sec:GMRF}. 
\medskip

\begin{breakbox}
\begin{verbatim}
Lambda = ones(n,n).*(X < 0); %random intensity
%imagesc(Lambda); set(gca,'YDir','normal')
Lambda = Lambda(:); %reshape as a column vector
N = poissrnd(3000);
P = rand(N,2); %generate homogenous PP
Pn = ceil(P*n); %PP scaled by factor n
K = (Pn(:,1)-1)*n + Pn(:,2); % indices of scaled PP
ind = find(Lambda(K)); %indices for which intensity is 1
Cox = P(ind,:);  %realization of the Cox process
plot(Cox(:,1),Cox(:,2),'.');
\end{verbatim}
\end{breakbox}
\medskip

Neyman and Scott \cite{NeymanScott} applied the following Cox process
to cosmology. Used to model the positions of stars in the universe, it
now bears the name {\bf Neyman--Scott process}. Suppose $\bC$ is a
homogeneous Poisson process in $\R^d$ with constant intensity $\kappa$. Let the random intensity $\Lambda$ be given by
\[
\Lambda(x)=\alpha \sum_{c\in \bC}k(x-c)
\]
for some $\alpha>0$ and some $d$-dimensional probability density function $k$. 
Such a Cox process is also a Poisson cluster process. Note that in
this case the cluster centers are not part of the Cox process. Given a
realization  of the cluster center process $\bC$,  the cluster of points
originating from $c \in \bC$ form a non-homogeneous Poisson process with intensity
$k(x-c), x \in \R^d$, independently of the other clusters.
 Drawing such a cluster via
Algorithm~\ref{alg:poi_rand_measure} 
simply means that (1) the number of points $N_c$ in the cluster has a
$\Poi(\alpha)$ distribution, and (2) these $N_c$ points are iid distributed
according to pdf $k(x -c)$.

A common choice for pdf $k$ is $k(x)\propto \I_{\{ \| x \| \leq 
r \}}$, first  proposed by Mat\'ern \cite{Matern}. The resulting process is
called a  {\bf Mat{\'e}rn
process}. Thus, for a  Mat{\'e}rn
process each point in the cluster with center $c$ is uniformly distributed within a  ball of
radius $r$ at  $c$. If instead a $\Nor(c,\sigma^2 I)$ distribution is
used, where $I$ is the $d$-dimensional identity matrix, then the
process is known as a (modified) {\bf Thomas process}
\cite{Jesper}. 

\begin{figure}[H]
\centerline{\includegraphics[width=0.5\linewidth]{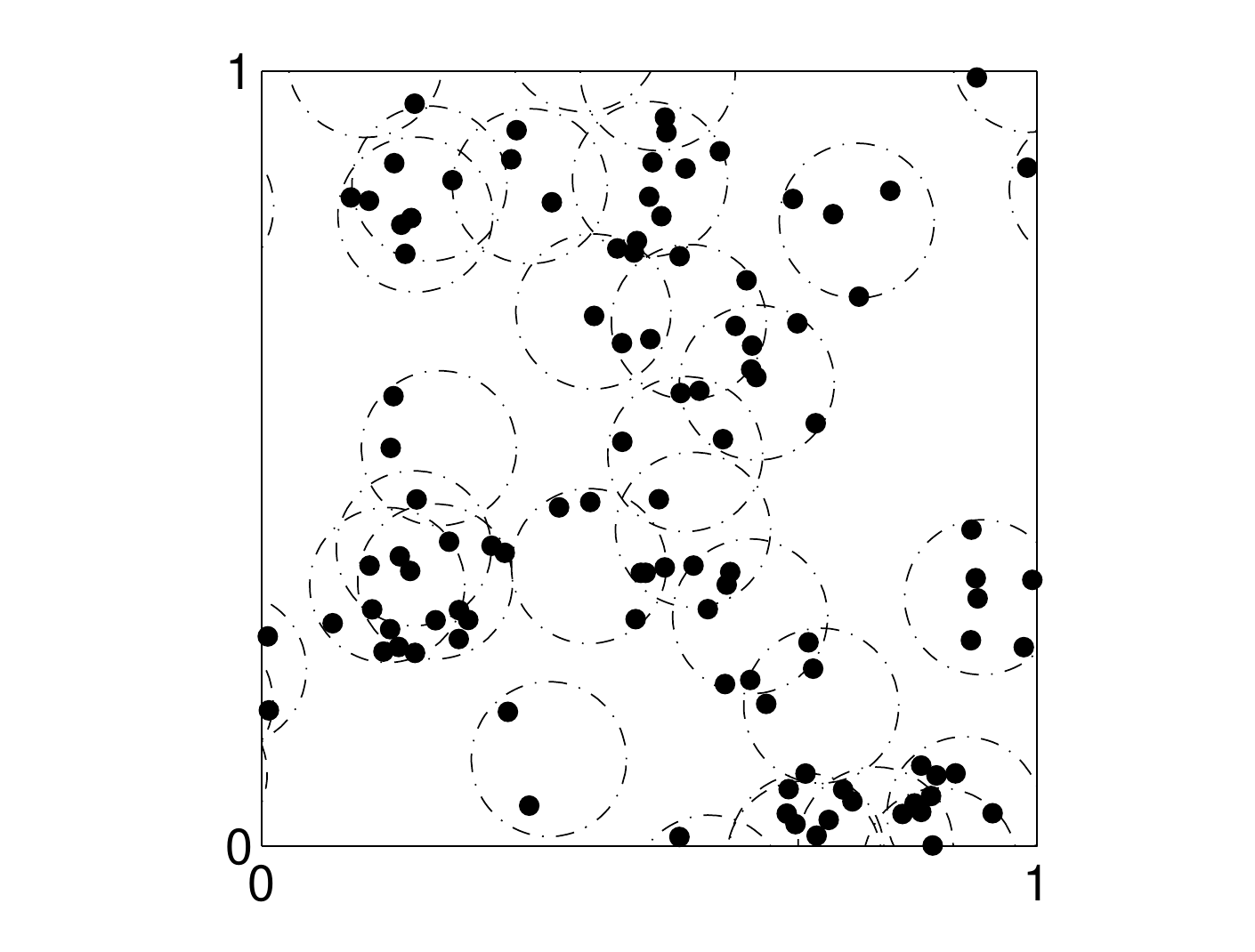}}
\caption{A realization of a Mat\'ern  process with parameters $\kappa
= 20$, $\alpha = 5$, and $r = 0.1$. The process extends beyond the
shown window. The cluster centers
(the centers of the circles) are
not part of the point pattern.}
\label{fig:Matern}
\end{figure}

Figure~\ref{fig:Matern} shows a realization of a Mat\'ern process with
parameters $\kappa = 20$, $\alpha = 5$ and $r = 0.1$ on 
$[0,1]\times[0,1]$. Note that the cluster centers are assumed to lie
on the whole of $\R^2$. To show a genuine outcome of the process
within the window $[0,1]\times[0,1]$ it suffices to consider only the
points that are generated from centers lying in square
$[-r,1+r]\times[-r,1+r]$. 

\newpage
The following $\matlab$ program was used.
\medskip

\begin{breakbox}
\begin{verbatim}
X = zeros(10^5,2);  %initialise the points
kappa = 20; alpha = 5; r = 0.1; %parameters
meanpts=kappa*(1 + 2*r)^2; 
N = poissrnd(meanpts);     %number of cluster centers
C = rand(N,2)*(1+2*r) - r;  ;%draw cluster centers                             
total_so_far = 0;
for c=1:N
    NC = poissrnd(alpha);  %number of points in cluster
    k = 0;
    while k < NC %draw uniformly in the n-ball via accept-reject
        Y = 2*r*rand(1,2) - r; %candidate point
        if norm(Y) < r   
            X(total_so_far+k+1,:) = C(c,:) + Y;
            k = k+1;
        end
    end
    total_so_far = total_so_far + NC;
end
X = X(1:total_so_far,:);    %cut off unused rows
plot(X(:,1),X(:,2),'.')
axis([0, 1,0, 1])
\end{verbatim}
\end{breakbox}
\medskip

A versatile generalization the Neyman--Scott process is the 
 {\bf Shot noise Cox process}   \cite{moller_2003}, where the
 random intensity is of the form
\[
\Lambda(x)=\sum_{(c_{j},\gamma_{j}) \in \bZ}\gamma_{j} \, k(c_{j},x)\;,
\]
and $\{(c_{j},\gamma_{j})\}$ are the points of an inhomogeneous
Poisson process $\bZ$ on $\R^d \times \R_+$ with intensity
function $\zeta$, and $k$ is a {\bf kernel function}; that is,
$k(c,\cdot)$ is a probability density function (on $\R^d$) for each $c
\in \R^d$. 
By Theorem~\ref{the:pois}, if there exists a function $K(c):\R^d\rightarrow \R_+$ such that
\[
K(c)=\int_{0}^{\infty}\zeta(c,\gamma) \, \di \gamma < \infty\;,
\]
then $\bC =\{ c : (c,\gamma) \in \bZ \}$ is a Poisson process with
intensity function $K$, and $\bZ$ is a marked Poisson process with
mark density $\zeta(c,\cdot)/K(c)$. This decomposition of the shot noise Cox process into a marked Poisson process suggests a method for simulation.
\newpage

\begin{alg}[Simulation of a Shot-Noise Cox Process]~
\begin{enumerate}
\item Draw the points $\bC$ of an inhomogeneous Poisson process with intensity function $K(c)$.
\item For each $c_{j}\in \bC$, draw the corresponding mark $\gamma_{j}$ from the density $\zeta(c_{j},\cdot)/K(c_{j})$.
\item For each $c_{j}\in \bC$, draw $N_{j}\sim {\Po}(\gamma_{j})$.
\item Draw for each $c_{j}\in C$, $N_{j}$ points from the kernel $k(c_{j},x)$. The collection of all points drawn at this step constitutes a realization from the shot-noise Cox process.
\end{enumerate}
\end{alg}

A special case of the shot-noise Cox process, and one that appears
frequently in the literature, is the {\bf shot-noise G Cox process}.
Here the intensity function $\zeta$ is of the form
\[
\zeta(c,\gamma)=\beta\gamma^{\alpha-1}\exp(-\lambda \gamma)/\Gamma(1+\alpha)\;,
\]
where $\beta>0,\alpha>0,\lambda>0$, and 
\[
K(c)=\int_{0}^{\infty}\zeta(c,\gamma)\, \di \gamma =\beta \lambda^{-\alpha}/\alpha< \infty\;.
\]
Hence, $\bZ$ is a marked Poisson process, where the intensity
function of the centers is $K(c)=\beta\lambda^{-\alpha}/\alpha$ and
the marks are independently and identically ${\sf
Gamma}(\alpha,\lambda)$ distributed; see  \cite{Jesper}
and 
 \cite{brix} for more information.

\subsection{Point Process Densities}
Let $\bX$ be a point process with mean measure $\mu$ on a bounded
region $E \subset \R^d$. We assume that the expected total number of
points $\mu(E)$ is finite. We may view $\bX$ as a random object taking
values in the
 space $\scX = \cup_{n=0}^{\infty} \left( \{n \}\times E^n\right)$, where
$E^n$ is the Cartesian product of $n$ copies of $E$. Note that in
this representation the coordinates of $\bX$ are ordered: $\bX =
(N,(X_1,\ldots,X_N))$. We  identify each vector $(N,\bX)$ with $\bX$.  
 If, for every
set $\{n\}\times A$ with $A$ a measurable set in $E^n$ we can write 
\[
\Pm(\bX \in \{n\} \times A) = \int_{A} f(x_1,\ldots,x_n) \, \di x_1
\ldots \di x_n \;, 
\]
then $f(\bx)$ is the probability density function or simply {\bf density} of $\bX$ on $\scX$
(with respect to the Lebesgue measure on $\scX$). Using the
identification $(n,\bx) = \bx$, we can view $f(\bx)$ as the joint
density of the random variable $N$ and the $N$-dimensional vector
$\bX$, where each component of $\bX$ takes values in $E$. Using a
Bayesian notation convention where all pdfs and conditional pdfs are
indicated by the same symbol $f$, we have  
$f(\bx)  = f(n,\bx) = f(n) f(\bx\gvn n)$, where $f(n)$ is the
(discrete) pdf of the random number of points $N$, and $f(\bx\gvn n)$
is the joint pdf of $X_1,\ldots,X_n$ given $N=n$.   
As an example, for the Poisson process on $E$ with intensity function
$\lambda(x)$ and mean measure $\mu$ we have, in correspondence to 
Algorithm~\ref{alg:poi_rand_measure},    
\begin{equation}\label{denspois}
f(\bx) = f(n)\,  f(\bx\gvn n) = \frac{\e^{-\mu(E)} \{\mu(E)\}^n}{n!}
\prod_{i=1}^n \frac{\lambda(x_i)}{\mu(E)} = \frac{\e^{-\mu(E)}}{n(\bx)!
}\prod_{i=1}^{n(\bx)} \lambda(x_i)\;, 
\end{equation}
where $n(\bx)$ is the number of components in $\bx$. Conversely, the 
expression for the 
pdf in \eqref{denspois} shows immediately how $\bX$ can be
generated; that is,  via Algorithm~\ref{alg:poi_rand_measure}.  

A general recipe for generating a point process $\bX$ is thus:
\begin{enumerate}
\item draw $N$ from the discrete pdf $f(n)$;  
\item given $N = n$, draw $(X_1,\ldots,X_n)$ from the conditional pdf $f(\bx\gvn n)$. 
\end{enumerate} 

Unfortunately, the pdf $f(n)$ and $f(\bx \gvn n)$ may not be available
explicitly. Sometimes $f(\bx)$ is known  up to an unknown  normalization
constant. In such cases one case use 
 {\bf Markov Chain Monte Carlo} (MCMC) to simulate from $f(\bx)$. 
The basic idea of MCMC is to run a 
 Markov chain long enough so that its limiting distribution
 is close to the target distribution. The most well-known MCMC algorithm
is the following; see, for example, \cite{mcmc:robcas04}.

 \index{Metropolis--Hastings algorithm} 
 \begin{alg}
{\bf (Metropolis--Hastings Algorithm)} \rm \label{alg.mcmc}  ~ 
 Given a {\bf transition density} $q(\by \gvn \bx)$, and 
starting from an initial state $\bX_0$, repeat the following steps from
$t = 1$ to $N$: 
 \begin{enumerate}
 \addtolength{\itemsep}{3pt}
 \item  Generate a candidate $\bY \sim q(\by \gvn \bX_t)$.
 \item Generate  $U\sim \U (0,1)$
 and set
 \begin{equation}   \label{mcmc4}
  \bX_{t+1}  = \begin{cases}
 \bY,  &  \ \text{ if }  \  U  \le  \alpha(\bX_t, \bY)  \\
  \bX_t &  \ \text{ otherwise},
 \end{cases}
 \end{equation}
  where $\alpha(\bx,\by)$ is the {\bf acceptance probability}, given by: 
  \begin{equation}
  \alpha(\bx, \by) = \min
 \left\{ \frac{f(\by)  \, q(\bx\gvn \by)}{f(\bx) \,  q(\by \gvn \bx)}, \: 1 \right\}. 
 \label{mcmc5} \end{equation}
  \end{enumerate}
  \end{alg}
This produces a sequence
 $\bX_1,\bX_2,\ldots$ of {\em dependent} random vectors, with $\bX_t$
 approximately distributed according to $f(\bx)$, for large $t$. 
Since Algorithm \ref{alg.mcmc} is of the acceptance--rejection type, its
efficiency depends on the acceptance probability $\alpha(\bx, \by)$.
Ideally, one would like the proposal transition density $q(\by \gvn\bx)$ to reproduce the desired pdf
$f(\by)$ as faithfully as possible. For a 
{\bf random walk sampler} the proposal state $\bY$, for a given 
current state $\bx$, is given by  $\bY = \bx + \bZ$, where $\bZ$ is typically generated from some spherically symmetrical
distribution. In that case the proposal
transition density 
pdf is
symmetric; that is $q(\by\gvn \bx) = q(\bx\gvn \by)$. It follows that
the acceptance probability is:
\begin{equation}\label{alphrw}
\alpha(\bx,\by) = \min\left\{ \frac{f(\by)}{f(\bx)}, \: 1\right\}.
\end{equation}

As a specific example, suppose we wish to generate a  {\bf
Strauss process}  \cite{Straussprocess1,Straussprocess2}. This is a point process with density of the
form
\[
f(\bx) \propto \beta^{n(\bx)}\gamma^{s(\bx)}\;,
\]
where $\beta,\gamma \geq 0$ and $s(\bx)$ is the number of pairs of
points where the two points are within distance $r$ of each other. As
before, $n(\bx)$ denotes the number of points. The process exists
(that is, the normalization constant is finite) 
if $\gamma \leq 1$; otherwise, it does not exist in
general \cite{Jesper}. 

We first consider simulating from $f(\bx \gvn n)$ for a fixed
$n$. Thus, $f(\bx \gvn n) \propto \gamma^{s(\bx)}$, where $\bx =
(x_1,\ldots,x_n)$. 
The following $\matlab$ program implements a Metropolis--Hastings
algorithm for simulating a (conditional) Strauss process with $n = 200$
points on the unit
square $[0,1]\times [0,1]$, using the
parameter values $\gamma = 0.1$ and $r = 0.2$. Given a current state
$\bx = (x_1,\ldots,x_n)$, the proposal state $\bY = (Y_1,\ldots,Y_n)$
is identical to $\bx$ except for the $J$-th component, where $J$ is
a uniformly drawn index from the set $\{1,\ldots,n\}$. Specifically,
$Y_J = x_J + Z$, where $Z \sim \Nor(0,(0.1)^2)$. The proposal $\bY$
for this random walk sampler is
accepted with probability $\alpha(\bx,\bY) =
\min\{\gamma^{s(\bY)}/\gamma^{s(\bx)}, \: 1\}$.  The function $s(\bx)$
is implemented below as {\tt numpairs.m}.  
\medskip
\begin{breakbox}
\begin{verbatim}
gam = 0.1;
r = 0.2;
n = 200;
x = rand(n,2); %initial pp
K = 10000;
np= zeros(K,1);
for i=1:K
    J = ceil(n*rand);
    y = x;
    y(J,:) = y(J,:) + 0.1*randn(1,2); %proposal
    if (max(max(y)) > 1 || min(min(y)) <0)
        alpha =0; %don't accept a point outside the region
    elseif (numpairs(y,r) < numpairs(x,r))
        alpha =1;
    else
        alpha = gam^numpairs(y,r)/gam^numpairs(x,r);
    end
    R = (rand < alpha);
    x = R*y + (1-R)*x; %new x-value
    np(i) = numpairs(x,r);
    plot(x(:,1),x(:,2),'.'); 
    axis([0,1,0,1])
    refresh; pause(0.0001);
end
\end{verbatim}
\end{breakbox}
\begin{breakbox}
\begin{verbatim}
function s = numpairs(x,r)
n = size(x,1);
D = zeros(n,n);
for i = 1:n
   D(i,:) = sqrt(sum((x(i*ones(n,1),:) - x).^2,2));
end
D = D + eye(n);
s = numel(find((D < r)))/2;
end
\end{verbatim}
\end{breakbox}
\medskip

\begin{figure}[H]
\centering
\includegraphics[width=0.45\linewidth]{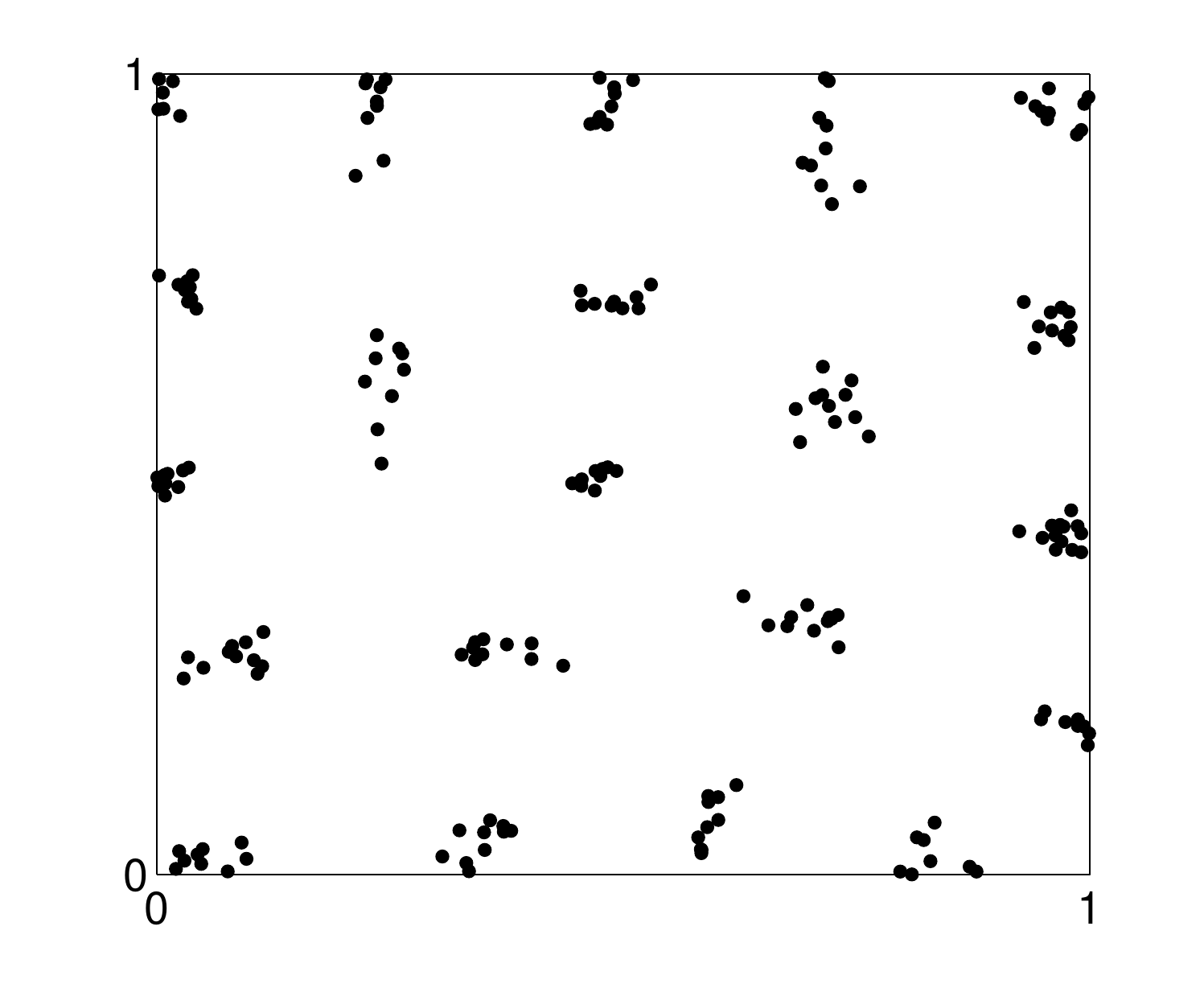}~~~
\includegraphics[width=0.45\linewidth]{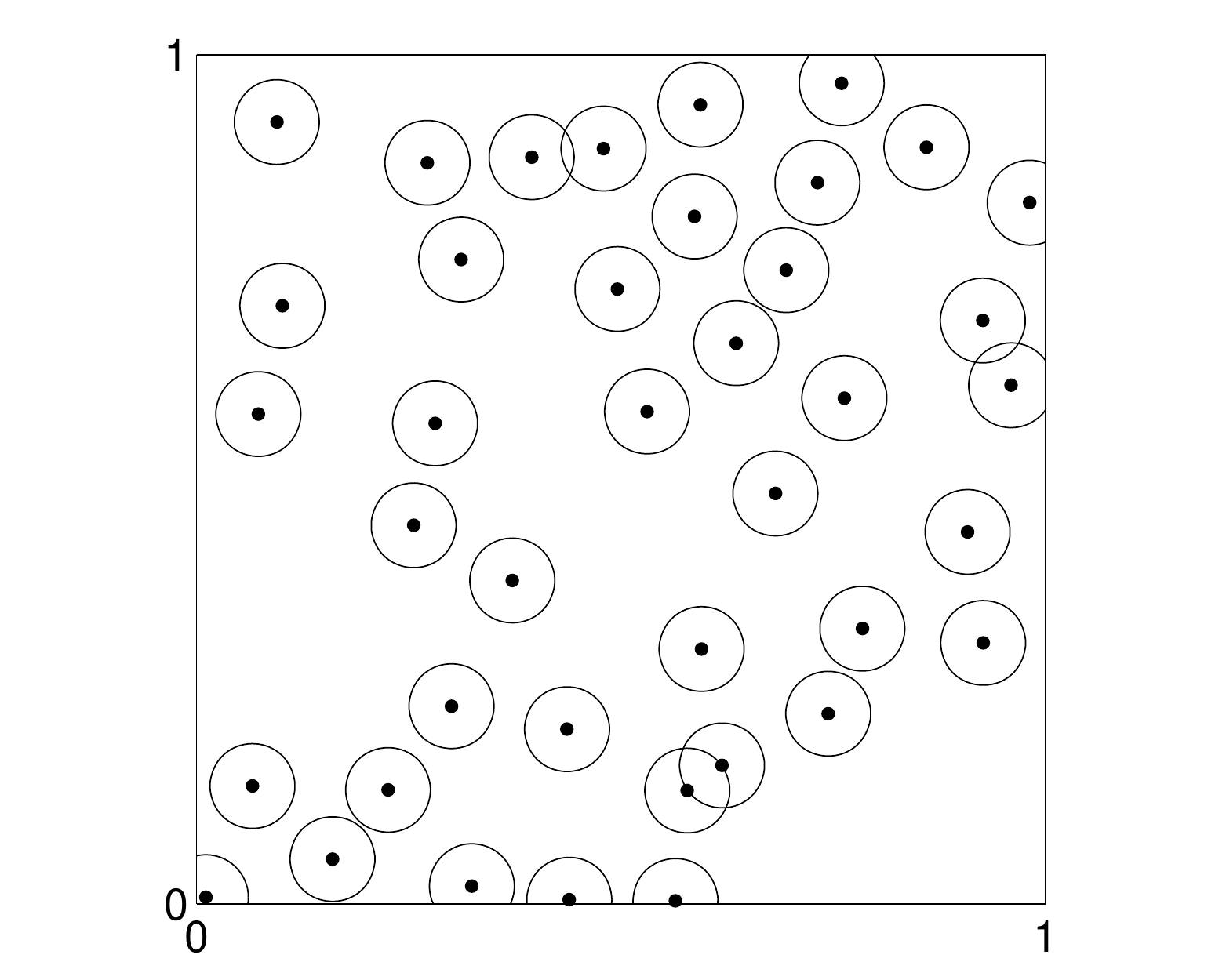}
\caption{Left: Conditional Strauss process with $n=200$ points and
parameters $\gamma = 0.1$ and $r = 0.2$. Right: Strauss
process with $\beta = 100$, $\gamma = 0.1$, and $r = 0.2$.}
\label{fig:Straussprocess}
\end{figure}

A typical realization of the conditional Strauss process is given in
 the left pane of Figure~\ref{fig:Straussprocess}. We see that the
 $n=200$ points are clustered together in groups. This pattern is
 quite different from a typical realization of the unconditional
 Strauss process, depicted in the right pane of
 Figure~\ref{fig:Straussprocess}. Not only are there typically far
 fewer points, but also these points tend to ``repel'' each other, so
 that the number of pairs within a distance of $r$ of each other is
 small. The radius of each circle in the figure is $r/2 = 0.1$. We see that  in
 this case $s(\bx) = 3$, because 3 circle pairs overlap. 



\newpage

To generate the (unconditional) Strauss process using the
Metropolis--Hastings sampler, the sampler needs to be able to 
``jump'' between
different dimensions $n$. 
The {\bf reversible jump sampler} \cite{mcmc:green,mcmc:robcas04} is 
an extension of the Metropolis--Hastings algorithm  designed
for this purpose. Instead of one transition density $q(\by \gvn
\bx)$, it requires a transition density for $n$, say $q(n \gvn m)$, and
for each $n$ a transition density $q(\by \gvn \bx, n)$ to jump from
$\bx$ to an $n$-dimensional vector $\by$. 

Given a current state $\bX_t$ of dimension $m$, Step~1 of Algorithm~\ref{alg.mcmc} is replaced with

\begin{enumerate}
\item[\em 1a.] Generate a candidate dimension $n \sim q(n \gvn m)$.
\item[\em 1b.] Generate an $n$-dimensional vector 
$\bY \sim q(\by \gvn \bX_t, n)$.  
\end{enumerate}
And in Step~2 the acceptance ratio in \eqref{mcmc5} is replaced with 
\begin{equation} 
\label{RJS acceptance}
\alpha(\bx,\by)
 =  \min\left\{
 \frac{ f(n,\by)\, q(m\gvn n )\, q(\bx\gvn\by, m)}{ f(m,\bx)\, q(n\gvn m)\, q(\by\gvn \bx, n)  }
, \: 1 \right\}\;.
\end{equation}

The $\matlab$ program below implements a simple version of the 
reversible jump sampler, suggested in 
\cite{geyerandmoller94}.
 From a current state $\bx$ of
dimension $m$, a candidate dimension is chosen to be either $m+1$ or
$m-1$, with equal probability. Thus, $q(m+1 \gvn m) = q(m-1 \gvn m) = 1/2, m=1,2,\ldots.$
The first transition corresponds to the birth of a
new point; the second to the death of a point. 
On the occasion of a birth, a candidate $\bY$ is generated by simply
appending a uniform point on $[0,1]^2$ to $\bx$. The corresponding
transition density is therefore $q(\by \gvn \bx,m) = 1$. On the
occasion of a death, the candidate $\bY$ is obtained from $\bx$ by
removing one of the points of $\bx$ at random and uniformly. The
transition density is thus $q(\by \gvn \bx,m) = 1/n(\bx)$.
This gives the acceptance ratios:
\begin{itemize}
\item {\em Birth}:
\[
\alpha(\bx, \by) = \min \left\{ \frac{\beta^{n(\by)} \gamma^{s(\by)} \frac{1}{n(\by)}}{\beta^{n(\bx)}
\gamma^{s(\bx)} 1}, \: 1 \right\} = \min\{ (\beta \gamma^{s(\by) -
s(\bx)})/n(\by), \:1 \}\;.
\] 
\item {\em Death}: 
\[
\alpha(\bx, \by) = \min \left\{\frac{\beta^{n(\by)} \gamma^{s(\by)} 1 }{\beta^{n(\bx)}
\gamma^{s(\bx)}  \frac{1}{n(\bx)}}, \: 1\right\} = \min\{(\gamma^{s(\by) - s(\bx)}
\, n(\bx))/\beta, \: 1\}\;. 
\] 
\end{itemize}

\begin{breakbox}
\begin{verbatim}
r = 0.1; gam = 0.2; beta = 100; %parameters
n = 200; x = rand(n,2); %initial pp
K = 1000; 
for i=1:K
    n = size(x,1);
    B = (rand < 0.5);
    if B %birth
        xnew = rand(1,2);
        y = [x;xnew];
        n = n+1;
    else %death
        y = setdiff(x,x(ceil(n*rand),:),'rows');
    end
    if (max(max(y)) > 1 || min(min(y)) <0)
        alpha =0; %don't accept a point outside the region
    elseif (numpairs(y,r) < numpairs(x,r))
        alpha =1;
    elseif B %birth
        alpha = beta*gam^(numpairs(y,r) - numpairs(x,r))/n;
    else  %death
        alpha = n*gam^(numpairs(y,r) - numpairs(x,r))/beta;
    end
    if (rand < alpha)
        x = y;
    end
    plot(x(:,1),x(:,2),'.'); 
    axis([0,1,0,1]); refresh; pause(0.0001);
\end{verbatim}
\end{breakbox}

\section{Wiener Surfaces}
\label{sec:GPBOTWP}

Brownian motion is one of the simplest  continuous-time stochastic  processes, 
and as such has found myriad applications in the physical sciences  \cite{lawler}. As a first step toward constructing  Brownian motion we introduce the one-dimensional
Wiener process $\{W_t,t\in\R_+\}$, which can be viewed  as a spatial process with a continuous index set on $\R_+$ and with a continuous state space $\mathbb{R}$.

\subsection{Wiener Process} 
\label{sec:Wiener process}
A one dimensional Wiener process is a stochastic process $\{W_t, t \geq 0\}$ characterized by the following properties: (1) the increments of $W_t$ are  stationary and normally distributed, that is, 
  $  W_t-W_s\sim \Nor(0,t-s) $
  for all $t\geq s\geq 0$; (2) $W$ has independent increments, that is, for  any $t_1<t_2\leq t_3<t_4$, the increments 
 $  W_{t_4}-W_{t_3}$ and $W_{t_2}-W_{t_1}$   are independent random variables (in other words,
 $W_{t}-W_s,\; t>s$ is independent of the past history of $\{W_u,\; 0\leq u\leq s\}$);
  (3) continuity of paths, that is, $\{W_t\}$ has continuous paths with $W_0=0$.

The simplest  algorithm for generating the process   uses the Markovian (independent increments) and Gaussian properties of the Wiener process.  Let $0=t_0 <t_1<t_2<\dots<t_n$ be the set of distinct times for which simulation of the process is desired. Then, the Wiener process is generated at times $t_1,\ldots,t_n$ via
\[ 
W_{t_k}=  \sum_{i=1}^k \sqrt{t_k-t_{k-1}}\; Z_i, \quad
k=1,\ldots,n\;,
\]
where $Z_1,\ldots,Z_n\simiid \Nor(0,1)$.  To obtain a continuous path approximation to the  path of the Wiener process, one could use linear interpolation on the points $W_{t_1},\dots,W_{t_n}$. 
A realization of  a Wiener process is given in the middle panel of  Figure~\ref{fig:fbm1d}.

Given the Wiener process $W_t$, we can now  define the 
$d$-dimensional \textbf{Brownian motion process}
via
\begin{equation}
\label{brownian motion}
\tilde \bX_t=\bmu \,t+\Sigma^{1/2} \bW_t,\quad \bW_t=(W_t^{(1)},\ldots,W_t^{(d)})^\T,\quad t\geq 0\;,
\end{equation}
where $W_t^{(1)},\ldots,W_t^{(d)}$ are independent Wiener processes and $\Sigma$ is a $d\times d$ covariance matrix. The parameter
 $\bmu\in \R^d$ is called the \emph{drift} parameter and $\Sigma$ is called the 
\emph{diffusion matrix}.
 
One approach to generalizing the Wiener process conceptually or  to higher spatial dimensions 
is to use  its characterization as a  zero-mean Gaussian process (see Section~\ref{sec:Gaussian property}) with   continuous sample paths and covariance function
$
 \Cov(W_t,W_s)=\min\{t,s\}
$
for $t,s>0$.
Since $\frac{1}{2}\left(|t|+|s|-|t-s|\right)=\min\{t,s\}$,
 we can consider the covariance function
 $\frac{1}{2}\left(|t|+|s|-|t-s|\right)$ as a basis for  generalization. The first generalization is obtained
 by considering a continuous zero-mean Gaussian process  with  covariance
 $\rho(t,s)=\frac{1}{2}\left(|t|^\alpha+|s|^\alpha-|t-s|^\alpha\right)$, where 
 $\alpha$ is a parameter such that $\alpha=1$ yields the  Wiener process. 
 This generalization gives rise to fractional Brownian motion discussed in the next section.
 \subsection{Fractional Brownian Motion}
 \label{sec:fbm}
 A continuous zero-mean Gaussian process $\{W_t, t\geq 0\}$ with
  covariance function 
 \begin{equation}
 \label{product form}
 \Cov(W_t,W_s)=\frac{1}{2}\left(|t|^\alpha+|s|^\alpha-|t-s|^\alpha\right),\quad t,s\geq 0
 \end{equation}
 is called \textbf{fractional Brownian motion} (fBm) with  roughness  parameter $\alpha\in(0,2)$. The process is  frequently parameterized with respect to $H=\alpha/2$,
 in which case $H\in(0,1)$ is called the \emph{Hurst} or \emph{self-similarity} parameter. The notion of self-similarity arises, because fBm
 satisfies the property that the rescaled process $\{c^{-H}\,W_{c\,t},t\geq 0\}$
 has the same distribution as $\{W_t, t\geq 0\}$ for all $c>0$.

\begin{figure}[htb]
\includegraphics[width=.34\linewidth]{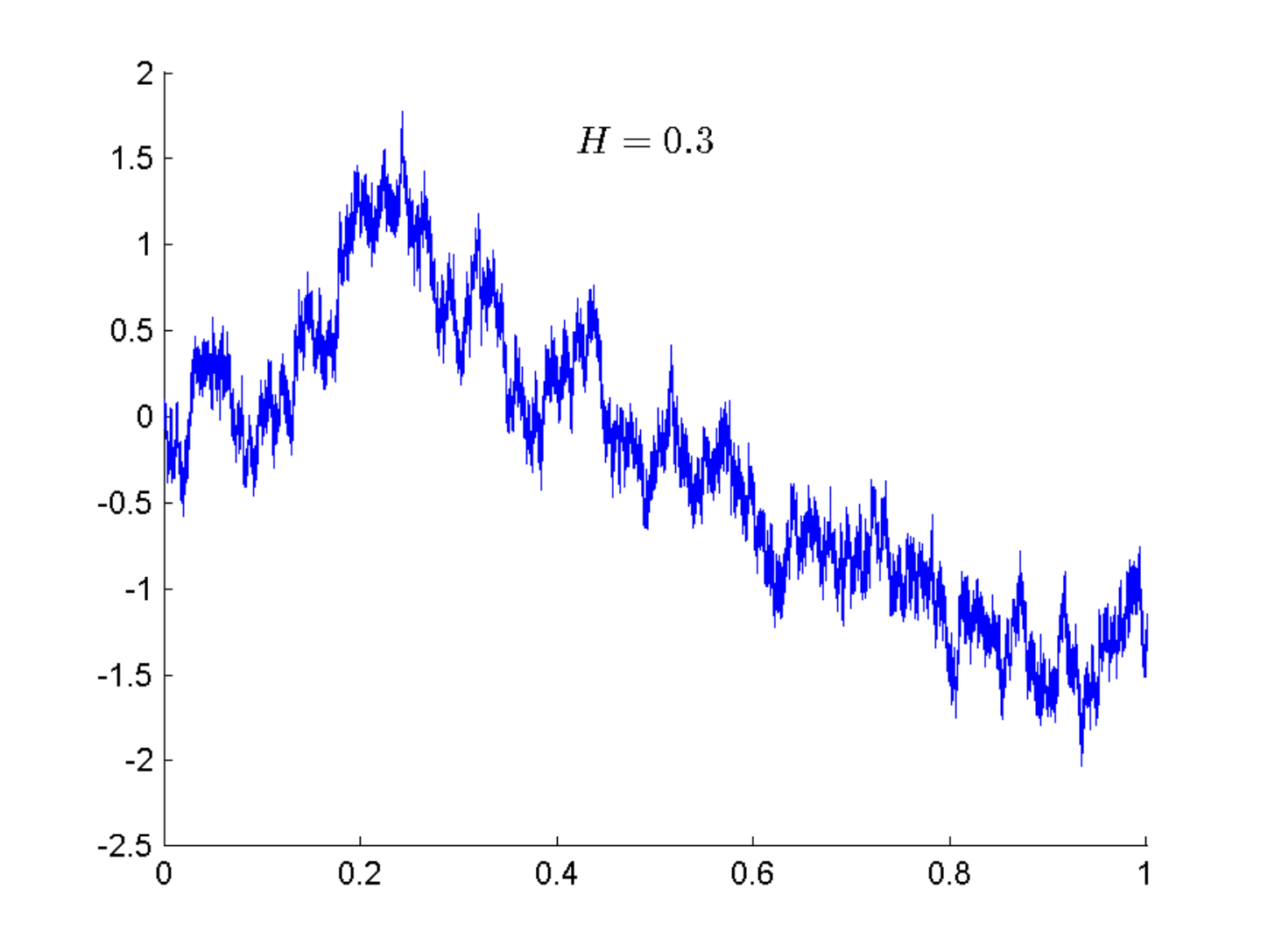}~
\includegraphics[width=.34\linewidth]{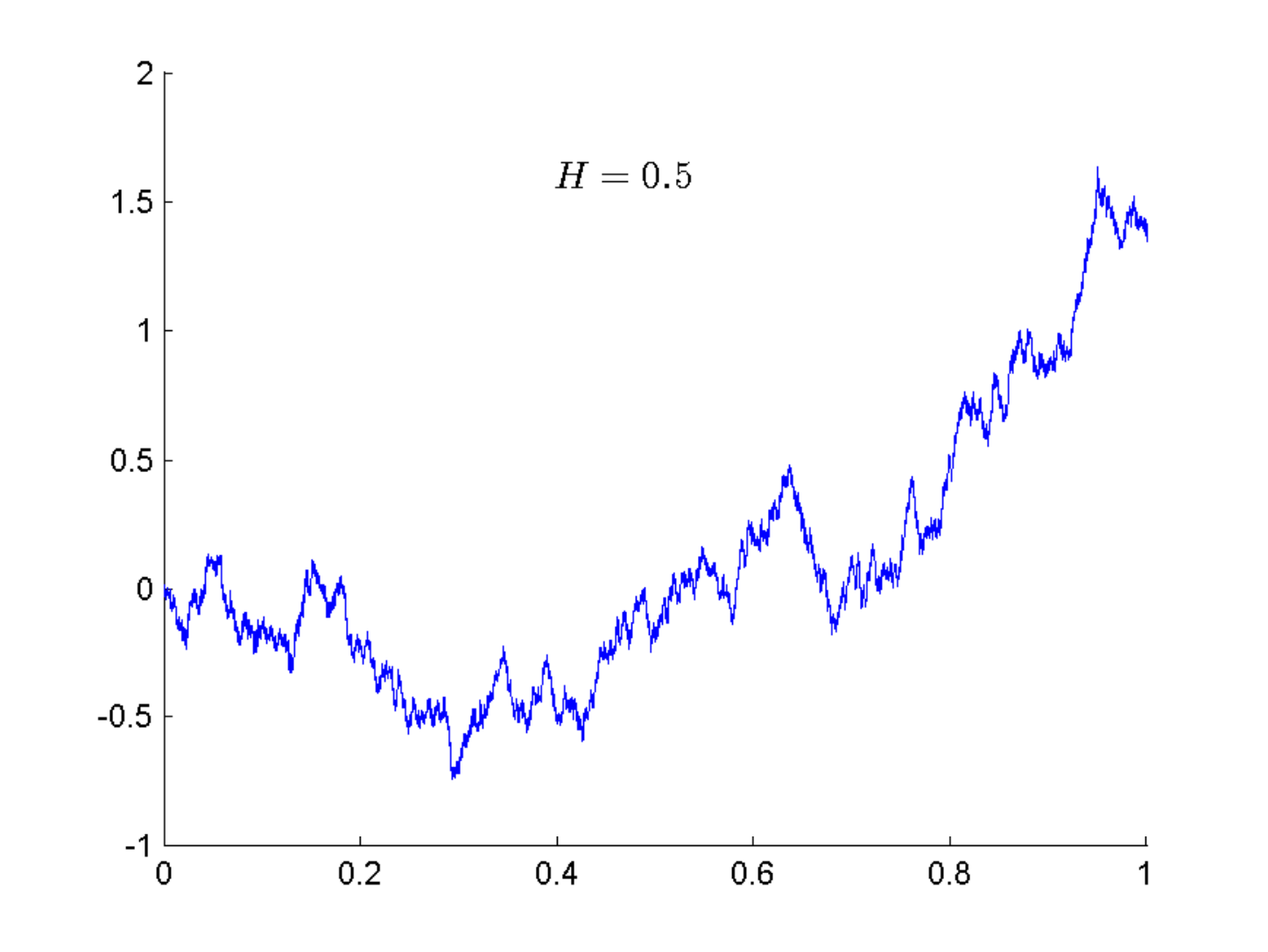}~
\includegraphics[width=.34\linewidth]{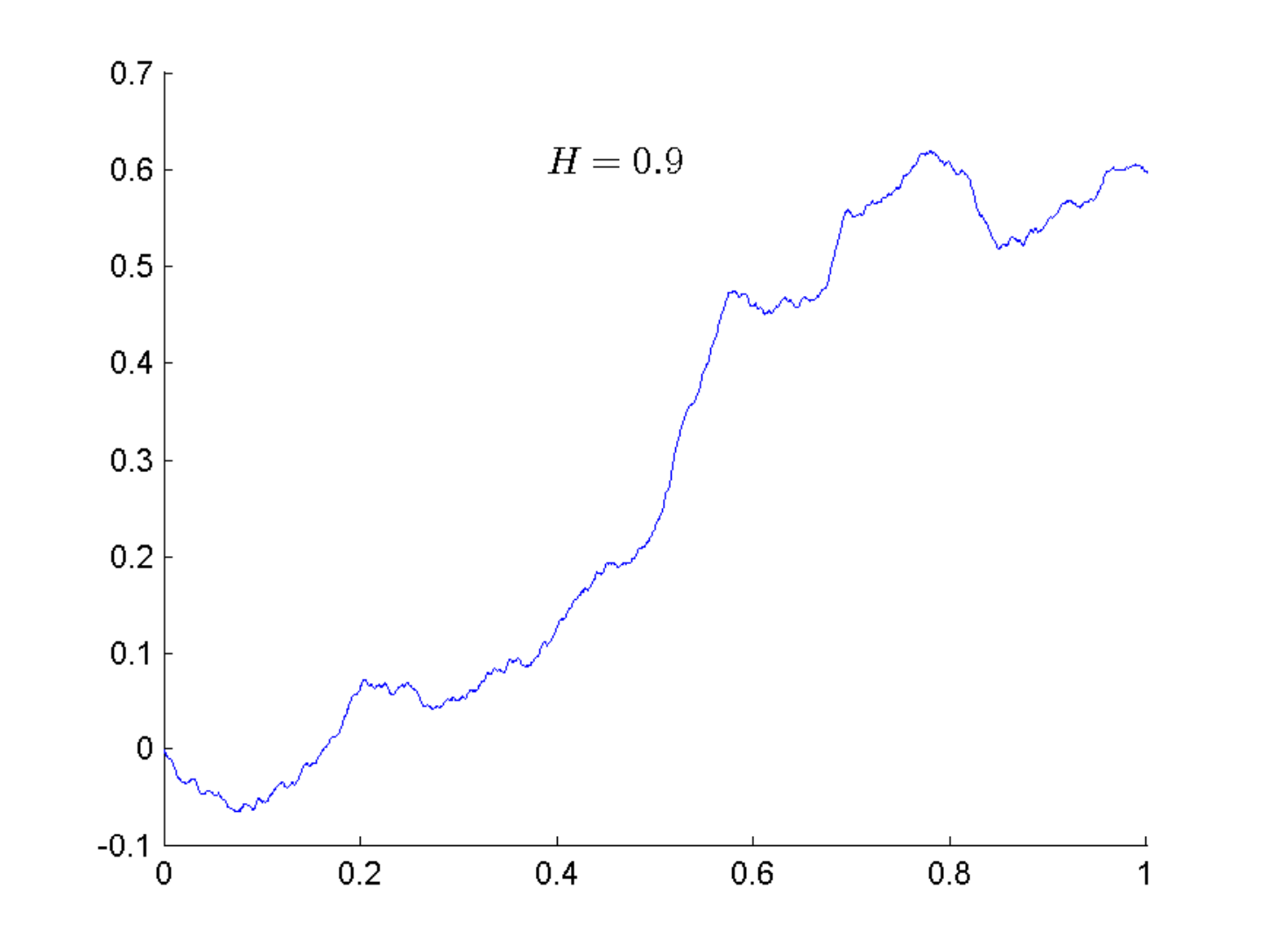}
\caption{Fractional Brownian motion for  different values of the Hurst parameter $H$. From left to right we have $H=0.3,0.5,0.9$.}
\label{fig:fbm1d}
\end{figure}

 Generation of fBm on the uniformly spaced grid $0=t_0 <t_1<t_2<\dots<t_n=1$ can be achieved by first generating the increment process $\{X_1,X_2,\ldots,X_n\}$, where
$X_i=W_i-W_{i-1}$,  and then delivering the cumulative sum 
\[
W_{t_i}=c^H\;\sum_{k=1}^{i}X_k,\;i=1,\ldots,n,\quad c=1/n\;. 
\] 
The increment process $\{X_1,X_2,\ldots,X_n\}$ is called \textbf{fractional Gaussian noise} and can be characterized as a discrete zero-mean stationary Gaussian process with covariance
\[
\Cov(X_i,X_{i+k})=\frac{1}{2}\left(|k+1|^\alpha-2|k|^\alpha+|k-1|^\alpha\right),\quad k=0,1,2,\ldots
\]
Since the  fractional Gaussian noise is stationary, we can generate it efficiently using
the circulant embedding approach in Section~\ref{sec:stationary gaussian}. 
First, we compute the first row $(r_1,\ldots,r_{n+1})$ of the  symmetric Toeplitz $(n+1)\times (n+1)$ covariance matrix $\Omega$ with elements $\Omega_{i+1,j+1}=\Cov(X_i,X_{j}),\;i,j=0,\ldots,n$. 
Second, we build the first row of the $2n\times 2n$ circulant matrix $\Sigma$, which embeds $\Omega$ in the upper left $(n+1)\times (n+1)$ corner. Thus, the first row of $\Sigma$ is given by $\br=(r_1,\ldots,r_{n+1},r_n,r_{n-1},\ldots,r_2)$.
We now seek a factorization of the form \eqref{diagonalization}.
Here, $\vect{\lambda}$ is the one-dimensional FFT of $\br$ defined as the linear transformation
$\vect{\lambda}=F\br$ with $F_{j,k}=\exp(-2\pi\ix jk/(2n))/\sqrt{2n},\;j,k=0,1,\ldots,2n-1$. Finally, the real and imaginary parts of the first $n+1$ components of 
$F^*\diag(\sqrt{\vect{\lambda}})\bZ$, where $\bZ$ is a $2n\times 1$ complex valued Gaussian vector, yield two independent realizations of fractional Brownian noise. Figure~\ref{fig:fbm1d} shows how the smoothness of fBm depends on the Hurst parameter. Note that $H=0.5$ corresponds to Wiener motion.

\noindent The following $\matlab$ code implements the circulant embedding method for fBm.

\medskip

\begin{breakbox}
\begin{verbatim}
n=2^15; % grid points
H = 0.9; %Hurst parameter
r=nan(n+1,1); r(1) = 1;
for k=1:n
    r(k+1) = 0.5*((k+1)^(2*H) - 2*k^(2*H) + (k-1)^(2*H));
end
r=[r; r(end-1:-1:2)]; % first rwo of circulant matrix
lambda=real(fft(r))/(2*n); % eigenvalues
W=fft(sqrt(lambda).*complex(randn(2*n,1),randn(2*n,1)));
W = n^(-H)*cumsum(real(W(1:n+1))); % rescale
plot((0:n)/n,W);
\end{verbatim}
\end{breakbox}
\medskip

\subsection{Fractional Wiener Sheet in $\R^2$}
\label{sec:fws}
A simple spatial generalization of the fractional Brownian motion 
is the fractional Wiener sheet in two dimensions.
The  {\bf fractional Wiener sheet} process on the unit square is the continuous zero-mean Gaussian process $\{W_\bt, \: \bt\in [0,1]^2\}$ with  covariance function 
\begin{equation}
\label{wiener sheet}
\cov(W_\bt, W_\bs) =\frac{1}{4}(| s_1|^\alpha+ | t_1|^\alpha-|s_1-t_1|^\alpha)(| s_2|^\alpha+ | t_2|^\alpha-|s_2-t_2|^\alpha)\;, 
\end{equation}
which is simply a product form extension of \eqref{product form}. For the special case of $\alpha=1$, we can also write the covariance  as $\cov(W_\bt, W_\bs)=\min\{t_1,s_1\} \,\min\{t_2,s_2\}$.

As in the one-dimensional case of fBm, we can consider the 
\textbf{two-dimensional fractional Gaussian noise} process $\{X_{i,j},\;i,j=1,\ldots,n\}$, which can be used to
construct a fractional Wiener sheet on a uniformly spaced square grid via the cumulative sum
\[
W_{t_i,t_j}=n^{-2H}\sum_{k=1}^{i}\sum_{l=1}^{j} X_{k,l}\;,\quad i,j=1,\ldots,n\;.
\]
Note that this process is self-similar in the sense that 
$(m\,n)^{-H}\sum_{k=1}^{m}\sum_{l=1}^n X_{k,l}$ has the same distribution as $X_{1,1}$ for all $m,n$, and $H$, see \cite{Mandelbrot_van_Ness_1968}.

\begin{figure}[htb]
\hspace{-0.5cm}
\includegraphics[width=.34\linewidth]{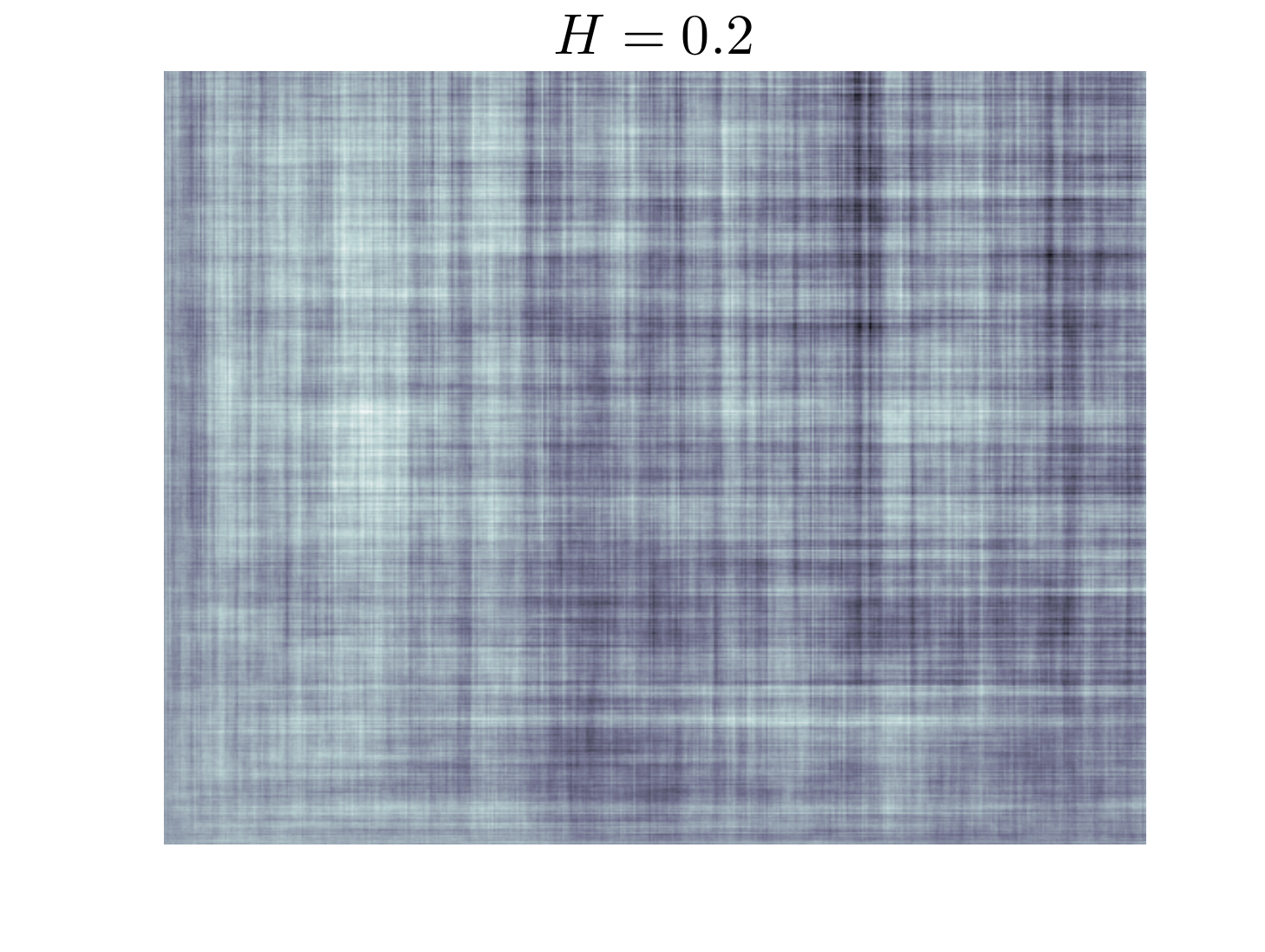}~
\includegraphics[width=.34\linewidth]{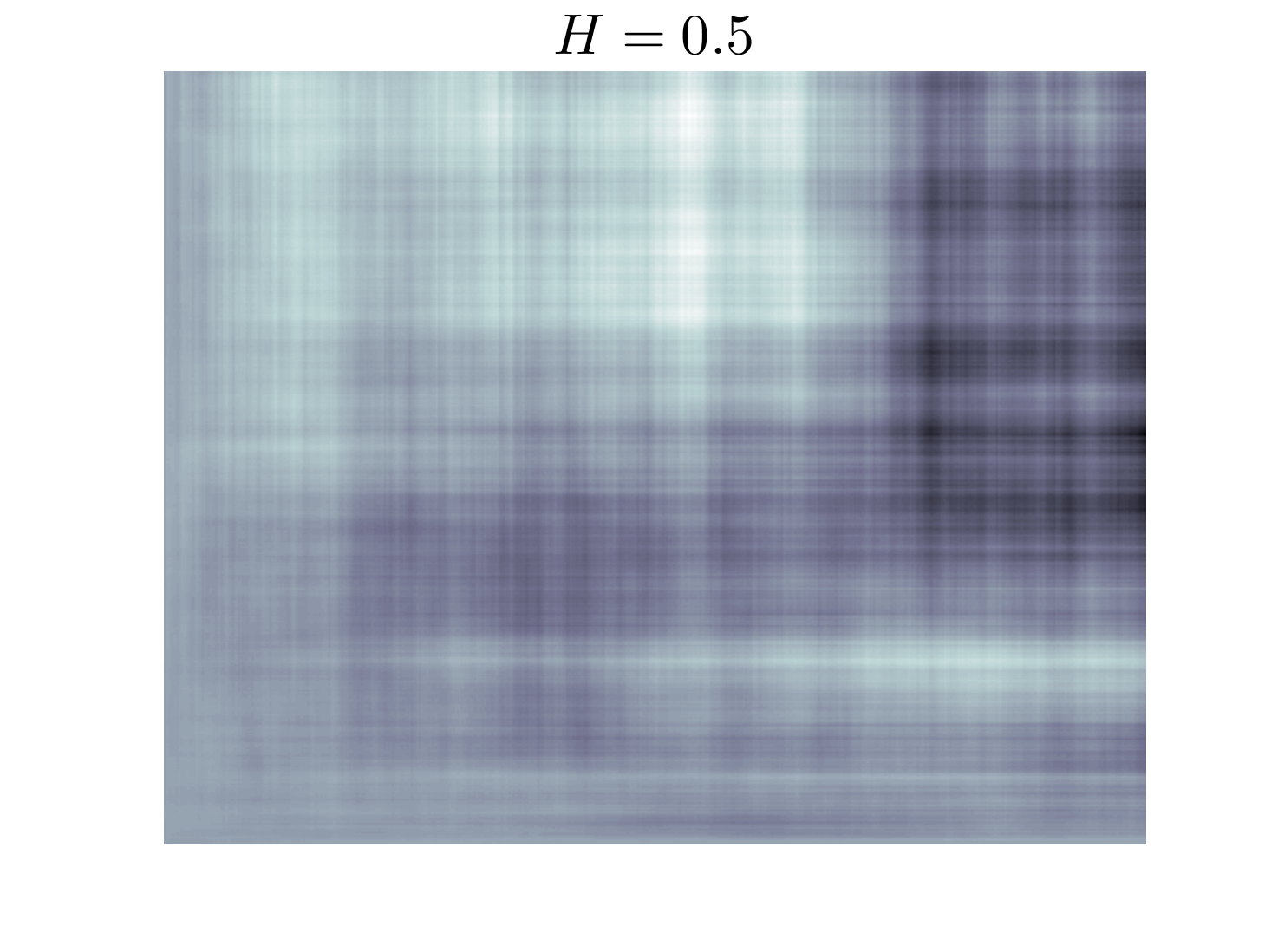}~
\includegraphics[width=.34\linewidth]{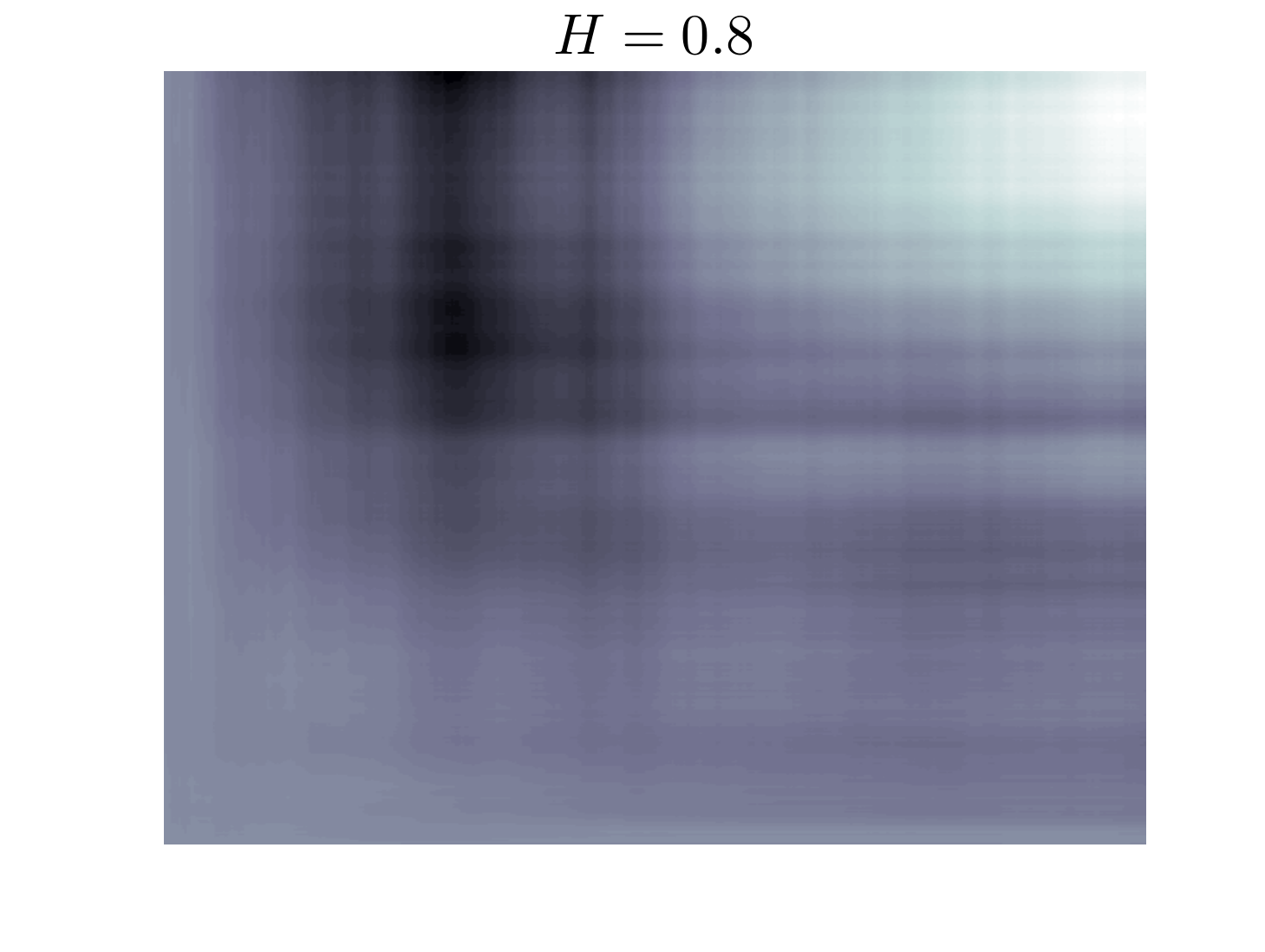}
\caption{Fractional Wiener sheets with different Hurst parameter $H$.}
\label{fig:brownsheet1}
\end{figure}

Generating a the two-dimensional fractional Gaussian noise process 
requires that we generate a zero-mean stationary Gaussian process with covariance \cite{Qian98}
\[
\Cov(X_{i,j},X_{i+k,j+l})=\frac{|k+1|^\alpha-2|k|^\alpha+|k-1|^\alpha}{2}\times\frac{|l+1|^\alpha-2|l|^\alpha+|l-1|^\alpha}{2}
\]
for  $k,l=0,1,\ldots,n$. 
We can thus proceed to generate this process using the  circulant embedding method in 
Section~\ref{sec:stationary gaussian}. 
 The generation of the Wiener sheet for 
$H=0.5$ is particularly easy since then all of the $X_{i,j}$ are independent standard normally distributed. Figure~\ref{fig:brownsheet1} shows  realizations
of  fractional Wiener sheets for $H=\frac{1}{2}\alpha\in\{0.2,0.5,0.8\}$ with $n=2^9$. 


 Two objects which are closely related to $W_\bt$ are the {\bf Wiener pillow} and the {\bf Wiener bridge}, which  are zero-mean Gaussian processes on $[0,1]^d$ with covariance functions $\Cov(W_\bt,W_\bs) = \prod_{i=1}^d(\min(t_i,s_i)-t_is_i)$ and $\prod_{i=1}^d \min(t_i,s_i)-\prod_{i=1}^d t_is_i$, respectively.


\subsection{Fractional Brownian Field}
\label{sec:FBS}
\textbf{Fractional Brownian surface} or field in two dimensions can be defined as the zero-mean Gaussian process
$\{\tilde X_\bt,\bt\in\mathbb{R}^2\}$ with nonstationary 
covariance function
\begin{equation}
\label{fbs}
\Cov(\tilde X_\bs,\tilde X_\bt)=\tilde\rho(\bs,\bt)=\| \bs\|^\alpha+ \| \bt\|^\alpha-\|\bs-\bt\|^\alpha\;.
\end{equation}
The parameter $H=\frac{\alpha}{2}\in (0,1)$ is  the Hurst parameter controlling the roughness of the random field or surface. Contrast this isotropic generalization of the covariance \eqref{product form} with the product form extension of the covariance in \eqref{wiener sheet}. 

\begin{figure}[H]
\includegraphics[clip=,width=.34\linewidth]{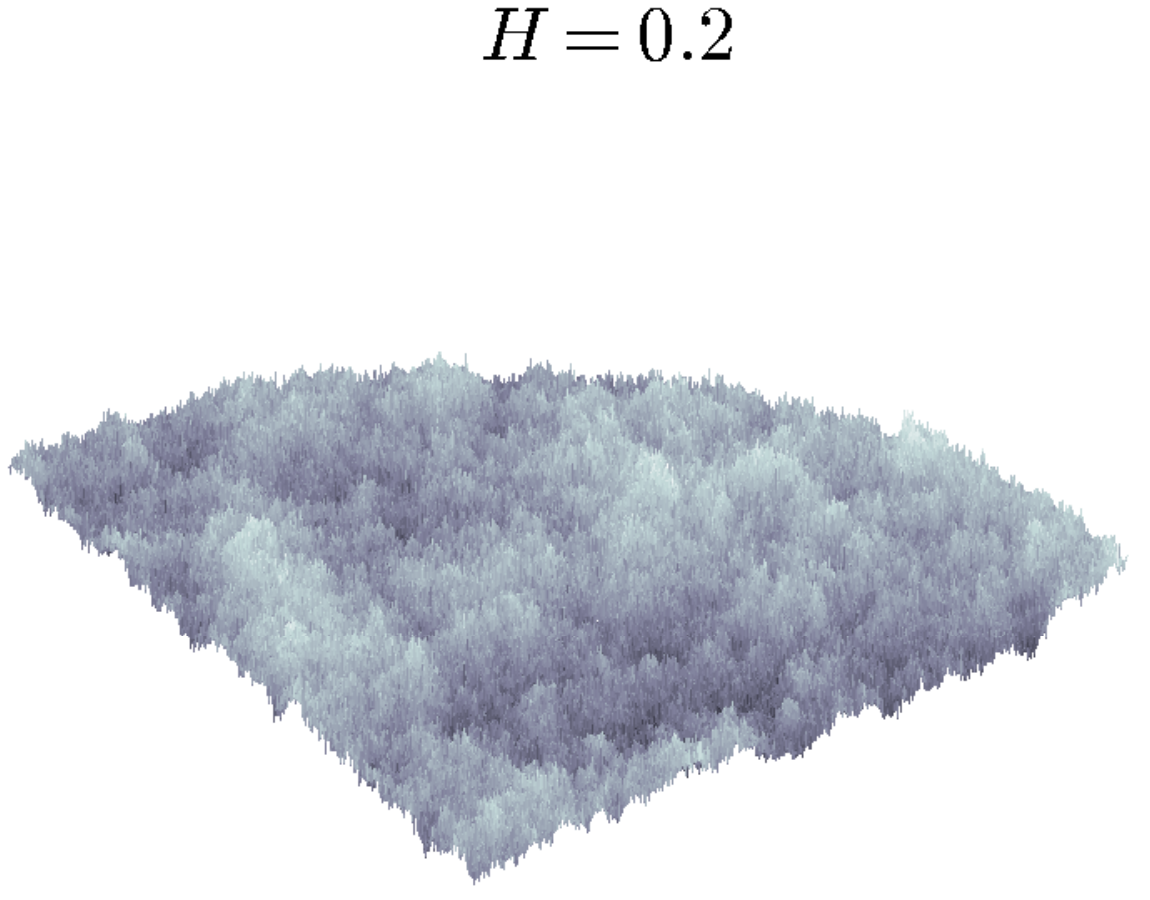}~
\includegraphics[clip=,width=.34\linewidth]{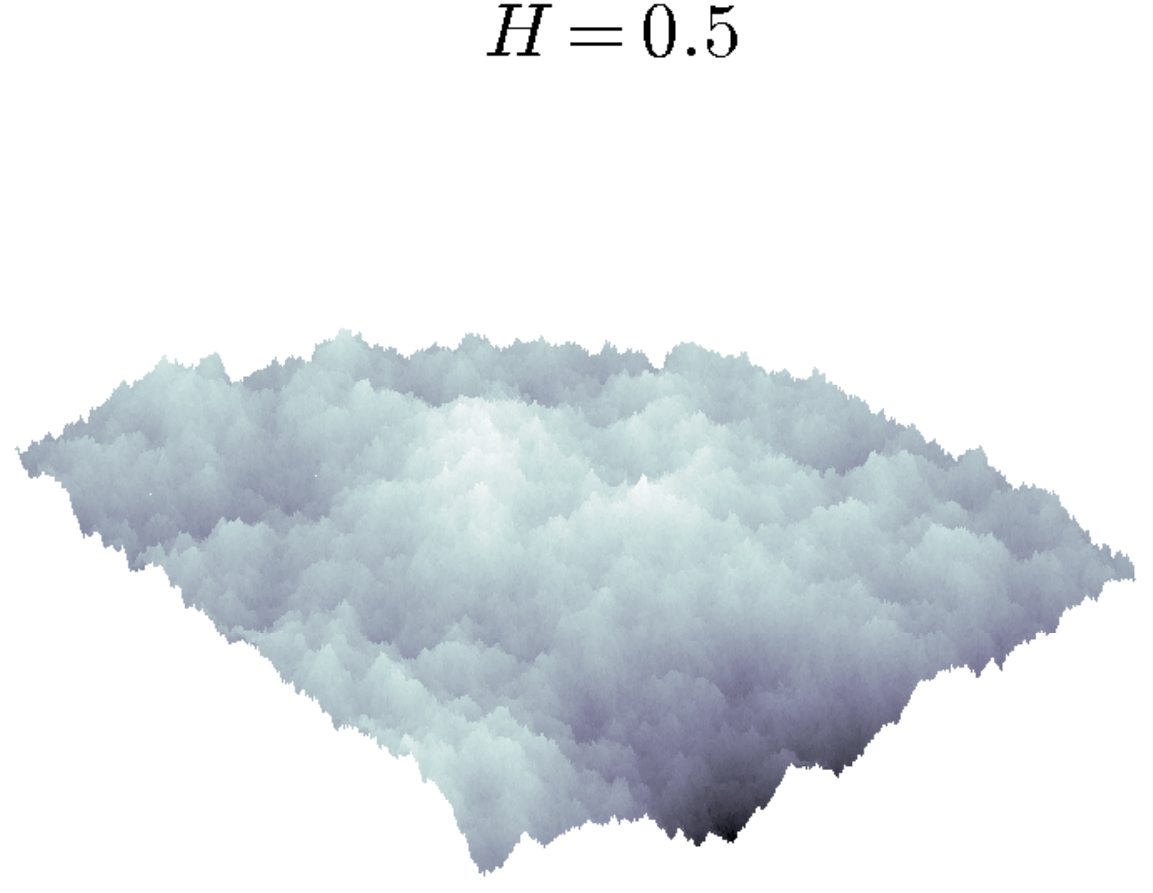}~
\includegraphics[clip=,width=.34\linewidth]{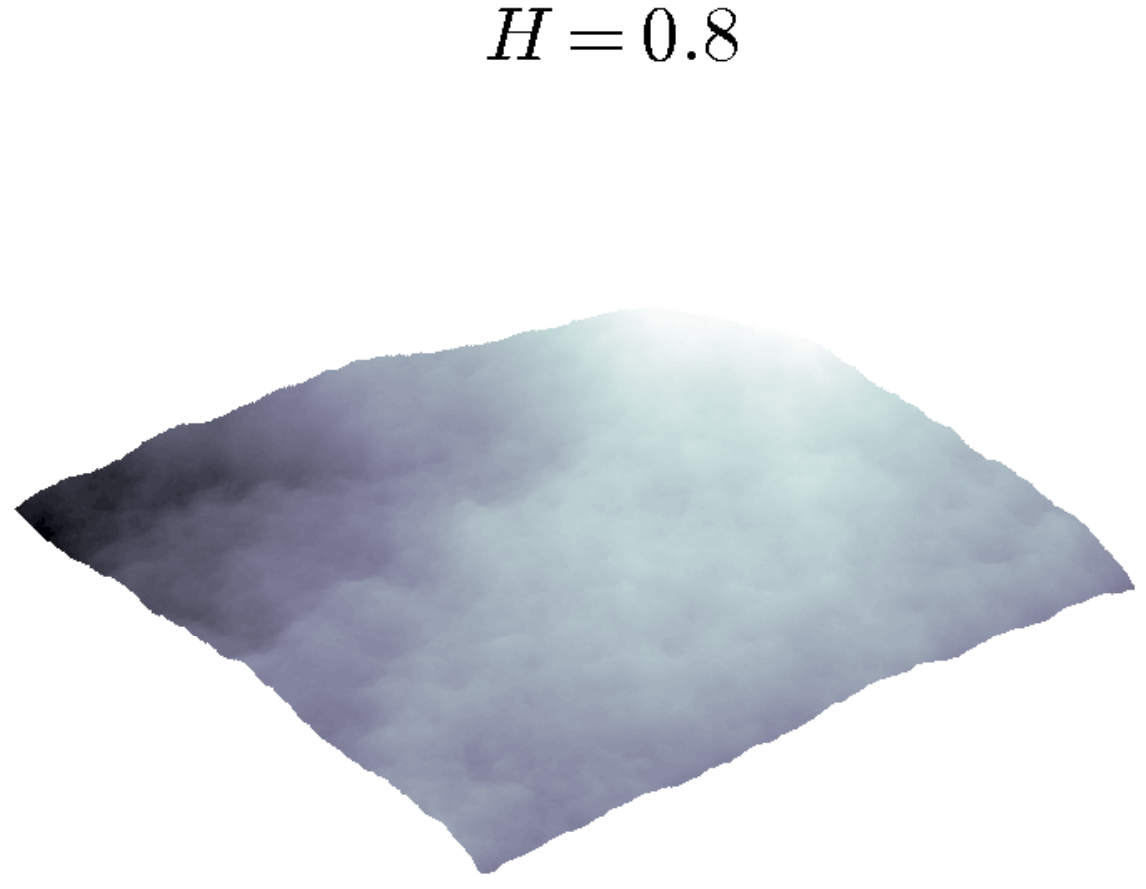}
\caption{Fractional Brownian fields with different roughness parameter $\alpha=2H$.}
\label{fig:fields}
\end{figure}

Generation of $\tilde X_\bt$ over a unit (quarter) disk in the first quadrant  involves the following steps. First,
we use Dietrich and Newsam's method (Section~\ref{sec:stationary gaussian})  to generate a stationary Gaussian field $\breve X_\bt$ with covariance
function over the quarter disk  $\{\bh: \|\bh\|\leq 1, \bh>\mathbf{0}\}$
\[
\breve\rho(\bs,\bt)=c_0+c_2\|\bs-\bt\|^2 -\|\bs-\bt\|^\alpha
\]
for some constants  $c_0, c_2\geq 0$ whose selection will  be discussed later. 
Once we have generated $\breve X_\bt$, the process $\tilde X_\bt$ is obtained
via the adjustment:
\[
\tilde X_\bt=\breve X_\bt-\breve X_\mathbf{0}+\sqrt{2c_2}\;\bt^\T \bZ,\quad \bZ=(Z_1,Z_2)^\T,\quad
Z_1,Z_2\simiid\Nor(0,1)\;.
\]
It is straightforward to verify that the covariance structure of $\tilde X_\bt$ over the disk $\{\bh: \|\bh\|\leq 1, \bh>\mathbf{0}\}$ is given by
 \eqref{fbs}. It now remains to explain how we  generate the process $\breve X_\bt$.
 The idea is to generate the process on $[0,R]^2, \, R\geq 1$ via  the  intrinsic embedding  of Stein (see \eqref{stein}) using the covariance function:
 \begin{equation}
 \label{stein_fbs}
  \psi(\bh)=\begin{cases} c_0+c_2\|\bh\|^2-\|\bh\|^\alpha, & \|\bh\|\leq 1\\
   \frac{\beta(R-\|\bh\|)^3}{\|\bh\|}, & 1\leq\|\bh\|\leq R\\
   0,& \|\bh\|\geq R
  \end{cases},
  \end{equation}  
 where depending on the value of $\alpha$, the constants $R\geq 1,\beta\geq 0,c_2>0,c_0\geq 0$ are defined in  Table~\ref{tab:param}.
\renewcommand{\arraystretch}{1.21}
\begin{table}[H]
\begin{center}
\begin{tabular}{c|c|c}
&$0<\alpha\leq 1.5$ & $1.5<\alpha< 2$\\
\hline
$R$ & 1 & 2\\
$\beta$ & 0 & $\frac{\alpha(2-\alpha)}{3R(R^2-1)}$ \\
$c_2$ & $\frac{1}{2}\alpha$ &  $\frac{\alpha-\beta(R-1)^2(R+2)}{2}$\\
$c_0$ & $1-c_2$ &  $\beta(R-1)^3+1-c_2$
\end{tabular}
\end{center}
\caption{Parameter values needed to ensure that   \eqref{stein_fbs} allows for a nonnegative circulant embedding. }
\label{tab:param}
\end{table}
Note that for $\alpha>1.5$, the  parameters needed for a nonnegative embedding are more complex, because a covariance function that is smoother close to the origin has to be even smoother elsewhere. In particular, for $\alpha>1.5$ the choice of constants ensures that $\psi$ is twice continuously differentiable as a function of $\|\bh\|$. Notice that while we generate the process 
$\tilde X_\bt$ over  the square grid $[0,R]^2$, we are only interested in $\tilde X_\bt$ restricted inside  the quarter disk with  covariance  $\breve\rho(\bs,\bt)$. Thus,  in order to reduce the computational effort,
we would like to have  $R\geq 1$ as close as possible to $1$. While the optimal choice   $R=1$  guarantees a nonnegative embedding for all  $\alpha\leq 1.5$,  in general we need  $R>1$ to ensure the existence of a minimal embedding for $\alpha>1.5$. The choice $R=2$ given in Table~\ref{tab:param} is the most conservative one that guarantees a nonnegative circulant embedding for $\alpha>1.5$. Smaller  values of $R$ that admit a nonnegative circulant embedding can be determined numerically \cite[Table 1]{stein02}. 
As a numerical example consider generating a fractional Brownian surface
with $m=n=1000$ and for Hurst parameter $H\in(0.2,0.5,0.8)$. Figure~\ref{fig:fields} shows the effect of the parameter on the smoothness of the surface, with larger values providing a smoother surface. 
We used the following $\matlab$ code for the generation of the surfaces.

\medskip

\begin{breakbox}
\begin{verbatim}
H=0.8;        % Hurst parameter
R=2;          % [0,R]^2 grid, may have to extract only  [0,R/2]^2
n=1000; m=n; % size of grid is m*n; covariance matrix is m^2*n^2
tx=[1:n]/n*R; ty=[1:m]/m*R; % create grid for field
Rows=zeros(m,n);
for i=1:n
    for j=1:m % rows of blocks of cov matrix
        Rows(j,i)=rho([tx(i),ty(j)],[tx(1),ty(1)],R,2*H);
    end
end
BlkCirc_row=[Rows, Rows(:,end-1:-1:2);
    Rows(end-1:-1:2,:), Rows(end-1:-1:2,end-1:-1:2)];
% compute eigen-values
lam=real(fft2(BlkCirc_row))/(4*(m-1)*(n-1));
lam=sqrt(lam);
% generate field with covariance given by block circular matrix
Z=complex(randn(2*(m-1),2*(n-1)),randn(2*(m-1),2*(n-1)));
F=fft2(lam.*Z);
F=F(1:m,1:n); % extract sub-block with desired covariance
[out,c0,c2]=rho([0,0],[0,0],R,2*H);
field1=real(F); field2=imag(F); % two independent fields 
field1=field1-field1(1,1); % set field zero at origin
% make correction for embedding with a term c2*r^2
field1=field1 + kron(ty'*randn,tx*randn)*sqrt(2*c2);
[X,Y]=meshgrid(tx,ty);
field1((X.^2+Y.^2)>1)=nan;
surf(tx(1:n/2),ty(1:m/2),field1(1:n/2,1:m/2),'EdgeColor','none') 
colormap bone
\end{verbatim}
\end{breakbox}
\medskip

The code  uses the function $\texttt{rho.m}$, which implements the embedding \eqref{stein_fbs}.

\medskip

\begin{breakbox}
\begin{verbatim}
function [out,c0,c2]=rho(x,y,R,alpha)
% embedding of covariance function on a [0,R]^2 grid
if alpha<=1.5 % alpha=2*H, where H is the Hurst parameter
    beta=0;c2=alpha/2;c0=1-alpha/2;
else % parameters ensure piecewise function twice differentiable
    beta=alpha*(2-alpha)/(3*R*(R^2-1)); 
    c2=(alpha-beta*(R-1)^2*(R+2))/2;
    c0=beta*(R-1)^3+1-c2;
end
% create continuous isotropic function
r=sqrt((x(1)-y(1))^2+(x(2)-y(2))^2);
if r<=1
    out=c0-r^alpha+c2*r^2;
elseif r<=R
    out=beta*(R-r)^3/r;
else
    out=0;
end
\end{verbatim}
\end{breakbox}
\medskip

\section{Spatial Levy Processes}
\label{sec:LP} 
 
 Recall from Section~\ref{sec:Wiener process} that the Brownian motion process  \eqref{brownian motion} can be characterized  as a continuous sample path process with stationary and independent Gaussian increments. 
The L\'evy process is one  of the simplest generalizations of the Brownian motion process, in cases where either the   assumption of normality of the increments, or the continuity of the sample path is not  suitable for modeling purposes. 

 
 \newcommand{\J}{\mathbf{J}}
 \subsection{L\'evy Process}
 A $d$-dimensional  \textbf{L\'evy process} $\{\bX_t,t\in \R_+\}$ with $\bX_0=0$ is a stochastic process with a continuous index set on $\R_+$ and with a continuous state space $\mathbb{R}^d$  defined by the following  properties: 
 (1) the increments of $\{\bX_t\}$ are stationary, that is,  $(\bX_{t+s}-\bX_t)$ has the same distribution as  $\bX_s$ for all $t,s\geq 0$;
 (2) the increments of $\{\bX_t\}$ are independent, that is, $\bX_{t_i}-\bX_{t_{i-1}},\; i=1,2,\ldots$ are independent for any
 $0\leq t_0<t_1<t_2<\cdots$; and
 (3) for any $\epsilon>0$, we have $\Pm(\|\bX_{t+s}-\bX_t\|\geq\epsilon)=0$ as $s\downarrow 0$.
 
 From the definition it is clear that
 Brownian motion \eqref{brownian motion} is an example of a L\'evy process,  with normally distributed increments. 
Brownian motion is the only L\'evy process with continuous sample paths. 
Other basic examples of  L\'evy processes include the  Poisson process $\{N_t,t\geq 0\}$ with intensity $\lambda>0$, where   $N_t\sim \poi(\lambda t)$ for each $t$, and the  \textbf{compound  Poisson  process} defined via
\begin{equation}
\label{compound}
\J_t=\sum_{k=1}^{N_t}  \delta \bX_k, \quad N_t\sim \Poi(\lambda t)\;,
\end{equation} 
where $  \delta \bX_1, \delta \bX_2,\ldots$ are  iid random variables,
independent of $\{N_t, t \geq 0\}$.  We can more generally express $\J_t$ as 
$\J_t=\int_0^t\int_{\R^d} \bx\, N(\di s,\di \bx)$, where $N(\di s,\di \bx)$
is a Poisson random counting measure on $\R_+\times \R^d$ (see
Section~\ref{PoissonProcesses}) with mean measure  $\Em[N([0,t] \times 
A)], \; A\in \cE,$ equal to the expected number of jumps of size  $A$ in the interval $[0,t]$.

 A crucial property of L\'evy processes is  infinite divisibility.
 In particular, if we define
 $\bY_j^{(n)}=\bX_{j\,t/n}-\bX_{(j-1)\,t/n}$, then using the stationarity and independence properties of the L\'evy process, we obtain that
for each $n\geq 2$ the $\{\bY_1^{(n)}\}$ are independent and 
 identically distributed random variables with the same distribution as $\bX_{t/n}$.
 Thus, for a fixed $t$  we can write
  \begin{equation}
  \label{divisibility inf}
 \bX_t\sim \bY_1^{(n)}+\cdots + \bY_n^{(n)}\;,  \textrm{ for any } n\geq 2\;,
 \end{equation}
and hence   by definition the random vector $\bX_t$
 is \textbf{infinitely divisible} (for a fixed $t$). The L\'evy--Khintchine theorem \cite{Sato} 
 gives  the most general form of the characteristic function 
 of an infinitely divisible random variable.  Specifically, 
 the logarithm of the characteristic function of $\bX_t$ (the
 so-called \emph{characteristic exponent}) is of the form  
 \begin{equation}
 \label{exponent}
 \log\Em[\e^{\ix \, \bs^\T \bX_t}]=\ix \,t\, \bs^\T\bmu-\frac{1}{2}\,t\,\bs^\T\Sigma\bs+t\int_{\R^d}\left(\e^{\ix \, \bs^\T  \bx}-1-\ix \, \bs^\T \bx\,\I_{\{\|\bx\|\leq 1\}}\right) \nu(\di \bx)\;,
 \end{equation}
for some $\bmu\in \mathbb{R}^d$, covariance matrix $\Sigma$ and measure $\nu$
such that $\nu(\{\mathbf{0}\})=0$,
\begin{equation} 
\label{integrability}
 \int_{\|\bx\|> 1}\!\!\!\nu(\di \bx)<\infty\;\textrm{ and }
 \int_{\|\bx\|\leq 1}\|\bx\|^2\nu(\di \bx)<\infty\; \Leftrightarrow
 \int_{\mathbb{R}^d}\min\{1,\|\bx\|^2\}\nu(\di \bx)<\infty \;.
\end{equation}
The triplet $(\bmu,\Sigma,\nu)$ is referred to as the \textbf{characteristic triplet} defining  the L\'evy process. The measure $\nu$ is referred to as the \textbf{L\'evy measure}. 
 Note that  for a general $\nu$ satisfying \eqref{integrability}, the integral $\int_{\mathbb{R}^d}\big(\e^{\ix \, \bs^\T  \bx}-1\big) \nu(\di \bx)$ in \eqref{exponent} does not converge separately. In this sense $\ix \,\bs^\T \bx\,\I_{\{\|\bx\|\leq 1\}}$ in \eqref{exponent} serves the purpose of   enforcing convergence under the very general integrability condition \eqref{integrability}. However, if in addition to \eqref{integrability} the measure $\nu$ satisfies $\int_{\R^d}\min\{1,\|\bx\|\}\nu(\di \bx)<\infty$ and $\nu(\R^d)<\infty$, then 
the integral in \eqref{exponent} can be separated as
$t\int_{\R^d}(\e^{\ix \, \bs^\T\bx}-1)\nu(\di \bx)-\ix \,t\, \bs^\T\int_{\|\bx\|\leq 1}\bx\,\nu(\di \bx)$, and the characteristic exponent simplifies to
\[
\underbrace{\ix \,t\,\bs^\T\bmu^* -\frac{1}{2}\,t\,\bs^\T\Sigma \,\bs}_{\textrm{Brownian motion term}}+\underbrace{t\int_{\R^d}\left(\e^{\ix \, \bs^\T  \bx}-1\right) \nu(\di \bx)}_{\textrm{Poisson process term}},
\]
 where $\bmu^*=\bmu-\int_{\|\bx\|\leq 1}\bx\,\nu(\di \bx)$. We can now recognize this characteristic exponent as the one corresponding to the process $\{\bX_t\}$ defined by $\bX_t=t\bmu^*+\Sigma^{1/2}\bW_t+\J_t$, where $t\bmu^*+\Sigma^{1/2}\bW_t$
 defines a Brownian motion  (see \eqref{brownian motion}) and $\{\J_t\}$ is a compound Poisson process \eqref{compound}
 with jump size  distribution $\delta \bX_1\sim\nu(\di \bx)/\lambda$. Thus,   $\nu(\di \bx)$ can be interpreted as the intensity of the jump sizes in this particular L\'evy process. In a  similar way, it can be shown  that the most general L\'evy process $\{\bX_t\}$ with characteristic triplet $(\bmu,\Sigma,\nu)$ and integrability condition  \eqref{integrability} can be represented as the limit  in probability of a process $\{\bX_t^{(\epsilon)}\}$ as $\epsilon\downarrow 0$, where $\bX_t^{(\epsilon)}$
 has the \textbf{L\'evy-\I t\^o decomposition} 
 \begin{equation}
 \label{levy decomposition}
 \bX_t^{(\epsilon)}=t\bmu+\Sigma^{1/2} \bW_t+\J_t+\left(\J_t^{(\epsilon)}-t\int_{\epsilon<\|\bx\|\leq 1} \,\bx\,\nu(\di \bx)\right)
 \end{equation}
with the following \emph{independent} components:
 \begin{enumerate}
 \item $\{t\bmu+\Sigma^{1/2} \bW_t\}$ is the Brownian motion \eqref{brownian motion}, which corresponds to the $\ix \,t\,\bs^\T\bmu -\frac{1}{2}\,t\,\bs^\T\Sigma\, \bs$ part of \eqref{exponent}.
 \item $\{\J_t\}$ is a compound Poisson process of the form
 \eqref{compound} with $\lambda=\int_{\|\bx\|>1}\nu(\di \bx)$ and increment distribution
 $\delta \bX_1\sim \nu(\di \bx)/\lambda$ over $\|\bx\|> 1$, which corresponds to the $\int_0^t\int_{\|\bx\|> 1}(\e^{\ix \, \bs^\T\bx}-1)\nu(\di \bx)\,\di t$ part in the characteristic exponent \eqref{exponent}. 
 \item $\{\J_t^{(\epsilon)}\}$ is a compound Poisson process with $\lambda=\nu(\epsilon<\|\bx\|\leq 1)$ and increment distribution $ \nu(\di \bx)/\lambda$ over $\epsilon<\|\bx\|\leq 1$, so that the \emph{compensated} compound Poisson process
 $\{\J_t^{(\epsilon)}-t\int_{\epsilon<\|\bx\|\leq 1} \bx\,\nu(\di \bx)\}$
corresponds to the $\int_0^t\int_{\|\bx\|\leq 1}\left(\e^{\ix \, \bs^\T \bx}-1-\ix \,\bs^\T \bx\right) \nu(\di \bx)\,\di t$ part of \eqref{exponent} in the limit  $\epsilon\downarrow 0$.
 \end{enumerate}
The L\'evy--\I t\^o decomposition immediately suggests an approximate generation method --- we generate
$\{\bX_t^{(\epsilon)}\}$ in \eqref{levy decomposition} for a given small $\epsilon$, where the Brownian motion part is generated via  the methods in Section~\ref{sec:Wiener process} and the compound 
Poisson process \eqref{compound} in the obvious way. We are thus throwing away the very small jumps of size less than $\epsilon$.
For $d=1$ it can then be shown \cite{Asmussen_Rosinski_2001} that the error process  $\{X_t-X_t^{(\epsilon)}\}$ for this approximation is a L\'evy process with
 characteristic triplet $(0,0,\nu(\di x)\I_{\{|x|<\epsilon\}})$ and  variance $\Var(X_t-X_t^{(\epsilon)})=t\int_{-\epsilon}^\epsilon x^2\nu(\di x)$. 

A given approximation $\bX_t^{(\epsilon_n)}$ can always be further refined by adding smaller jumps of size $[\epsilon_{n+1},\epsilon_n]$ to obtain: \[
\bX_t^{(\epsilon_{n+1})}=\bX_t^{(\epsilon_n)}+\J_t^{(\epsilon_{n+1})}-t\int_{\epsilon_{n+1}<\|\bx\|\leq \epsilon_{n}} \bx\,\nu(\di \bx)\;,\quad \epsilon_n>\epsilon_{n+1}>0\;,
\] 
where the compound Poisson process 
$\{\J_t^{(\epsilon_{n+1})}\}$ has increment distribution given by $\nu(\di \bx)/\nu(\epsilon_{n+1}<\|\bx\|\leq \epsilon_{n})$ for all $\epsilon_{n+1}<\|\bx\|\leq \epsilon_{n}$. For a more sophisticated method of refining the approximation see \cite{Asmussen_Rosinski_2001}. 
 
As an example consider the L\'evy process $\{X_t\}$ with 
characteristic triplet $(\mu,0,\nu)$, where $\nu(\di x)=\alpha\,
\e^{-x}/x \, \di x$ for $\alpha,x>0$ and 
$\mu=\int_{|x|\leq 1} x\,\nu(\di x)=\alpha(1-\e^{-1})$.  Here, in addition to 
\eqref{integrability}, the infinite measure $\nu$ satisfies the stronger integrability condition $\int_\R\min\{1,|x|\}\nu(\di x)<\infty$ and hence we can write
$X_t=t\mu-t\int_{|x|\leq 1} x\,\nu(\di x)+\int_0^t\int_\R x\,N(\di s,\di x)=\int_0^t\int_{\R_+} x\,N(\di s,\di x)$, where $\Em [N(\di s,\di x)]=\di s\,\nu(\di x)$. 

\begin{figure}[H]
 \hspace{-0.5cm}
\includegraphics[clip=,width=.34\linewidth]{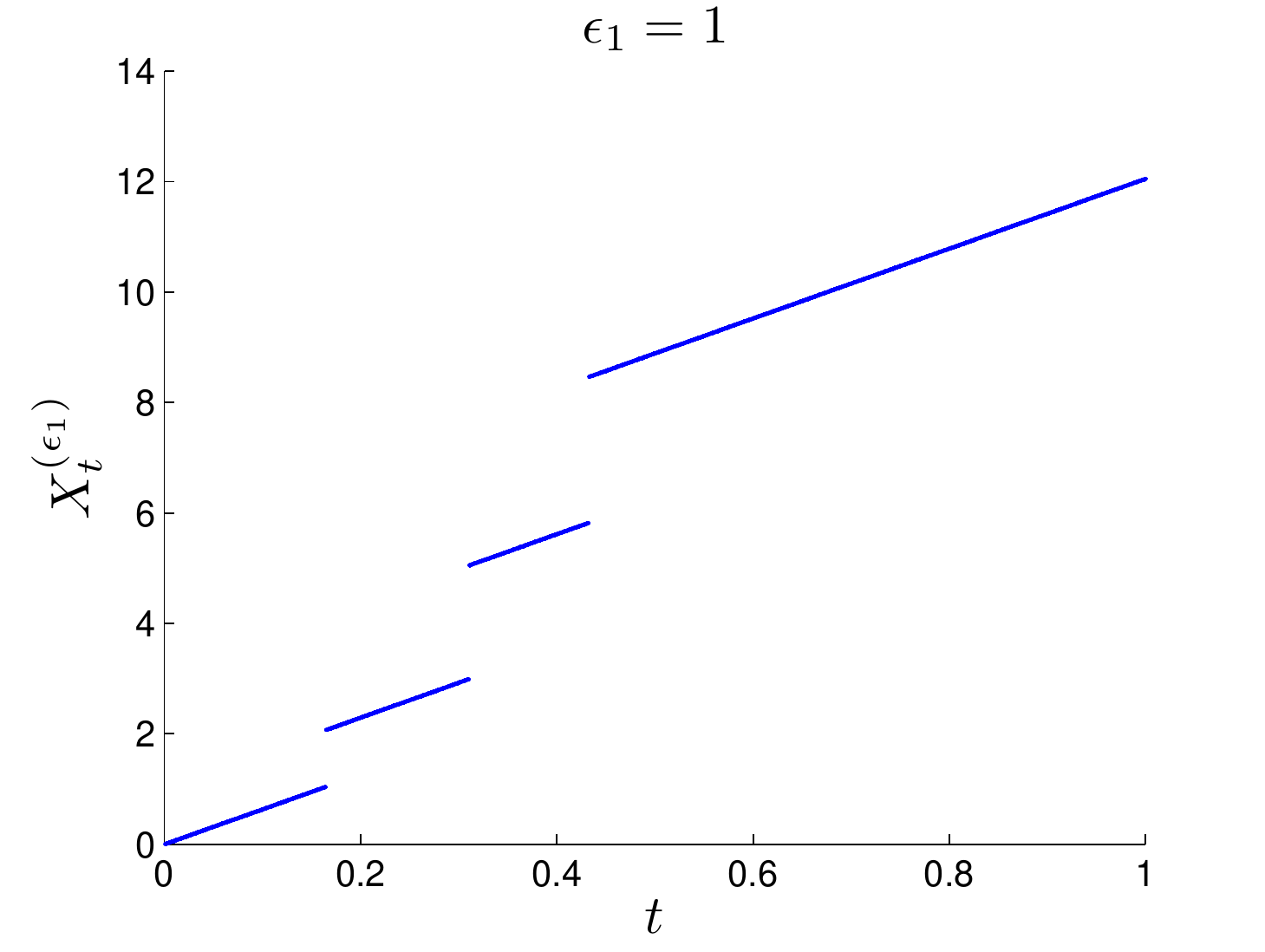}~
\includegraphics[clip=,width=.34\linewidth]{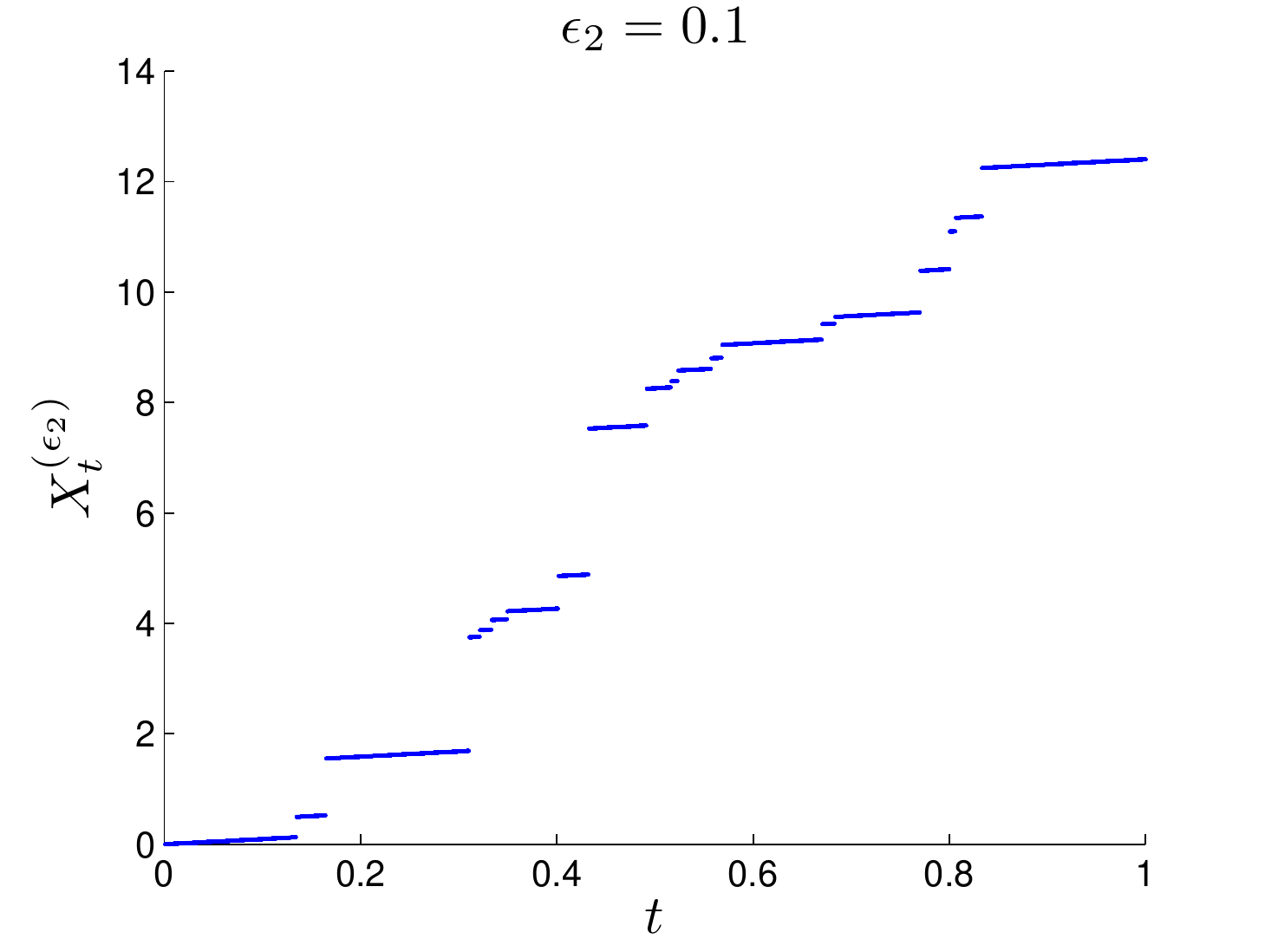}~
\includegraphics[clip=,width=.34\linewidth]{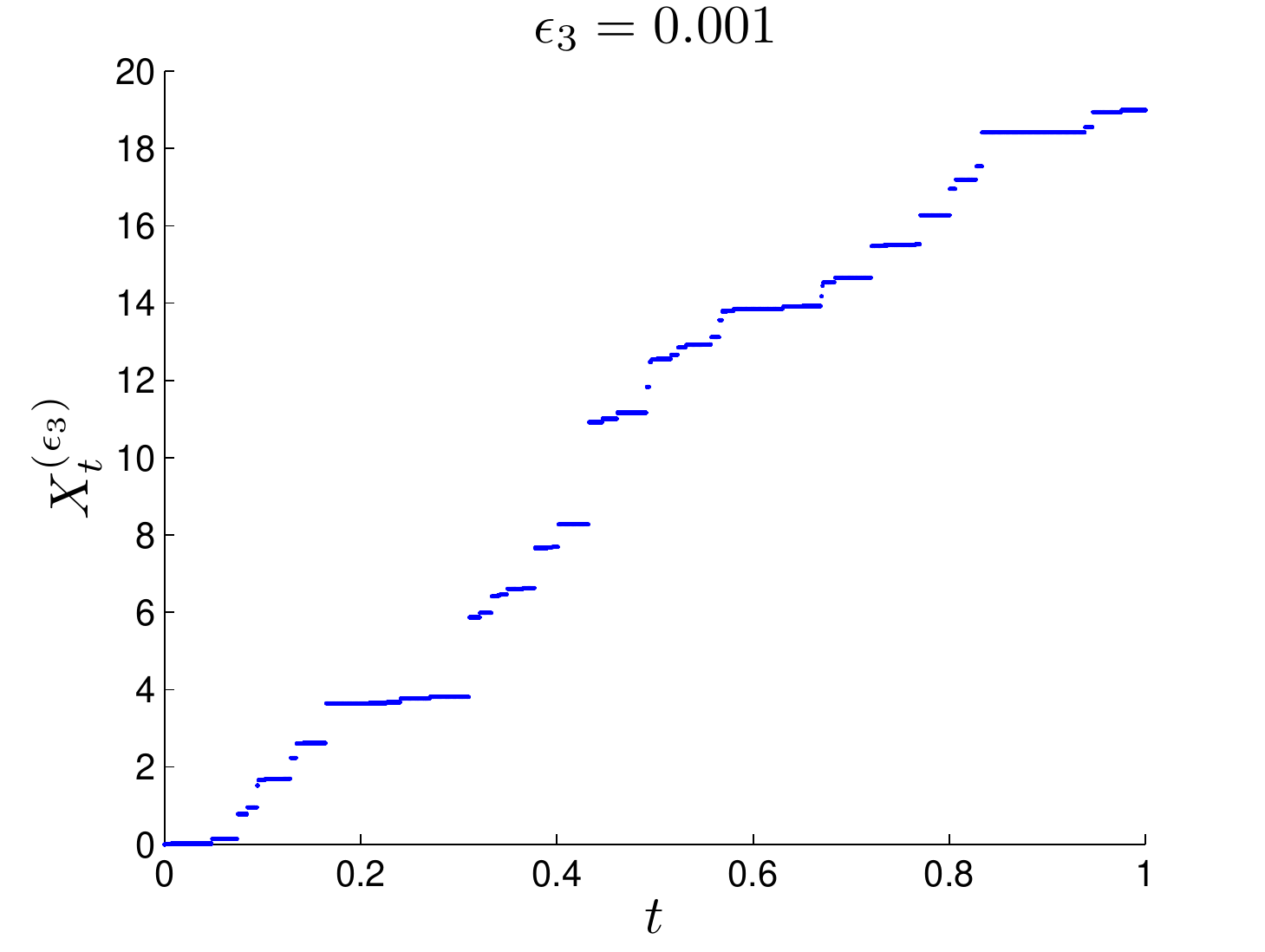}
\caption{Approximate gamma process  realizations obtained by throwing away jumps smaller than  $\epsilon_i$.}
\label{fig:Levy paths}
\end{figure}

The leftmost panel of Figure~\ref{fig:Levy paths} shows an outcome of the approximation $X_t^{(\epsilon_1)}$ in \eqref{levy decomposition} over $t\in[0,1]$ for $\epsilon_1=1$ and $\alpha=10$. The middle and rightmost panels show the  refinements $X_t^{(\epsilon_2)}$ and 
$X_t^{(\epsilon_3)}$, respectively, where 
 $(\epsilon_1,\epsilon_2,\epsilon_3)=(1,0.1,0.001)$. Note that  the refinements add finer and finer jumps to the path. 
 
This process is called a {\bf gamma process}  and can  
be generated more simply using the fact that the increments $\{X_{jt/n}-X_{(j-1)t/n}\}$ have a 
known  gamma distribution \cite[Page 212]{handbook_dirk}. The 
gamma process is an example of an  increasing L\'evy process 
(that is, $X_t\geq X_s$ almost surely for all $t\geq s$) called a 
\textbf{L\'evy subordinator}. A L\'evy subordinator $\{X_t,t\geq 
0\}$ has characteristic triplet $(\mu,0,\nu)$ satisfying 
the positive jump property $\nu((-\infty,0])=0$ and the positive 
drift property $0\leq \mu-\int_0^1 x\nu(\di x)< \infty$.




 \subsection{L\'evy Sheet in $\mathbb{R}^2$}
One way of generalizing the  L\'evy process to the spatial case is to insist on the preservation of the 
infinite divisibility property. 
In this generalization,  a  {\bf spatial L\'{e}vy process}
or simply a {\bf L\'evy sheet} $\{X_\bt,\bt\in\R^2\}$ possesses the property that
$(X_{\bt_1},\ldots,X_{\bt_n})$ is infinitely divisible for all  indexes
$\bt_1,\ldots,\bt_n\in\R^2$ and any integer $n$ (see \eqref{divisibility inf}). To  construct such an infinitely divisible  process, consider stochastic integration with respect to a \textbf{random infinitely divisible
 measure} $\Lambda$  
 defined as the stochastic process $\{\Lambda(A),\; A\in\cE\}$ with the following properties:
 \begin{enumerate}
 \item  For any set $A\in\cE$ the random variable $\Lambda(A)$ has an infinitely divisible distribution with 
 $\Lambda(\emptyset)=0$ almost surely.
 \item For any disjoint sets $A_1,A_2,\ldots \in \cE$,  the random variables $\Lambda(A_1),\Lambda(A_2),\ldots$ are independent and $\Lambda(\cup_i A_i)=\sum_i\Lambda(A_i)$.
\end{enumerate}
An example of a random infinitely divisible measure is  the Poisson random measure \eqref{prm} in Section~\ref{PoissonProcesses}.
 As a consequence of the independence and infinite divisibility properties, the characteristic  exponent of $\Lambda(A),\; A\in\cE$ has 
  the L\'evy--Khintchine representation \eqref{exponent}:
  \[
\log\Em[\e^{\ix \, s \Lambda(A)}]=\ix \, s\,\tilde\mu(A)-\frac{1}{2}\,s^2(\tilde\sigma(A))^2+\int_{\R}\left(\e^{\ix \, s  x}-1-\ix \, s\, x\,\I_{\{|x|\leq 1\}}\right) \tilde\nu(\di x,A)\;,  
  \]
  where $\tilde\mu$ is an additive set function, $\tilde\sigma$
  is a measure on the Borel sets $\cE$, the measure $\tilde\nu(\cdot,A)$   is a L\'evy measure for each fixed $A\in\cE$ (so that $\tilde\nu(\{0\},A)=0$ and $\int_\R \min\{1,x^2\}\tilde\nu(\di x,A)<\infty$),    and $\tilde\nu(\di x,\cdot)$ is a Borel measure  for each fixed $\di x$. For example,  $X_t=\Lambda((0,t])$ defines a  one-dimensional L\'evy process. 
  
  We can then  construct  a L\'evy sheet
 $\{X_\bt,\bt\in \R^2\}$  via the stochastic integral
 \begin{equation}
 \label{levy integral}
 X_\bt=\int_{\R^d} \kappa_\bt(\bx)\, \Lambda(\di \bx),\quad \bt\in\R^2\;,
 \end{equation}
where   $\kappa_\bt: \R^d\rightarrow \R$ is a H\"older continuous
\emph{kernel function} for all $\bt\in\R^2$, which is integrable  with respect to the random infinitely divisible measure $\Lambda$.
Thus, the L\'evy  sheet \eqref{levy integral} is a stochastic integral with a deterministic kernel function  as integrand (determining the spatial structure) and a random infinitely divisible measure as integrator \cite{Karcher13}.

Consider simulating the L\'evy sheet \eqref{levy integral} over $\bt\in[0,1]^2$  for $d=2$. Truncate the region of integration to a bounded domain, say $[0,1]^2$, so that $\kappa_\bt(\bx)=0$ 
for each $\bx \not\in [0,1]^2$ and $\bt\in[0,1]^2$. Then, one way of simulating   $\{X_\bt,\bt\in[0,1]^2\}$  
is to consider the approximation
 \begin{equation}
 \label{lattice}
 X_\bt^{(n)}=\sum_{i=0}^{n-1}\sum_{j=0}^{n-1}  \kappa_\bt(i/n,j/n)\;\Lambda(\triangle_{i,j}),\quad \triangle_{i,j}\equiv \left[\frac{i}{n},\frac{i+1}{n}\right]
 \times\left[\frac{j}{n},\frac{j+1}{n}\right]\;,
 \end{equation}
 where all $\Lambda(\triangle_{i,j})$ are  independent 
infinitely divisible random variables with characteristic triplet
$(\tilde\mu(\triangle_{i,j}),\tilde\sigma(\triangle_{i,j}),\tilde\nu(\,\cdot\,,\triangle_{i,j}))$. Under some technical conditions \cite{Karcher13}, it can be shown that $ X_\bt^{(n)}$ converges to $X_\bt$ in probability  as $n\uparrow\infty$. 
 
\begin{figure}[H]
\includegraphics[clip=,width=.5\linewidth]{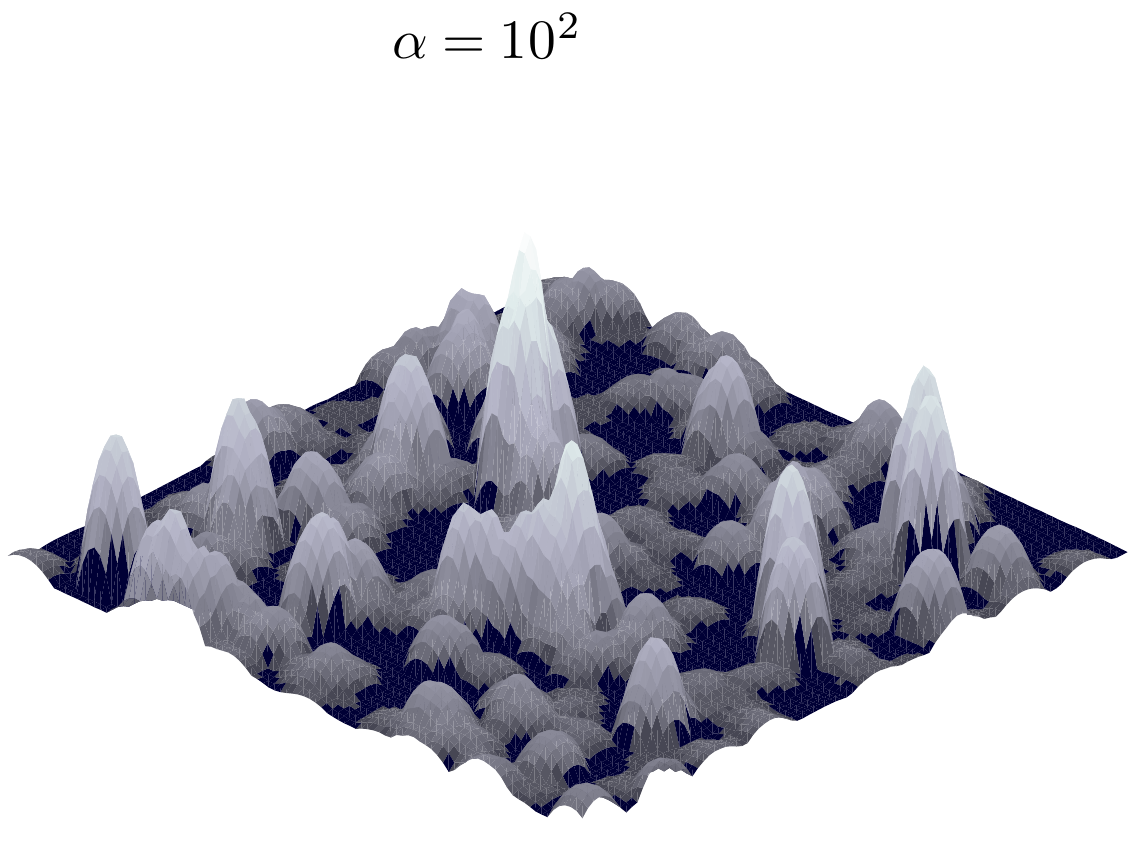}~
\includegraphics[clip=,width=.5\linewidth]{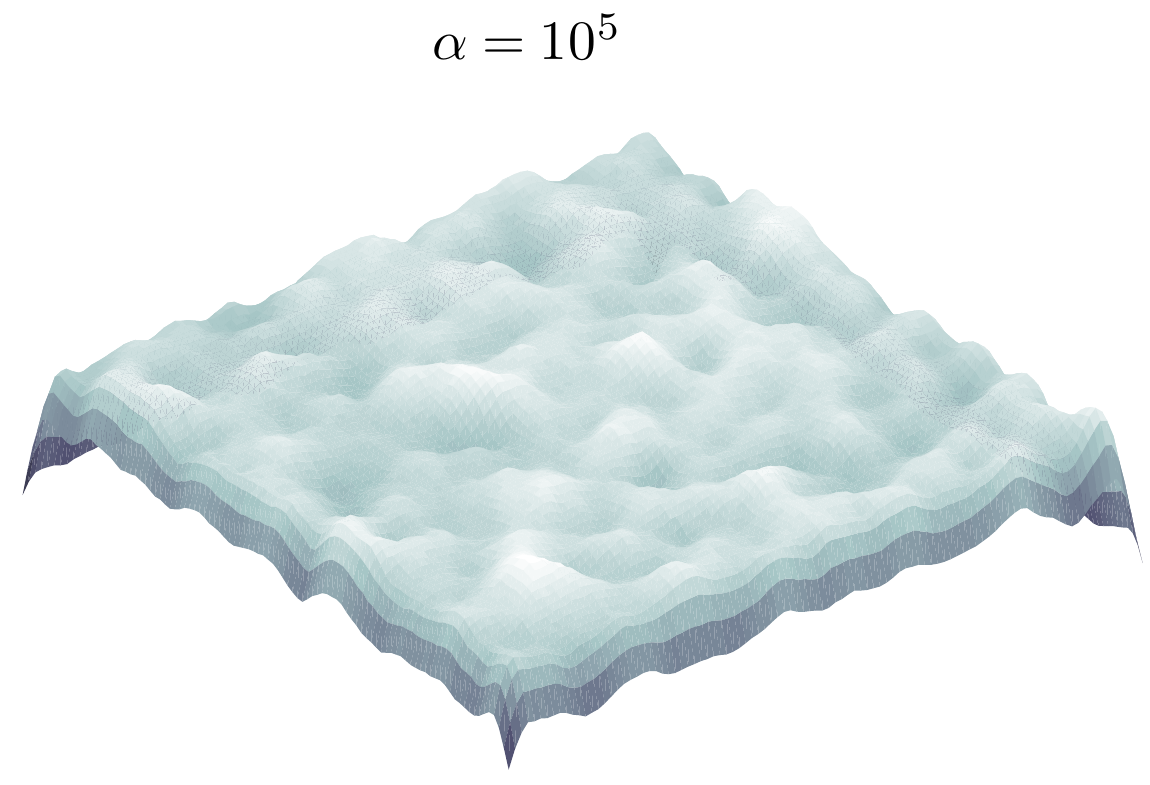}~
\caption{ Gamma L\'evy random sheet realizations for different values of the shape parameter $\alpha$.}
\label{fig:Levy sheet}
\end{figure}

As an example, consider generating \eqref{lattice}
on the square grid $\{(i/m,j/m),\;i,j=0,\ldots,m-1\}$
with the kernel function 
 \[
 \kappa_\bt(x_1,x_2)= (r^2-\|\bx-\bt\|^2)\;\I_{\{\|\bx-\bt\|\leq r\}},\quad \bt\in[0,1]^2\;,
 \]
and  $\Lambda(\triangle_{i,j})\simiid {\sf Gamma}(\alpha|\triangle_{i,j}|,\beta),\; |\triangle_{i,j}|=1/n^2$ for all $i,j$, where 
${\sf Gamma}(\alpha,\beta)$ denotes the density of the gamma distribution with pdf $\beta^\alpha x^{\alpha-1}\e^{-\beta x}/\Gamma(\alpha),$ $x\geq0$. The corresponding limiting process $\{X_\bt\}$ is called a \textbf{gamma L\'evy sheet}. 
Figure~\ref{fig:Levy sheet}
shows  realizations of \eqref{lattice} for $m=n=100$ and $r=0.05,\;\alpha=\beta\in\{10^2,10^5\}$, so that we have the scaling $\Em[\Lambda(\triangle_{i,j})]=\alpha/(\beta\,n^2)=1/n^2$.

 Note that the sheet exhibits more bumps for smaller values of $\alpha$ than for  large values of $\alpha$. For a method of approximately generating \eqref{levy integral} using wavelets see \cite{Karcher13}.

\subsubsection*{Acknowledgement}This work was  supported by the Australian Research
Council under grant number DP0985177.
\bibliographystyle{plain}
\bibliography{MCSpatial}

\begin{thebibliography}{10}

\bibitem{Asmussen_Rosinski_2001}
S.~Asmussen and J.~Rosi\'nski.
\newblock Approximations of small jumps of {L}\'evy processes with a view
  towards simulation.
\newblock {\em Journal of Applied Probability}, 38(2):482--493, 2001.

\bibitem{Baddeley07}
A.~Baddeley.
\newblock {\em Spatial Point Processes and their Applications}.
\newblock Springer Lecture Notes in Mathematics (1892), 2007.

\bibitem{CaseStudies}
A.~Baddeley, P.~Gregori, J.~Mateu, R.~Stoica, and D.~Stoyan.
\newblock {\em Case Studies in Spatial Point Process Modeling}.
\newblock Springer Lecture Notes in Statistics (185), 2006.

\bibitem{Barnett90}
S.~Barnett.
\newblock {\em Matrices, Methods and Applications}.
\newblock Oxford Applied Mathematics and Computing Sciences Series, 1990.

\bibitem{brereton_etal}
T.~Brereton, D.~P. Kroese, O.~Stenzel, V.~Schmidt, and G.~Baumier.
\newblock Efficient simulation of charge transport in deep-trap media.
\newblock In C.~Laroque, J.~Himmelspach, R.~Pasupathy, O.~Rose, and A.~M.
  Uhrmacher, editors, {\em Proceedings of the 2012 Winter Simulation
  Conference}, 2012.

\bibitem{brix}
A.~Brix.
\newblock Generalized gamma measures and shot-noise cox processes.
\newblock {\em Advances in Applied Probability}, 31:929--953, 1999.

\bibitem{chan_wood}
G.~Chan and A.~T.~A. Wood.
\newblock Simulation of stationary {G}aussian vector fields.
\newblock {\em Statistics and Computing}, 9:265--268, 1999.

\bibitem{cox_1955}
D.~Cox.
\newblock Some statistical methods connected with series of events.
\newblock {\em Journal of the Royal Statistical Society, B}, 17(24):129--164,
  1955.

\bibitem{Mathematicalpointprocesses}
D.~J. Daley and D.~Vere-Jones.
\newblock {\em An Introduction to the Theory of Point Processes Volumes 1 and
  2}.
\newblock Springer Series in Statistics, 2003.

\bibitem{Dietrich_Newsam_1997}
C.~R. Dietrich and G.~N. Newsam.
\newblock Fast and exact simulation of stationary {G}aussian processes through
  circulant embedding of the covariance matrix.
\newblock {\em SIAM Journal on Scientific Computing}, 18(4):1088--1107, 1997.

\bibitem{Digglebook}
P.~J. Diggle.
\newblock {\em Statistical Analysis of Spatial Point Patterns}.
\newblock Oxford University Press, London, New York, 2003.

\bibitem{geyerandmoller94}
Charles~J. Geyer and Jesper Møller.
\newblock Simulation procedures and likelihood inference for spatial point
  processes.
\newblock {\em Scandinavian Journal of Statistics}, 21(4):pp. 359--373, 1994.

\bibitem{Gneiting2012}
T.~Gneiting, H.~Seveikova, D.~B. Percival, M.~Schlather, and Y.~Jiang.
\newblock Fast and exact simulation of large {G}aussian lattice systems in
  $\mathbb{R}^2$: Exploring the limits.
\newblock {\em Journal of Computational and Graphical Statistics},
  15(3):483--501, 2006.

\bibitem{golub87}
G.~H. Golub and C.~F.~Van Loan.
\newblock {\em Matrix computations}.
\newblock Johns Hopkins University Press, third edition, 1996.

\bibitem{mcmc:green}
P.~J. Green.
\newblock Reversible jump {{Markov}} chain {Monte Carlo} computation and
  {Bayesian} model determination.
\newblock {\em Biometrika}, 82(4):711--732, 1995.

\bibitem{Karcher13}
W.~Karcher, H.-P. Scheffler, and E.~Spodarev.
\newblock Simulation of infinitely divisible random fields.
\newblock {\em Communications in Statistics -- Simulation and Computation},
  42:215--246, 2013.

\bibitem{Straussprocess1}
F.~P. Kelly and B.~D. Ripley.
\newblock A note on {Strauss}{'}s model for clustering.
\newblock {\em Biometrika}, 63(2):357--360, 1976.

\bibitem{Kendall_Ord_1990}
M.~Kendall and J.~K. Ord.
\newblock {\em Time Series}.
\newblock Oxford University Press, Oxford, third edition, 1990.

\bibitem{handbook_dirk}
D.~P. Kroese, T.~Taimre, and Z.~I. Botev.
\newblock {\em Handbook of Monte Carlo Methods}.
\newblock Wiley, New Jersey, 2011.

\bibitem{lawler}
G.~F. Lawler.
\newblock {\em Introduction to Stochastic Processes}.
\newblock Chapman \& Hall/CRC, FL, 2006.

\bibitem{moller_2003}
J.~M\o ller.
\newblock Shot noise cox process.
\newblock {\em Advances in Applied Probability}, 35(3):614--640, 2003.

\bibitem{Mandelbrot_van_Ness_1968}
B.~B. Mandelbrot and J.~W. van Ness.
\newblock Fractional {Brownian} motions, fractional noises and applications.
\newblock {\em SIAM Review}, 10(4):422--437, 1968.

\bibitem{Matern}
B.~Mat{\'e}rn.
\newblock Spatial variation.
\newblock {\em Meddelanden fr{\aa}n Statens Skogsforskningsinstitut},
  49:1--144, 1960.

\bibitem{Jesper}
J.~M{\o}ller and R.~P. Waagepetersen.
\newblock {\em Statistical Inference and Simulation for Spatial Point
  Processes}.
\newblock Chapman \& Hall/CRC, 2004.

\bibitem{NeymanScott}
J.~Neyman and E.~L. Scott.
\newblock Statistical approach to problems of cosmology.
\newblock {\em Journal of the Royal Statistical Society. Series B
  (Methodological)}, 20(1):1--43, 1958.

\bibitem{Qian98}
H.~Qian, G.~M. Raymond, and J.~B. Bassingthwaighte.
\newblock On two-dimensional fractional {B}rownian motion and fractional
  {B}rownian random field.
\newblock {\em Journal of Physics A: Mathematical and General}, 31(28), 1998.

\bibitem{mcmc:robcas04}
C.~P. Robert and G.~Casella.
\newblock {\em {Monte Carlo} Statistical Methods}.
\newblock Springer-Verlag, New York, second edition, 2004.

\bibitem{rueheld}
H.~Rue and L.~Held.
\newblock {\em {G}aussian Markov Random Fields: Theory and Applications}.
\newblock Chapman \& Hall, London, 2005.

\bibitem{Sato}
K.~Sato.
\newblock {\em L\'evy Processes and Infinitely Divisible Distributions}.
\newblock Cambridge University Press, Cambridge, 1999.

\bibitem{stein02}
M.~L. Stein.
\newblock Fast and exact simulation of fractional {B}rownian motion.
\newblock {\em Journal of Computational and Graphical Statistics},
  11(3):587--599, 2002.

\bibitem{stenzeletal}
O.~Stenzel, D.~P.~Kroese D.~Westhoff, I.~Manke, and V.~Schmidt.
\newblock Graph-based simulated annealing: A hybrid approach to stochastic
  modeling of complex microstructures.
\newblock {\em submitted}, 2012.

\bibitem{Straussprocess2}
D.~J. Strauss.
\newblock Analysing binary lattice data with the nearest-neighbor property.
\newblock {\em Journal of Applied Probability}, 12(4):702--712, 1975.

\bibitem{mcmc:swendsen}
R.~H. Swendsen and J.-S. Wang.
\newblock Nonuniversal critical dynamics in {Monte Carlo} simulations.
\newblock {\em Physical Review Letters}, 58(2):86--88, 1987.

\bibitem{wood_chan}
A.T.A. Wood and G.~Chan.
\newblock Simulation of stationary {G}aussian process in $[0,1]^d$.
\newblock {\em Journal of Computational and Graphical Statistics}, 3:409--432,
  1994.

\end{thebibliography}
\end{document}